\documentclass{aa}  

\usepackage{txfonts}
\usepackage{graphicx, nicefrac}	
\usepackage{amsmath}	
\usepackage{amssymb}	
\usepackage{multirow}
\usepackage{makecell}   

\usepackage[utf8]{inputenc}
\usepackage[export]{adjustbox}
\usepackage{wrapfig}
\usepackage{dutchcal}
\usepackage{braket}
\usepackage{grffile} 
\usepackage{xspace} 
\usepackage{pdflscape}
\usepackage{booktabs}

\usepackage{caption}
\usepackage{subcaption}

\usepackage{longtable}
\usepackage{pdflscape}

\usepackage{ulem}
\usepackage[framemethod=tikz]{mdframed}

\usepackage{tablefootnote}

\graphicspath{ {images/} }
\usepackage[colorlinks=true,linkcolor=blue,allcolors=blue]{hyperref}

\begin{document}

   \title{The miniJPAS \& J-NEP surveys: Identification and characterization of the Ly$\alpha$\xspace Emitter population and the Ly$\alpha$\xspace Luminosity Function at redshift $2.05<z<3.75$}
   
   \titlerunning{miniJPAS\&J-NEP Ly$\alpha$\xspace Emitters}

   \subtitle{}

   \author{
    Alberto Torralba-Torregrosa\inst{\ref{inst1}, \ref{inst2}}\thanks{E-mail: alberto.torralba@uv.es}
    \and Siddhartha Gurung-López\inst{\ref{inst1}, \ref{inst2}}
    \and Pablo Arnalte-Mur\inst{\ref{inst1}, \ref{inst2}}
    \and Daniele Spinoso\inst{\ref{inst7}}
    \and David Izquierdo-Villalba\inst{\ref{inst5}, \ref{inst6}}
    \and Alberto Fernández-Soto\inst{\ref{inst8}, \ref{inst9}}
    \and Raúl Angulo\inst{\ref{inst3}, \ref{inst4}}
    \and Silvia Bonoli\inst{\ref{inst3}, \ref{inst4}}
    \and Rosa M. González Delgado\inst{\ref{inst11}}
    \and Isabel Márquez\inst{{\ref{inst11}}}
    \and Vicent J. Martínez\inst{\ref{inst1}, \ref{inst2}, \ref{inst9}}
    \and P. T. Rahna\inst{\ref{inst15}}
    \and José M. Vílchez\inst{\ref{inst11}}
    \and Raul Abramo\inst{\ref{inst16}}
    \and Jailson Alcaniz\inst{\ref{inst10}}
    \and Narciso Benitez\inst{\ref{inst11}}
    \and Saulo Carneiro\inst{\ref{inst13}}
    \and Javier Cenarro\inst{\ref{inst15}}
    \and David Cristóbal-Hornillos\inst{\ref{inst15}}
    \and Renato Dupke\inst{\ref{inst10}}
    \and Alessandro Ederoclite\inst{\ref{inst15}}
    \and Antonio Hernán-Caballero\inst{\ref{inst15}}
    \and Carlos López-Sanjuan\inst{\ref{inst15}}
    \and Antonio Marín-Franch\inst{\ref{inst15}}
    \and Claudia Mendes de Oliveira\inst{\ref{inst12}}
    \and Mariano Moles\inst{\ref{inst15}}
    \and Laerte Sodré Jr.\inst{\ref{inst12}}
    \and Keith Taylor\inst{\ref{inst14}}
    \and Jesús Varela\inst{\ref{inst15}}
    \and Héctor Vázquez Ramió\inst{\ref{inst15}}
    }

   \institute{
    Observatori Astron\`omic de la Universitat de Val\`encia, Ed. Instituts d’Investigaci\'o, Parc Cient\'ific. C/ Catedr\'atico Jos\'e Beltr\'an, n2, 46980 Paterna, Valencia, Spain\label{inst1}
    \and Departament d’Astronomia i Astrof\'isica, Universitat de Val\`encia, 46100 Burjassot, Spain\label{inst2}
    \and Department of Astronomy, MongManWai Building, Tsinghua University, Beijing 100084, China\label{inst7}
    \and Dipartimento di Fisica ``G. Occhialini'', Universit\`{a} degli Studi di Milano-Bicocca, Piazza della Scienza 3, I-20126 Milano, Italy, Universit\`{a} degli Studi di Milano-Bicocca\label{inst5}
    \and INFN, Sezione di Milano-Bicocca, Piazza della Scienza 3, 20126 Milano, Italy\label{inst6}
    \and Instituto de F\'{\i}sica de Cantabria (CSIC-UC), Avda. Los Castros s/n, 39005 Santander, Spain\label{inst8}
    \and Unidad Asociada ``Grupo de Astrof\'{\i}sica Extragal\'actica y Cosmolog\'{\i}a'', IFCA-CSIC/Universitat de Val\`encia, Val\`encia, Spain\label{inst9}
    \and Donostia International Physics Center. Paseo Manuel de Lardizabal, 4, 20018 Donostia-San Sebasti\'an (Gipuzkoa), Spain\label{inst3}
    \and IKERBASQUE, Basque Foundation for Science, 48013, Bilbao, Spain\label{inst4}
    \and Instituto de Astrofísica de Andalucía - CSIC, Apdo 3004, E-18080, Granada, Spain\label{inst11}
    \and Centro de Estudios de Física del Cosmos de Aragón (CEFCA), Plaza San Juan, 1, E-44001, Teruel, Spain\label{inst15}
    \and Departamento de Física Matemática, Instituto de Física, Universidade de Sa\~o Paulo, Rua do Mat\~ao 1371, CEP 05508-090, S\~ao Paulo, Brazil\label{inst16}
    \and Observatório Nacional, Rua General José Cristino, 77, São Cristóvão, 20921-400, Rio de Janeiro, RJ, Brazil\label{inst10}
    \and Instituto de Física, Universidade Federal da Bahia, 40210-340, Salvador, BA, Brazil\label{inst13}
    \and  Departamento de Astronomia, Instituto de Astronomia, Geofísica e Ciências Atmosféricas, Universidade de São Paulo, São Paulo, Brazil\label{inst12}
    \and Instruments4, 4121 Pembury Place, La Canada Flintridge, CA 91011, U.S.A.\label{inst14}
    }

   \date{Accepted XXX. Received YYY; in original form ZZZ}

 
    \abstract{
    We present the Lyman-$\alpha$ (Ly$\alpha$\xspace) Luminosity Function (LF) at $2.05<z<3.75$, estimated from a sample of 67 Ly$\alpha$-emitter (LAE) candidates in the Javalambre Physics of the Accelerating Universe { Astronomical} Survey (J-PAS) Pathfinder surveys: miniJPAS and J-NEP. These two surveys cover a total effective area of $\sim 1.14$ deg$^2$ with 54 Narrow Band (NB) filters (FWHM $\sim 145$ \AA) across the optical range, with typical limiting magnitudes of $\sim 23$. This set of NBs allows to probe Ly$\alpha$\xspace emission in a wide and continuous range of redshifts. We develop a method for detecting Ly$\alpha$\xspace emission for the estimation of the Ly$\alpha$\xspace LF using the whole J-PAS filter set. We test this method by applying it to the miniJPAS and J-NEP data. In order to compute the corrections needed to estimate the Ly$\alpha$\xspace LF and to test the performance of the candidates selection method, we build mock catalogs. These include representative populations of Ly$\alpha$\xspace Emitters at $1.9<z<4.5$ as well as their expected contaminants, namely low-$z$ galaxies and $z<2$ quasars (QSOs). We show that our method is able to provide the Ly$\alpha$\xspace LF at the intermediate-bright range of luminosity ($43.5\lesssim\log_{10}(L_{\mathrm{Ly}\alpha} / \text{erg\,s$^{-1}$\xspace})\lesssim 44.5$) combining both miniJPAS and J-NEP. The photometric information provided by these surveys suggests that our samples are dominated by bright, Ly$\alpha$-emitting Active Galactic Nuclei (i.e., AGN/QSOs). At $\log_{10}(L_{\mathrm{Ly}\alpha} / \text{erg\,s$^{-1}$\xspace})<44.5$, we fit our Ly$\alpha$\xspace LF to a power-law with slope $A=0.70\pm0.25$. We also fit a Schechter function to our data, obtaining: $\log_{10} (\Phi^* / \text{Mpc$^{-3}$})=-6.30^{+0.48}_{-0.70}$, $\log_{10} (L^*/ \mathrm{erg\,s}^{-1})=44.85^{+0.50}_{-0.32}$, $\alpha=-1.65^{+0.29}_{-0.27}$. Overall, our results confirm the presence of an AGN component at the bright-end of the Ly$\alpha$\xspace LF. In particular, we find no significant contribution of star-forming LAEs to the Ly$\alpha$\xspace LF at $\log_{10}(L_{\mathrm{Ly}\alpha} / \text{erg\,s$^{-1}$\xspace})>43.5$. This work serves as a proof-of-concept for the results that can be obtained with the upcoming data releases of the J-PAS survey.
    }

   \keywords{Methods: observational --
        Quasars: emission lines --
        Galaxies: luminosity function --
        Galaxies: high-redshift --
        Line: identification
        }

   \maketitle
%


\section{Introduction}


The Lyman-$\alpha$ (Ly$\alpha$\xspace) emission line ($\lambda_0=1215.67$ \AA) is among the brightest lines in the UV spectrum of astrophysical sources \citep{Partridge1967, Pritchet1994, VandenBerk2001, Nakajima2018}. Due to its intrinsic strength, Ly$\alpha$\xspace constitutes a fundamental probe for the high-$z$ Universe, allowing us to identify very faint objects at the optical and near-infrared, sometimes even without an explicit detection of the continuum (e.g., \citealt{Bacon2015}). The Ly$\alpha$\xspace line can be seen redshifted into the optical range at $z\sim 2\text{--}7$. Several works have searched for Ly$\alpha$\xspace emission in this range: using blind spectroscopy (e.g., \citealt{Martin2004, Cassata2011, Song2014, Cassata2015, McCarron2022, Liu2022b}), using narrow band (NB) imaging (e.g., \citealt{Cowie1998, Hu1998, Gronwall2007, Ouchi2008, Ciardullo2012, Yamada2012, Shibuya2012, Matthee2015, Santos2016, Konno2018, Ono2021, Santos2021}) and using Integral Field Unit spectroscopy (e.g., \citealt{Blanc2011, Adams2011, Bacon2015, Karman2015, Drake2017}).

One of the main drivers of Ly$\alpha$\xspace surveys is to measure the Ly$\alpha$\xspace luminosity function (LF) over a specific redshift interval. The LF is a statistical measurement of the abundance of Ly$\alpha$\xspace emitters (LAEs), defined as the number density of LAEs per unit comoving volume as a function of the Ly$\alpha$\xspace luminosity ({$L_{\mathrm{Ly}\alpha}$}). Many works have managed to estimate the Ly$\alpha$\xspace LF for different redshift ranges \citep[e.g.,][]{Konno2016, Sobral2018, Spinoso2020, Zhang2021, Liu2022b}. Generally, the observed LAE population is divided in two main kinds of sources: quasi-stellar objects (QSO) with an active galactic nucleus (AGN) and star forming galaxies (SFG). It was found that the population that dominates the low luminosity regime of the Ly$\alpha$\xspace LF ($\log_{10}(L_{\mathrm{Ly}\alpha} / \text{erg\,s$^{-1}$\xspace}) \lesssim 43.5$) is that of SFGs (e.g., \citealt{Guaita2011, Drake2017}). These objects are typically low-mass galaxies with a high star formation rate, low dust content, small rest-frame half-light radius and in general, faint emission lines except Ly$\alpha$\xspace \citep[see e.g.,][]{ArrabalHaro2020,Santos2020}. In SFGs, Ly$\alpha$\xspace emission is produced through recombination processes in the inter-stellar medium (ISM), heated by recent star formation events (e.g., \citealt{Charlot1993, Pritchet1994, ArrabalHaro2020}). Meanwhile the brightest part of the LF ($\log_{10}(L_{\mathrm{Ly}\alpha} / \text{erg\,s$^{-1}$\xspace}) \gtrsim 43.5$) is populated mainly by QSOs, where the recombination processes are triggered by the action of the AGN (e.g., \citealt{Calhau2020}).

Identifying the LAEs population and characterizing its luminosity census is a crucial step in order to understand a multitude of processes in the high-$z$ Universe. SFG LAEs are thought to be analogous to the progenitors of many galaxies that we observe in the nearby Universe, for this reason they provide useful insight about early phases of galaxy evolution (e.g., \citealt{Gawiser2007, Ouchi2010}). Furthermore, at high-$z$, these objects constitute a probe of high-$z$ galaxy clustering and the large structure formation history (e.g., \citealt{Guaita2010, Khostovan2019}). On the other hand, the characterization of the galactic features of LAEs are key to understanding processes such as the AGN fueling and feedback and their effects on star formation (e.g., \citealt{Bridge2013}). In addition, through the study of the fraction of ionizing photons in the ISM of LAEs, it is possible to measure the ionization state of the high-$z$ Universe, shortly after the cosmic {epoch of reionization \citep[EoR, $z\sim 6$--$7$; see e.g.,][]{Bouwens2012, Nakajima2014, Jaskot2014}.}

The evolution of the Ly$\alpha$\xspace LF with redshift is another interesting topic. Past studies claim that the SFG Ly$\alpha$\xspace LF grows substantially from $z\sim 0.3$ up to $z\sim2$--$3$ and remains broadly constant up to $z\sim 6$--$7$. At higher redshift, the SFG Ly$\alpha$\xspace LF shows a strong evolution, resulting in a decrease of the LAE observed number density. This is generally interpreted as an indirect probe of the Universe's reionization progress, since the higher fraction of neutral hydrogen would efficiently absorb Ly$\alpha$\xspace radiation, {hindering the detectability of $z\sim 6$--$7$} SFG LAEs (e.g., \citealt{Malhotra2004, Kashikawa2006, Clement2012, Dijkstra2016, Ning2022}). On the other hand, the Ly$\alpha$\xspace LF of AGN/QSOs shows an evolution compatible with the progress of AGN activity through cosmic history (see e.g., \citealt{Hasinger2005, Miyaji2015, Sobral2018}).

Many works in the last decade have sought to estimate the Ly$\alpha$\xspace LF at different luminosity regimes. The works of \cite{Konno2016, Sobral2017} and \cite{Sobral2018} estimate the Ly$\alpha$\xspace LF at various redshifts, in the faint and intermediate regime using deep NB imaging. In all of these works, it was found that the LF deviates from a Schechter function \citep{Schechter1976} to a power-law like shape for $43.5\lesssim\log_{10}(L_{\mathrm{Ly}\alpha} / \text{erg\,s$^{-1}$\xspace})\lesssim 44.5$. 
The analysis of the X-ray counterparts in \cite{Ouchi2008, Konno2016} and \cite{Sobral2018} revealed that essentially every LAE with $\log_{10}(L_{\mathrm{Ly}\alpha} / \text{erg\,s$^{-1}$\xspace})>43.5$  is associated with X-ray emission, typically interpreted as a signature of AGN activity.

More recently, a few works have explored the brightest end of the Ly$\alpha$\xspace LF. In \cite{Spinoso2020}, the Ly$\alpha$\xspace LF is built from $\sim 10^3$ deg$^2$ of data from the J-PLUS survey \citep{Cenarro2019}, at four redshifts defined by 4 NB-filters ($z=2.24, 2.38, 2.54, 3.23$) and focusing on the bright regime of $\log_{10}(L_{\mathrm{Ly}\alpha} / \text{erg\,s$^{-1}$\xspace})\gtrsim 44.5$. On the other hand, \cite{Zhang2021} combined the spectroscopic data from HETDEX \citep{Gebhardt2021} with the $r$-band images of Subaru/HSC to obtain the Ly$\alpha$\xspace LF, covering $\sim 11.4$ deg$^2$ of sky at $2<z<3$. Later, \cite{Liu2022b} obtained analogous results for the Ly$\alpha$\xspace AGN LF from the spectroscopic QSO sample of HETDEX, over $30.61$ deg$^2$. The latter work showed great agreement with the J-PLUS LF Schechter fit, while covering a wider range of luminosity ($42.3\lesssim\log_{10}(L_{\mathrm{Ly}\alpha} / \text{erg\,s$^{-1}$\xspace})\lesssim 45.9$). While the LAEs sample of \cite{Zhang2021} relies on the $r$-band detection of HSC, in \cite{Liu2022b} the selection is made using purely HETDEX blind spectroscopy, allowing to obtain a more complete sample over a broader area. Overall, the Ly$\alpha$\xspace LF at the full range of luminosity ($41\lesssim\log_{10}(L_{\mathrm{Ly}\alpha} / \text{erg\,s$^{-1}$\xspace})\lesssim 46$) can be well fit by a double Schechter curve, making evident the contributions of both, SFG and AGN/QSO populations (e.g., \citealt{Zhang2021, Spinoso2020}).

In this work, we develop a method for detecting Ly$\alpha$\xspace emission in the photometric data of multi-NB surveys such as the \textit{Javalambre-Physics of the Accelerating Universe { Astronomical Survey}} (J-PAS\footnote{\url{www.j-pas.org}}; \citealt{benitez2014}) and its pathfinder surveys, namely miniJPAS \citep{bonoli2020} and J-NEP \citep{Hernan-Caballero2023}. More specifically, we develop our method on the already-observed fields of miniJPAS and J-NEP, in order to pave the ground for the upcoming J-PAS survey.
{
Multi-NB photometric surveys such as J-PAS allow to perform blind search of LAEs at various redshifts over { a wide} field of observation, with a more efficient selection function than typical spectroscopic surveys \citep[e. g.][]{Zhang2021, Liu2022b}. Furthermore, the availability of a photospectrum { ($\lambda\sim 3750$--$9000$ \AA)} for each source in the survey catalog can be used for contaminant identification.
}

We characterize the performance of our selection method by building mock-survey data and we ultimately build the Ly$\alpha$\xspace LF at $2.05<z<3.75$ { in the medium-bright luminosity range ($43.5\lesssim\log_{10}(L_{\mathrm{Ly}\alpha} / \text{erg\,s$^{-1}$\xspace})\lesssim 45$)}. This range is yet poorly constrained by photometric surveys, due to the small areas probed and {the} limited {number} of NBs {used} \citep[e.g.,][]{Konno2016, Sobral2017, Matthee2017b, Sobral2018, Spinoso2020}.
{ At the same time, Ly$\alpha$ is an excellent tracer for $z>2$ AGNs due to its intrinsic brightness. The estimation of the Ly$\alpha$ LF is motivated by the need to constrain the evolution of the AGN population across cosmic times. For instance, the role played by AGNs at the EoR is still debated. While some works point out that AGNs could be a significant source to the ionizing photon budget \citep[e.g.,][]{Giallongo2015, Dayal2020}, others conclude that AGNs could only make a marginal contribution, in favor of star formation activity \citep[e.g.,][]{Qin2017, Hassan2018}. At lower redshifts, being able to trace the fraction of active galaxies as well as the AGN luminosity distribution is useful to shed light on processes such as AGN-driven feedback and its effect on the host galaxy \citep[e.g.,][]{Brownson2019,mezcua-suh-civano2019,jin-wang-kong2023} or the build-up of scaling relations between active super-massive BHs and their host galaxies \citep[see e.g.,][for a review]{reines-volonteri2015}. Therefore, developing reliable methods to estimate the Ly$\alpha$ LF allows systematic studies of AGN populations since the EoR down to cosmic noon ($z\sim2$). Furthermore, optical multi-NB surveys offer the possibility to perform these analyses in a tomographic fashion across redshift.
}

This paper is structured as follows. In Sect.~\ref{sec:observations} we describe the observations used to obtain the scientific results of this work. In Sect.~\ref{sec:mocks} we define the procedure to build mock catalogs that mimic the observations, which we use to assess the performance of our method. In Sect.~\ref{sec:methods} we explain our LAE candidate selection method, the procedure used to estimate the Ly$\alpha$\xspace LF and present the LAE catalog for miniJPAS\&J-NEP. In Sect.~\ref{sec:results} we present the Ly$\alpha$\xspace LF in different intervals of redshift and estimate the QSO/SFG fraction of our candidates. Finally, Sect.~\ref{sec:summary} summarizes the content of this work.

Throughout this work we use a $\Lambda$CDM cosmology as described by \texttt{PLANCK18}\ \citep{Planck18}, with $\Omega_\Lambda=0.69$, $\Omega_\text{M}=0.31$, $H_0=67.7$ km\,s$^{-1}$\,Mpc$^{-1}$; unless specified otherwise. All the magnitudes are given in the AB system \citep{Oke1983}.

\section{Observations}\label{sec:observations}

\subsection{J-PAS: Javalambre-Physics of the Accelerating Universe Astronomical Survey}

J-PAS is a ground-based survey that will be performed by the JST/T250 telescope at the \textit{Javalambre Astrophysical Observatory} at Teruel (Spain). It is planned to observe $\sim8500$ deg$^2$ of the northern sky via narrow-band imaging with the JPCam instrument. The JPCam is a 1.2 Gpixel multi-CCD camera composed of an array of 14 CCDs, with a field of view of $\sim 4.2$ deg$^2$ (see \citealt{Taylor2014, Marin-Franch2017}).

In the context of this work, the most relevant feature of J-PAS is its filter-set. This set is composed of 54 narrow bands with FWHM of $\sim$145 \AA, covering the optical range of the electromagnetic spectrum. In addition, this filter-set features two medium bands, respectively at the blue and red ends of the optical range, and four broad bands (BBs) equivalent to those used by the SDSS survey: $u$, $g$, $r$ and $i$ \citep{York2000}. These technical features make J-PAS particularly suitable to detect line emitters (e.g., \citealt{Martinez-Solaeche2021, Martinez-Solaeche2022, Iglesias-Paramo2022}). Indeed, the NB set provides a wide and continuous coverage of the optical range ($\sim3\,500$--$10\,000$ \AA), allowing the development of algorithms for photometric source identification (e.g., \citealt{baqui2021, Gonzalez_Delgado2021}), and precise determination of photometric redshifts \citep{Hernan-Caballero2021, Laur2022}. The same filter set was used by the pathfinder surveys miniJPAS and J-NEP.

\subsection{The pathfinder surveys of J-PAS: miniJPAS \& J-NEP}\label{sec:obs_pathfinder}

The miniJPAS survey \citep{bonoli2020} is a scientific project designed to pave the ground for J-PAS data analysis. The observations of miniJPAS were carried out between May and September 2018 using the JPAS-\textit{Pathfinder} camera mounted in the JST/T250. The JPAS-\textit{Pathfinder} camera is an instrument composed of one single CCD with an effective field of view of 0.27\,deg$^2$. The miniJPAS data cover a total of $\sim 1$\,deg$^2$ (effective area after masking 0.895 deg$^2$) of the AEGIS field \citep{Davis2007}, in the northern galactic hemisphere. This field is covered by miniJPAS in 4 pointings (AEGIS001--AEGIS004). AEGIS is a widely studied region of the sky, located within the Extended Groth Strip, for which a plethora of multi-band and spectroscopic observations are available in the literature. For instance, the entirety of the miniJPAS area is covered by the \textit{Sloan Digital Sky Survey} (SDSS; \citealt{Blanton2017}), granting spectroscopic counterparts to many sources in the miniJPAS catalogs. The outcome of miniJPAS serves as a demonstration of the potential of J-PAS and allows us to make a forecast about the results that will be possible to achieve once the survey delivers the first set of data.

J-NEP (for \textit{Javalambre North Ecliptic Pole}) is the second data release obtained using the JST/T250 and the \textit{Pathfinder} camera, covering the \textit{James Webb Space Telescope} North Ecliptic Pole Time-Domain Field (JWST-TDF; \citealt{Jansen2018}). This survey was carried out in a single pointing, with an effective area of $\sim 0.24$ deg$^2$ \citep{Hernan-Caballero2023}. The JWST-TDF will be covered by JWST via a dedicated program in the near future. J-NEP has slightly longer exposure times than miniJPAS, reaching deeper magnitudes.

In Table~\ref{tab:filter_properties}, we list the limiting 5$\sigma$ magnitudes of both surveys for all relevant filters for this work. We use 20 NBs to probe for Ly$\alpha$\xspace emission covering a redshift range of $z=2.05$--$3.75$, the choice of this range is discussed in Sect.~\ref{sec:puricomp1d}. The number in the J-PAS NB names makes reference to the approximate pivot wavelength ($\lambda_\mathrm{pivot}$) in nanometers. 

Throughout this work we use the dual mode catalogs of miniJPAS and J-NEP described in \cite{bonoli2020} and \cite{Hernan-Caballero2023}, respectively. These catalogs are generated using the  ``dual-mode'' of the \texttt{SExtractor} code { \citep{Bertin1996}}. In this operating mode, \texttt{SExtractor} performs a first source-detection on a specific band ($r$ for J-PAS, since this is the deepest among the BBs). Then, the positions of these $r$-band detected sources is used to perform forced-photometry in the images obtained with all the remaining filters. We use the 3\arcsec\xspace forced aperture photometry fluxes and magnitudes. { The PSF FWHM of the miniJPAS\&J-NEP images varies in the range 0.6--2\arcsec}. The total effective area, after masking, combining miniJPAS and J-NEP is 1.14 $\deg^2$.

\begin{table}
    \centering
    \caption{Limit 5$\sigma$ magnitudes of the BBs and NBs used to select LAEs.}
    \label{tab:filter_properties}
    \resizebox{\linewidth}{!}{
    \begin{tabular}{@{\extracolsep{-4pt}}ccccccc}
    \toprule
        \multirow{3}{*}{Filter} & \multicolumn{5}{c}{$5\sigma$ limit magnitudes (magAB)} & \multirow{3}{*}{$\Delta z$ (Ly$\alpha$\xspace)} \\
         & \multicolumn{4}{c}{miniJPAS} & \multirow{2}{*}{J-NEP} &  \\
         \cline{2-5}
         & \thead{AEGIS\\001} & \thead{AEGIS\\002} & \thead{AEGIS\\003} & \thead{AEGIS\\004} & & \\
         \midrule
        J0378 & 23.04 & 23.32 & 22.91 & 22.66 & 22.64 & 2.05-2.18 \\
        J0390 & 24.24 & 23.86 & 23.71 & 23.72 & 23.05 & 2.15-2.27 \\
        J0400 & 23.74 & 23.38 & 23.19 & 23.35 & 23.53 & 2.23-2.35 \\
        J0410 & 23.03 & 23.02 & 23.33 & 22.57 & 23.68 & 2.32-2.44 \\
        J0420 & 23.12 & 22.69 & 22.53 & 22.38 & 23.34 & 2.40-2.52 \\
        J0430 & 23.88 & 23.33 & 23.12 & 23.33 & 23.59 & 2.48-2.60 \\
        J0440 & 23.72 & 23.40 & 23.56 & 23.79 & 23.96 & 2.56-2.68 \\
        J0450 & 22.46 & 22.36 & 22.04 & 22.44 & 22.98 & 2.64-2.77 \\
        J0460 & 24.09 & 23.84 & 23.80 & 24.07 & 22.99 & 2.73-2.85 \\
        J0470 & 23.62 & 23.43 & 23.27 & 23.50 & 23.64 & 2.81-2.93 \\
        J0480 & 23.37 & 22.69 & 23.34 & 23.26 & 23.78 & 2.89-3.01 \\
        J0490 & 23.09 & 22.69 & 22.47 & 22.46 & 23.38 & 2.97-3.10 \\
        J0500 & 23.36 & 23.22 & 23.01 & 23.27 & 23.62 & 3.05-3.18 \\
        J0510 & 23.60 & 23.31 & 23.44 & 23.56 & 23.89 & 3.13-3.25 \\
        J0520 & 22.44 & 22.41 & 22.38 & 22.49 & 23.04 & 3.22-3.34 \\
        J0530 & 23.94 & 23.53 & 23.38 & 23.55 & 22.16 & 3.29-3.42 \\
        J0540 & 23.22 & 23.19 & 23.06 & 23.01 & 23.48 & 3.37-3.50 \\
        J0550 & 23.09 & 22.75 & 23.09 & 22.97 & 23.57 & 3.46-3.58 \\
        J0560 & 22.93 & 22.33 & 22.22 & 22.18 & 22.86 & 3.54-3.66 \\
        J0570 & 22.96 & 22.82 & 22.53 & 22.86 & 23.14 & 3.63-3.75 \\
        uJPAS & 23.00 & 22.96 & 22.78 & 22.66 & 22.68 & - \\
        gSDSS & 23.99 & 24.04 & 24.04 & 23.97 & 24.64 & - \\
        rSDSS & 24.01 & 23.82 & 23.78 & 23.91 & 24.33 & - \\
        iSDSS & 23.02 & 23.14 & 23.28 & 23.42 & 23.53 & - \\
    \bottomrule
    \end{tabular}
    }
    \tablefoot{The values are given for all four pointings of miniJPAS and the single J-NEP field. These values correspond to 3\arcsec\xspace aperture photometry. The last column shows the corresponding Ly$\alpha$\xspace redshift interval of the NBs.}
\end{table}

\section{Mock catalogs}\label{sec:mocks}

{
Given our wavelength coverage (3\,700\AA--5\,700\AA) our sample is prone to be contaminated by sources with prominent emission lines other than Ly$\alpha$\xspace , such as \ion{C}{IV} ($\lambda\,1549$ \AA), \ion{C}{III}] ($\lambda\,1908$ \AA), \ion{Mg}{II} ($\lambda\,2799$ \AA) and \ion{Si}{IV} ($\lambda\,1397$ \AA) AGN lines; and galactic emission lines associated to star formation at low-z such as H$_\beta$ ($\lambda\,4861$ \AA), [\ion{O}{III}] ($\lambda\lambda\,4959,5007$ \AA) and [\ion{O}{II}] ($\lambda\lambda\,3727,3729$ \AA).
We have designed mock catalogs of LAEs and its main contaminants in order to estimate completeness and purity of our selection methodology as well as the uncertainty on our measured Ly$\alpha$\xspace LF. 
}

\subsection{Characterization of the photometry flux uncertainties}\label{sec:mock_errors}

\begin{figure}
    \centering
    \resizebox{\hsize}{!}{\includegraphics{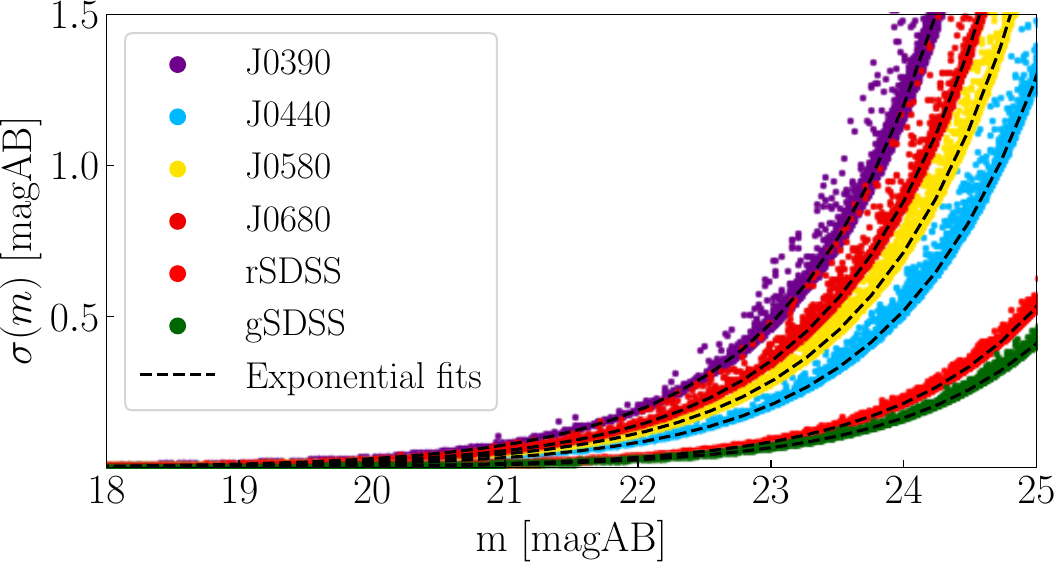}}
    \caption{Exponential fits of magnitude errors according to Eq.~\ref{eq:magerr}. For brevity, we show the magnitudes of 4 NBs and 2 BBs (colored dots) of J-NEP and their corresponding fits (dashed black lines). We get highly similar results for every filter in J-NEP and all pointings of miniJPAS.}
    \label{fig:mag_fit_err}
\end{figure}

The first step in building the mock catalogs is to characterize the photometric uncertainty distribution of the survey we want to emulate. We assume that the distribution of measured magnitude errors ($\sigma[m]$) in each observed pointing can be modeled as a simple exponential:
\begin{equation}
    \sigma[m] = A \cdot \exp\left[B \cdot m + C\right]\,.
    \label{eq:magerr}
\end{equation}
We perform a fit for the parameters $A$, $B$ and $C$ for every NB in every pointing of miniJPAS\&J-NEP. Following this fit, we add Gaussian uncertainties in magnitude to our mock objects in order to mimic the JPAS-Pathfinder observations. Next, all magnitudes are converted to fluxes ($f^\lambda$). The bands with a flux below the 5$\sigma$ limiting flux of that band $f^\lambda_{5\sigma}$, are assigned an error equal to $f^\lambda_{5\sigma}/5$. The reason for this is that assuming Gaussian magnitude uncertainties is only valid for $m\lesssim m_{5\sigma}$, as some parts of many of the synthetic spectra have fluxes compatible with zero, and the Gaussian approximation of the magnitude errors is no longer valid due to the logarithmic nature of the magnitude system. A few examples of these fits can be found in Fig.~\ref{fig:mag_fit_err}. We generate five versions of our mocks, each one with the uncertainty distribution corresponding to each field of miniJPAS\&J-NEP.

\subsection{Star Forming Galaxies mock}\label{sec:SFmock}

In order to reproduce the population of SFGs, we generate a set of synthetic spectra of galaxies. We use the stellar population models from \cite{bruzual2003}, where the stellar continuum of a galaxy is described by three parameters: metallicity, age and extinction (\texttt{MET}, \texttt{AGE} and \texttt{EXT}, respectively). In order to generate a realistic LAEs population, we start from a sample of 397 spectra of LAEs at $2 < z < 5$ from the VIMOS VLT Deep Survey (VVDS) and VIMOS Ultra-Deep Survey (VUDS) \citep{Cassata2011, Cassata2015}. We convert all the spectra to the rest-frame, using the spectroscopic redshifts, then stack them to obtain a composite spectrum. We fit the stacked spectrum to a grid of templates in \texttt{MET}, \texttt{AGE} and \texttt{EXT} using a \textit{Markov Chain Monte Carlo} algorithm (MCMC). \footnote{For the fit we use the Python package \texttt{emcee} \citep{Foreman-Mackey2013}.} The positions of the walkers in the final steps of the chain in the parameter space describe a disperse distribution of the most likely combinations of \texttt{MET}, \texttt{AGE} and \texttt{EXT} to reproduce the continuum of a SFG LAE. First we use the triplets of parameters sampled from this distribution to interpolate the \cite{bruzual2003} templates and generate the normalized spectral continua of our mock catalog of SFG LAEs.

In a second step, we add the Ly$\alpha$\xspace emission lines to the spectra, following the expected distributions for $L_{\mathrm{Ly}\alpha}$ and equivalent width. We sample values of $L_{\mathrm{Ly}\alpha}$ from the best Schechter fit in \cite{Sobral2018}: $\log_{10} (\Phi^* / \text{Mpc$^{-3}$})=-3.45$, $\log_{10} (L^*/ \mathrm{erg\,s}^{-1})=42.93$, $\alpha=-1.93$. The Ly$\alpha$\xspace LF has been proven to show little variation with redshift at $z=2.5$--$7$ \citep{Sobral2017, Drake2017, Sobral2018, ouchi2020}. In Appendix~\ref{sec:prior_LF_choice} we discuss the effect of assuming this prior LF at $z=2$--$2.5$. We sample values of Ly$\alpha$\xspace EW$_0$ from an exponential distribution,
\begin{equation}
    \rm N[EW_0]=N_0\cdot\exp[EW_0/\alpha]\,,
    \label{equation:EW_exponential_model}
\end{equation}
where $N_0$ is a normalizing factor and $\alpha=129$ \AA\ (see \citealt{Zheng2013, Santos2020, Kerutt2022}). The fluxes of each object are re-scaled applying a multiplicative factor so that the integrated line flux $F_\text{Ly$\alpha$\xspace}$ and the observed equivalent width (EW) follow the definition:
\begin{equation}
    \mathrm{EW} = \int\frac{f^\lambda_{\mathrm{Ly}\alpha}}{f^\lambda_\mathrm{cont}}\text{d}\lambda \approx F_{\mathrm{Ly}\alpha} / f^\lambda_\mathrm{cont.}\,,
    \label{eq:ew}
\end{equation}
where $f^\lambda_{\mathrm{Ly}\alpha}$ and $f^\lambda_\mathrm{cont.}$ are the flux densities of the Ly$\alpha$\xspace line and the continuum, respectively. The approximation at the rightmost part of Eq.~\ref{eq:ew} assumes a flat continuum over the width of the emission line. The relation between the observed EW and the rest-frame equivalent width is $\text{EW}=\text{EW}_0\cdot (1 + z)$. The redshift values are sampled from a distribution within $z\in[1.9, 4.5]$ such as the number density per unit volume is constant. Finally, the Ly$\alpha$\xspace line is added as a gaussian profile with $\sigma=5$ \AA\ \citep[see e.g.,][]{Gurung-Lopez2022, McCarron2022, Davis2023} and the adequate integrated flux to match the required $L_{\mathrm{Ly}\alpha}$.

The result is a sample of synthetic spectra of SFG LAEs at $2<z<4$ that mimics the Ly$\alpha$\xspace LF measured by \cite{Sobral2018} over 400 deg$^2$. All these spectra are convolved with the transmission curves of the J-PAS filters in order to obtain a mock catalog of fluxes. Then, the uncertainties are added as detailed in Sect.~\ref{sec:mock_errors}.

\subsection{QSO mock}\label{sec:QSO_mock}

\begin{figure}
    \centering
    \resizebox{\hsize}{!}{\includegraphics{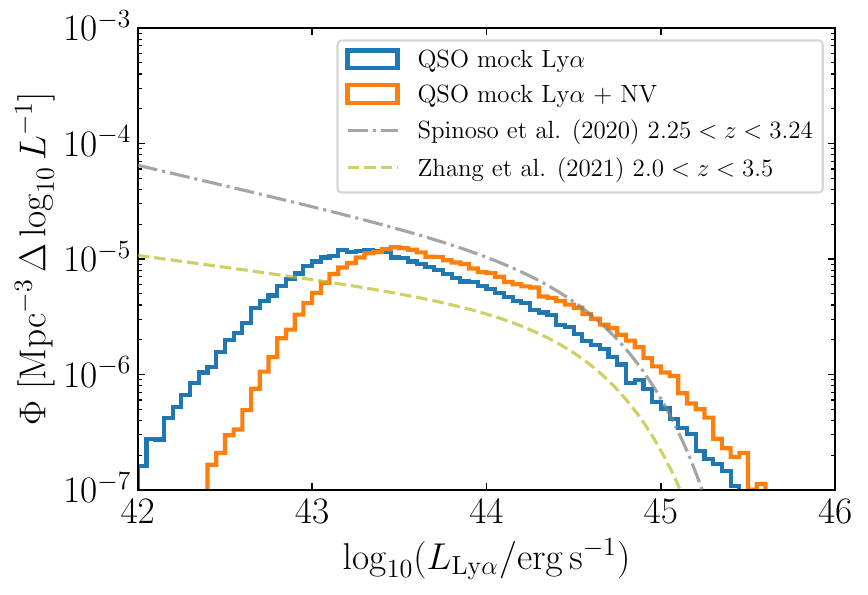}}
    \resizebox{\hsize}{!}{\includegraphics{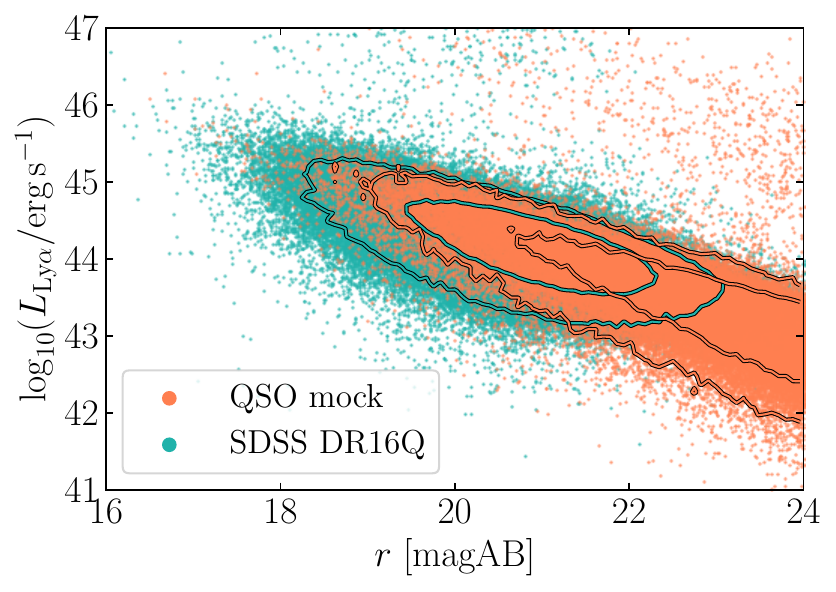}}
    \caption{Basic properties of the QSO mock. Top: Output Ly$\alpha$\xspace LF of the QSO mock at $z>2$. The measurement of the Ly$\alpha$\xspace line { can often be} affected by the presence of the \ion{N}{V} line, for this reason we also show the \ion{Ly}{$\alpha$}+\ion{N}{V} LF. The bias correction in the $L_{\mathrm{Ly}\alpha}$ measurement caused by \ion{N}{V} is addressed in Sect.~\ref{sec:Llya_comp}. We compare with the fits for the AGN/QSO Ly$\alpha$\xspace LF in \cite{Spinoso2020} and \cite{Zhang2021}, using a Schechter and a double power-law, respectively. Bottom: $L_{\mathrm{Ly}\alpha}$ as a function of $r$ magnitude for the sources in the QSO mock (orange) and the QSOs of SDSS DR16Q with a $L_{\mathrm{Ly}\alpha}$ S/N > 5 (blue). The contour lines mark the regions containing the 68\% and 95\% of the objects.
    }
    \label{fig:mock_L_dist}
\end{figure}

For the construction of our QSO mock we follow a very similar procedure to that used in \citealt{Queiroz2022}. In their work, they provide mock catalogs of QSOs ($0 <z<4.2$), morphologically point-like galaxies and stars for miniJPAS, based on the SDSS DR12Q Superset \citep{Paris2017}. We build a new QSO mock catalog following \cite{Queiroz2022} instead of using the already available mock for various reasons. In the first place, our mock needs to accurately represent the $L_{\mathrm{Ly}\alpha}$ distribution of the QSO population at $z>2$. Secondly, we add the flux uncertainties according to Eq.~\ref{eq:magerr} in order to be consistent with the rest of the populations in our mocks. Finally, we need to substantially increase the size of the mock sample in order to obtain significant statistics, as explained below.

Our aim is to generate a set of QSOs with redshifts $z=0$--$4.5$. For this, we use spectra from the SDSS DR16Q Superset \citep{Lyke2020}. We select all sources with good median signal-to-noise over all pixels (\texttt{SN\_MEDIAN\_ALL}>0), so we can neglect the errors of the spectroscopy when performing the synthetic photometry; no redshift warning flags (\texttt{ZWARNING}=0); and classified as QSO by the SDSS pipeline (\texttt{IS\_QSO\_FINAL}=1). We sample values of $z$ and $r$ magnitude from the 2D PLE+LEDE model in \cite{Palanque-Delabrouille2016}. This model predicts the number counts of detected QSOs in a photometric survey as a function of magnitude and redshift per unit area. We compute the total number of objects to include in the mock by integrating the \cite{Palanque-Delabrouille2016} model over an area of 400 deg$^2$. For QSOs with $\log_{10}(L_{\mathrm{Ly}\alpha} / \text{erg\,s$^{-1}$\xspace}) > 44$, due to the exponential drop of sources at this luminosity, we use a 10 times bigger area for better statistics. For each pair of values ($z$, $r$), a source is selected randomly from the SDSS DR16Q within a redshift interval smaller than 0.06, then the spectral flux is corrected by a multiplicative factor in order to match the sampled value of $r$.

As shown in the top panel of Fig.~\ref{fig:mock_L_dist}, { the} resulting QSO mock yields a $2<z<4$ distribution in line with a Schechter function for Ly$\alpha$\xspace line luminosity. The depth of SDSS is lower than that of miniJPAS and J-NEP, and their catalog is only complete up to $r\sim 20.5$. Hence, to obtain sources up to $r=24$, we need to largely correct some objects under the assumption of a weak dependecy of the QSO properties with luminosity (for a similar procedure and discussion see \citealt{Abramo2012, Queiroz2022}). The bottom panel of Fig.~\ref{fig:mock_L_dist} shows that the distribution of $L_{\mathrm{Ly}\alpha}$ of the QSO mock extrapolates reasonably at $r\gtrsim 22$, far out of the range of SDSS.

\subsection{Low-$z$ Galaxy mock}

As stated at the beginning of Sect.~\ref{sec:mocks}, a significant part of the contaminants are expected to be low-z galaxies ($z\sim0$--$1$) with prominent emission lines, especially at the faintest regime of the Ly$\alpha$\xspace LF. To reproduce this population, we generate a synthetic miniJPAS observation, analogous to that designed by \cite{Izquierdo-Villalba2019} for the J-PLUS survey. In our case, the mock-observation is built over a total area of 3 deg$^2$, by employing the \texttt{L-Galaxies} semi-analytic model \citep[][]{Guo2011,henriques2015} to predict the continuum features of galaxies over the halos of the \texttt{Millennium} N-body dark matter simulation \citep{Springel2005}, selecting the line of sight oriented at RA = 58.9 deg, DEC = 56.3 deg. This simulated observation directly produces synthetic photometry of all the 60 miniJPAS filters for a catalog of 144\,183 galaxies with magnitude $r<24$ and $z<4$ ($\sim90\%$ of which are at $z<1$). For simplicity, hereafter we refer to this mock-observation as ``lightcone''.

 The nebular emission lines of the galaxies in the lightcone are computed using the method of \cite{Orsi2014}, which employs the \cite{Levesque2010} model for \ion{H}{II} regions in order to compute the output line fluxes of simulated galaxy spectra. Several emission lines are considered, including the potential interlopers of a LAEs sample (e.g., \ion{H}{$\beta$}, [\ion{O}{III}] and [\ion{O}{II}], as stated in Sect.~\ref{sec:mocks}). These line fluxes are corrected with an empirical dust attenuation model in order to reproduce the \ion{H}{$\alpha$}, \ion{H}{$\beta$}, [\ion{O}{II}] and [\ion{O}{III}] luminosity functions from several observations. This dust model performs well for almost every galactic line for a wide range of redshift. However, as discussed in \cite{Izquierdo-Villalba2019}, this dust model tends to overcorrect the line flux in the case of [\ion{O}{II}] for $z\leq 0.5$, the interval in which this line has particular relevance for our work. In order to avoid underestimating the fraction of contaminants, we remove the dust attenuation coefficient from the [\ion{O}{II}] lines in our mock for $z<0.5$. By doing so, the [\ion{O}{II}] LF is better reproduced in the lightcone, for this specific redshift interval.

\section{Methods}\label{sec:methods}

In this section, we describe our methods to obtain a LAE sample in miniJPAS\&J-NEP and estimate the Ly$\alpha$\xspace LF. The parameters used in this pipeline were chosen to optimize the selection of LAEs, after extensive testing using the mocks described in Sect.~\ref{sec:mocks}. In this section we also characterize the candidate sample obtained from the observations and provide the catalog of LAEs.

\subsection{Candidate selection method}\label{sec:selection_method}

In this subsection, we describe the procedure we use to select sources from the miniJPAS\&J-NEP catalogs, and classify them as LAEs. In the first place, in Sect.~\ref{sec:parent_sample} we define a parent sample from which to perform the selection. Second, in Sect.~\ref{sec:contest} we explain how the continuum flux is computed. Finally, in Sect.~\ref{sec:nbexsel} we enumerate the criteria for selecting candidates based on NB flux excess with respect to the continuum.

\subsubsection{Parent sample}\label{sec:parent_sample}

Our LAE candidate selection is based on the dual mode catalogs of miniJPAS and the J-NEP field (see Sect.~\ref{sec:obs_pathfinder}). We remove every source flagged by the catalog masks. The masks cover the window frames, artifacts, bright stars and objects near them. We also remove objects marked with \texttt{SExtractor} photometry flags. After this first cut, we are left with a total of 63\,923 objects: 46\,477 in miniJPAS and 17\,446 in J-NEP.

We continue the preliminary cuts by requiring $17\leq r\leq 24$. Sources fainter than this threshold may have very low signal-to-noise to be classified reliably; { $r\sim 24$ is the 5$\sigma$ detection limit for miniJPAS\&J-NEP (see Table \ref{tab:filter_properties})}. {On the other hand, we expect the number of LAEs at $z\geq 2$ with magnitudes brighter than $r=17$ to be very low (see Fig.~\ref{fig:mock_L_dist}). At these bright magnitudes, the number counts will be dominated by stars \citep[see, e.g., Fig.~18 of][]{bonoli2020}. For this reason, we remove these bright sources which are very likely to be stars. In any case, the exact value for this bright cut is somehow arbitrary.}

Objects showing significant proper motion or parallax are likely to be stars. We remove these objects making use of the cross-match tables of the miniJPAS\&J-NEP dual mode catalogs with the the Gaia survey Early Data Release 3 (EDR3) \citep{Brown2021}. Among all the non-flagged sources in the miniJPAS\&J-NEP dual catalogs, only 2\,739 (4.3\%) have a counterpart in Gaia EDR3. In the spectroscopic follow up program of \cite{Spinoso2020}, it was found that stars constituted a non-negligible part of the NB emitters sample from J-PLUS. Therefore, we remove secure stars following \citealt{Spinoso2020}, imposing
\begin{equation}
    \sqrt{\sigma^2_\mathrm{pmdec} + \sigma^2_\mathrm{pmra} + \sigma^2_\mu} < \sqrt{3^2 + 3^2 + 3^2}\,,
\end{equation}
where $\sigma_\mathrm{pmdec}$, $\sigma_\mathrm{pmra}$ and $\sigma_\mu$ are the relative errors of the proper motion in declination and right ascension and parallax, respectively.

After these cuts the dual-mode catalogs we are left with  36\,026 sources in total (28\,447 in miniJPAS and 7\,549 in J-NEP). This constitutes our starting sample for the selection of LAE candidates.

\subsubsection{Continuum estimation}\label{sec:contest}

In order to find emission lines within the sources of the miniJPAS\&J-NEP catalogs, we look for NBs with a reliable flux excess with respect to the continuum flux at the central wavelength of those NBs. The continuum flux density can be estimated using the information from the filters near the narrow band of interest. In particular, for a given NB filter $n$, we compute the continuum estimate $f^\lambda_{\rm cont}$ by considering an equal number $k$ of NBs both at bluer and redder wavelengths than $n$. We obtain $f^\lambda_{\rm cont}$ as the weighted average of this set of $2k$ NBs, after excluding the two NBs directly adjacent to $n$, on each side. The reason for excluding these two NBs is that emission lines can be broad enough to be detected in more than one NB at a time, as in many cases of QSO's \ion{Ly}{$\alpha$} lines (see e.g., \citealt{Greig2016}). Narrow emission lines can also contribute to more than one NB due to the overlap of the transmission curves of the J-PAS filters.

We chose to set $k=6$, so that our $f^\lambda_{\rm cont}$ estimate is based on 12 NBs. With this number of NBs, we cover the widest wavelength range possible without contamination of other luminous lines near Ly$\alpha$\xspace (\ion{O}{VI}+\ion{Ly}{$\beta$} and \ion{C}{IV}). We underline that the 7 NBs at the bluest-end of the miniJPAS filter set do not have enough NBs on their bluer side. In these cases we still use the same computation described before, but only with the available filters; this leads to a bias in the line luminosity estimation that will be corrected later on, as detailed in Sect.~\ref{sec:Llya_comp}.

We note that by estimating $f^\lambda_{\rm cont}$ as the average flux around the wavelength of the emission line we are implicitly assuming that the continuum has an anti-symmetric shape with respect to that wavelength. However, at bluer wavelengths than the observed $\lambda_{\mathrm{Ly}\alpha}$ the effect of the \textit{Lyman-alpha forest} comes into play. The Ly$\alpha$\xspace forest is a series of absorption lines caused by the scattering of the Ly$\alpha$\xspace photons by neutral hydrogen in the inter-galactic medium (IGM) \citep[see e.g.,][]{Weinberg2003, Gurung-Lopez2020}. The Ly$\alpha$\xspace forest cannot be resolved through NB photometry, but its overall effect is a significant attenuation of the measured flux in a given band. The effective transmission of the IGM due to the Ly$\alpha$\xspace forest can be approximated with an exponential law,
\begin{equation}
T_\mathrm{IGM}\,[\lambda_\mathrm{obs}]=
\begin{cases}
    \exp\left[{a\cdot\left(\frac{\lambda_\mathrm{obs}}{\lambda_{\mathrm{Ly}\alpha}}\right)^b}\right],& \lambda_\mathrm{obs}\leq\lambda_{\mathrm{Ly}\alpha}\\
    1,& \lambda_\mathrm{obs}>\lambda_{\mathrm{Ly}\alpha}
\end{cases}\,,
\end{equation}
with $a=-0.001845$ and $b=3.924$, as found by \cite{Faucher-Giguere2008}. Having this in mind, we can compensate the attenuation due to the \ion{Ly}{$\alpha$} forest on our continuum estimate. We do this for each of the NBs $i$ which are at bluer wavelengths than $n$. In particular, we divide the flux $f_i$ by the IGM transmission computed at the central wavelength of the $i$-th NB ($\lambda_i$). That is: $t_i=T_\mathrm{IGM}(\lambda_i)$. Then, the continuum flux density is estimated as
\begin{equation}
    f^\lambda_\mathrm{cont}=\frac{\sum\limits_{i}f^\lambda_{i}/t_i\cdot\sigma_i^{-2}}{\sum\limits_{i}\sigma_i^{-2}}\,,
    \label{eq:fcontIGM}
\end{equation}
where $f^\lambda_{i}$ is the flux of the $i^\mathrm{th}$ NB, $\sigma_i$ its associated uncertainty and $i\in\left[n-k-1,\,\dots,\,n-2,\,n+2,\,\dots,\,n+k+1\right]$.

Correcting for the average IGM transmission allows us to both: i) improve our continuum estimate and reduce the bias on our Ly$\alpha$\xspace luminosity estimate and ii) discard low-z contaminants from our selection. Indeed, the latter do not suffer from the Ly$\alpha$\xspace forest effect, therefore our correction produces an artificial over-estimation of their continua. This translates into an under-estimate of their measured EW, pushing these sources out of our selection cut.

\subsubsection{LAE candidate selection criteria}\label{sec:nbexsel}

\begin{figure}
    \centering
    \resizebox{\hsize}{!}{\includegraphics{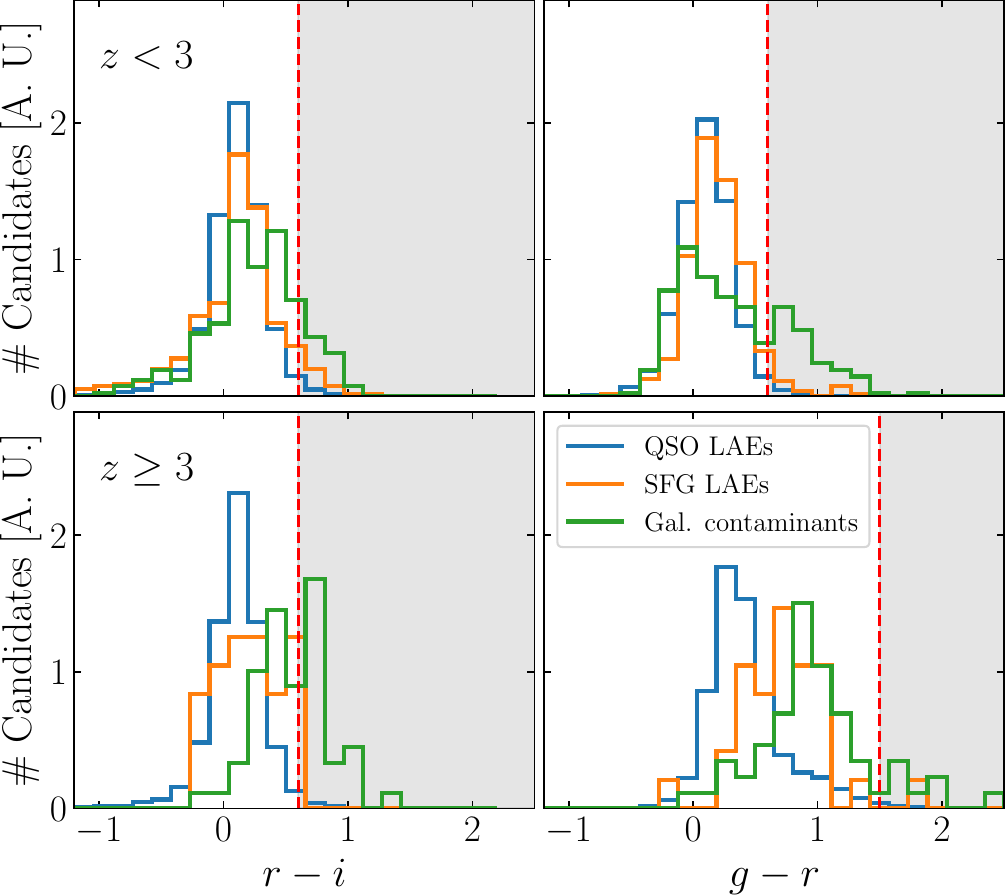}}
    \caption{Color distributions of the selected candidates in our mock before applying the color cut. The objects in the shaded area are removed from the sample after applying the cut. Through these BB color cuts we remove objects showing significant red colors, which are likely to be nearby galaxies.}
    \label{fig:color_hists}
\end{figure}

After the estimation of the continuum for every source at the central wavelength of every NB, we check every filter for a reliable excess that is compatible with a Ly$\alpha$\xspace emission line. The criteria of this selection are the following:
\begin{itemize}
    \item \textbf{3$\sigma$ flux excess:} The NB flux density, $f^\lambda_\mathrm{NB}$, must show an excess with respect to the continuum $f^\lambda_\mathrm{cont}$ larger than a 3$\sigma$ confidence interval (see e.g., \citealt{Bunker1995, Fujita2003, Sobral2009, Bayliss2011}). That is:
    \begin{equation}
        f^\lambda_\mathrm{NB} - f^\lambda_\mathrm{cont} > 3 \cdot \sqrt{\sigma_\mathrm{NB}^2 + \sigma_\mathrm{cont}^2}\,,
    \end{equation}
    where $\sigma_\text{NB}$ and $\sigma_\text{cont}$ are the uncertainties of the NB and continuum fluxes. When multiple NBs satisfy this condition in one source (either in adjacent or non-contiguous NBs), we consider as a candidate Ly$\alpha$\xspace emission only the NB with the highest measured flux. { We adopt this criterion under the assumption that Ly$\alpha$ is the most luminous line in the optical range for $z>2$ QSOs, and the only relevant line in the case of SFGs.}
    Then, we assign a redshift $z_\mathrm{NB}$ assuming $\lambda_\mathrm{pivot}$ of the detection NB as the observed Ly$\alpha$\xspace wavelength.
    
    \item \textbf{Minimum S/N:} In addition to the NB-excess significance, we impose a minimum signal-to-noise ratio of S/N $>6$ for the NB where we identified the line detection. This ensures that the photometry in the selected filter is clean and reliable. Lowering this threshold significantly increases the number of spurious detections due to random fluctuations of the photometric fluxes.
    
    \item \textbf{EW$_0$ cut:} \ion{Ly}{$\alpha$} has a large intrinsic EW$_0$ in comparison to other galactic emission lines \citep{VandenBerk2001, Nakajima2018}. Many past works have imposed a minimum EW$_0$ in order to reduce the number of contaminants (e.g., \citealt{Fujita2003, Gronwall2007, Ouchi2008, Santos2016, Sobral2018, Spinoso2020}). Following these approaches, we impose: EW$_0>$ EW$_0^\mathrm{min}$. From the definition of EW  we can derive
    \begin{equation}
        \frac{f^\lambda_\mathrm{NB}}{f^\lambda_\mathrm{cont}} > 1 + \frac{(1+z_\text{NB})\cdot\mathrm{EW}_0^\mathrm{min}}{\mathrm{FWHM}_\mathrm{NB}}\,,
    \end{equation}
    where $z_\text{NB}$ is the Ly$\alpha$\xspace redshift associated with the selected NB. We choose a cut at EW$_0^\mathrm{min}=30$ \AA . Lowering the value of EW$_0^\mathrm{min}$ significantly increases contamination without a meaningful increase in completeness.

    \item \textbf{Multiple line combinations:} Some sources of our catalog show multiple NB excesses compatible with emission lines. The ratios between the observed wavelengths of the multiple lines in a given source can be used to identify contaminants or to confirm true positive LAE detections. Indeed, SFG LAEs are not expected to show relevant line emission features other than \ion{Ly}{$\alpha$} in the rest-frame UV \citep{Nakajima2018}. On the other hand, QSOs are likely to present extra emission lines which can only appear in specific combinations. After the Ly$\alpha$\xspace line search, we check for other NBs with 5$\sigma$ significant excesses, with an observed equivalent width EW$_\mathrm{obs} > 100$ \AA. For the detection of these additional lines we estimate the spectral continuum without applying the IGM correction, which is only correct assuming the position of a Ly$\alpha$\xspace line. In particular, we check if these additional flux excesses are compatible with the most prominent QSO lines: \ion{O}{VI}, \ion{Si}{IV}, \ion{C}{IV}, \ion{C}{III}] or \ion{Mg}{II} (see, e.g., \citealt{Matthee2017b, Spinoso2020}). The sources showing multiple NB excesses which do not follow a compatible QSO emission pattern are discarded from our LAE candidate sample.
    
    \item \textbf{Color cuts:} In most cases, the continuum of both QSO and SFG LAEs can be well fitted by a power law \citep{VandenBerk2001, Nakajima2018}. Therefore, as shown in Fig.~\ref{fig:color_hists}, both classes of LAEs are likely to present bluer broad-band colours than the low-$z$ galaxy contaminants.
    This allows us to define a set of colour cuts to remove part of these contaminants. We keep only sources with $r - i < 0.6 \wedge g - r < 0.6$ when selecting candidates with $z_\text{\ion{Ly}{$\alpha$}}<3$. At $z_\text{\ion{Ly}{$\alpha$}}>3$ the Ly$\alpha$\xspace \textit{forest} affects the flux of the $g$ band and this propagates into the expected colors. The color cut at $z_\text{\ion{Ly}{$\alpha$}}>3$ is defined as  $r - i < 0.6 \wedge g - r < 1.5$. However, if a source has multiple line detection compatible with QSO lines, this color cut does not apply and the object is classified as a true QSO LAE.

\end{itemize}

\subsection{Ly$\alpha$\xspace luminosity estimation}\label{sec:Llya_comp}

The flux of a NB selected as a Ly$\alpha$\xspace emission-line candidate contains the contribution of both the line flux and the continuum. Therefore, we estimate the Ly$\alpha$\xspace integrated flux as:
\begin{equation}
    F_{\mathrm{Ly}\alpha} \approx \left(f^\lambda_\mathrm{NB} - f^\lambda_\mathrm{cont}\right) \cdot \text{FWHM}_\text{NB}\,.
\end{equation}
This equation implicitly assumes that the NB transmission curve can be reasonably approximated by a squared top-hat filter \citep[as in the case of the J-PAS NBs, see][]{bonoli2020}. Then, the Ly$\alpha$\xspace luminosity is obtained as
\begin{equation}
    L_{\mathrm{Ly}\alpha} = F_{\mathrm{Ly}\alpha} \cdot 4\pi d_\text{L}^2\,,
\end{equation}
where $d_\text{L}$ is the luminosity distance corresponding to the Ly$\alpha$\xspace redshift associated to the wavelength of the detection NB, according to our cosmology.

There are several factors that can affect the estimation of $L_{\mathrm{Ly}\alpha}$: the variable width of the Ly$\alpha$\xspace line, the uncertain position of the line center with respect to the NB transmission boundaries, the chosen photometric aperture and the uncertainty on the continuum estimate, among others. In addition to that, QSOs often show rather strong \ion{N}{V} emission lines ($\lambda\, 1240$ \AA) that contaminate the Ly$\alpha$\xspace measurement. While in most spectroscopic surveys it is possible to resolve the Ly$\alpha$\xspace and \ion{N}{V} line profiles separately, both lines cannot be disentangled with NB imaging, thus \ion{N}{V} significantly affects the Ly$\alpha$\xspace flux measurement in QSOs. As a consequence, our NB-estimated Ly$\alpha$\xspace line flux actually includes the sum of both contributions: $F_{\mathrm{Ly}\alpha + \mathrm{\ion{N}{V}}}$. We account for these biases on our measured $L_{\mathrm{Ly}\alpha}$ by computing the median offset between the estimated and real values for our mock LAEs ($L_{\mathrm{Ly}\alpha}^{\rm intrinsic}$), as a function of $L_{\mathrm{Ly}\alpha}$ ,
\begin{figure}
    \centering
    \resizebox{\hsize}{!}{\includegraphics{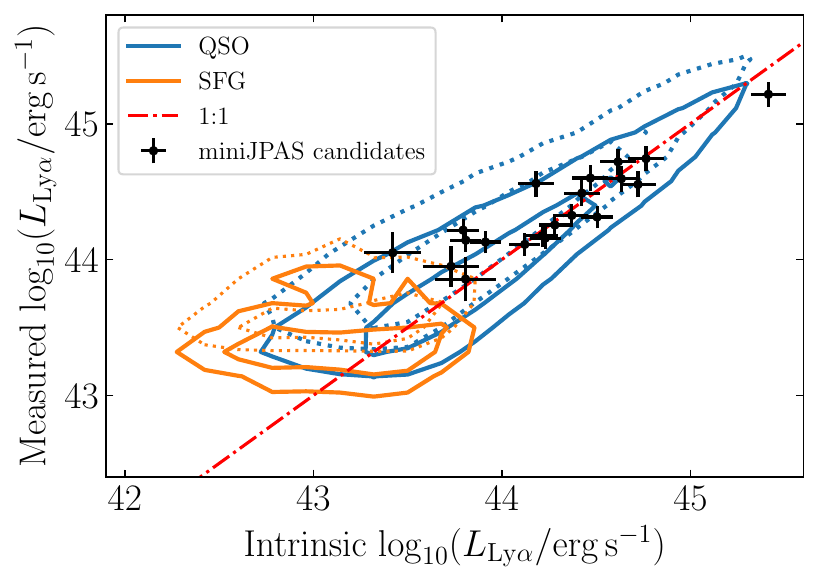}}
    \caption{Contours showing the areas encompassing the 68\% and 95\% of the LAEs in the mock showing the retrieved (observed) Ly$\alpha$\xspace luminosity as a function of the real value in the catalog. The dotted and solid line contours show the values of $L_{\mathrm{Ly}\alpha}$ before and after applying the bias subtraction, respectively (see Sect.~\ref{sec:Llya_comp}). Black dots represent the observational candidates from miniJPAS with a spectroscopic counterpart (see Sect.~\ref{sec:spec_xmatch}).
    }
    \label{fig:L_lya_contours}
\end{figure}

\begin{equation}
    \Delta\log_{10} L = \mathrm{median}[\log_{10}L_{\mathrm{Ly}\alpha} - \log_{10}L_{\mathrm{Ly}\alpha}^{\rm intrinsic}]\,,
\end{equation}
This bias is computed in bins of $\log_{10}(L_{\mathrm{Ly}\alpha} / \text{erg\,s$^{-1}$\xspace})$\ and subtracted from the $L_{\mathrm{Ly}\alpha}$ measurement. Fig.~\ref{fig:L_lya_contours} shows the measured $\log_{10}(L_{\mathrm{Ly}\alpha} / \text{erg\,s$^{-1}$\xspace})$ distribution from the mock as a function of the intrinsic luminosity. Since there is no clear way to systematically disentangle QSOs from SFG LAEs in our sample, we apply the same correction indistinctly.

\subsection{Purity and completeness}\label{sec:puricomp1d}

We compute the purity $P$ of our mock sample as
\begin{equation}
    P = \frac{\mathrm{TP}}{\mathrm{TP} + \mathrm{FP}}
    \label{eq:purity}
\end{equation}
and the completeness C as
\begin{equation}
    C = \frac{\mathrm{TP}}{\mathrm{TP} + \mathrm{FN}}\,,
    \label{eq:completeness}
\end{equation}
where TP, FP and FN are the number of true positive, false positive and false negative detections, respectively. After applying the selection method to our mocks, we can estimate the purity and completeness curves of the selected sample for each filter, and for the whole set as a function of $L_{\mathrm{Ly}\alpha}$.

\begin{figure}
    \centering
        \resizebox{\hsize}{!}{\includegraphics{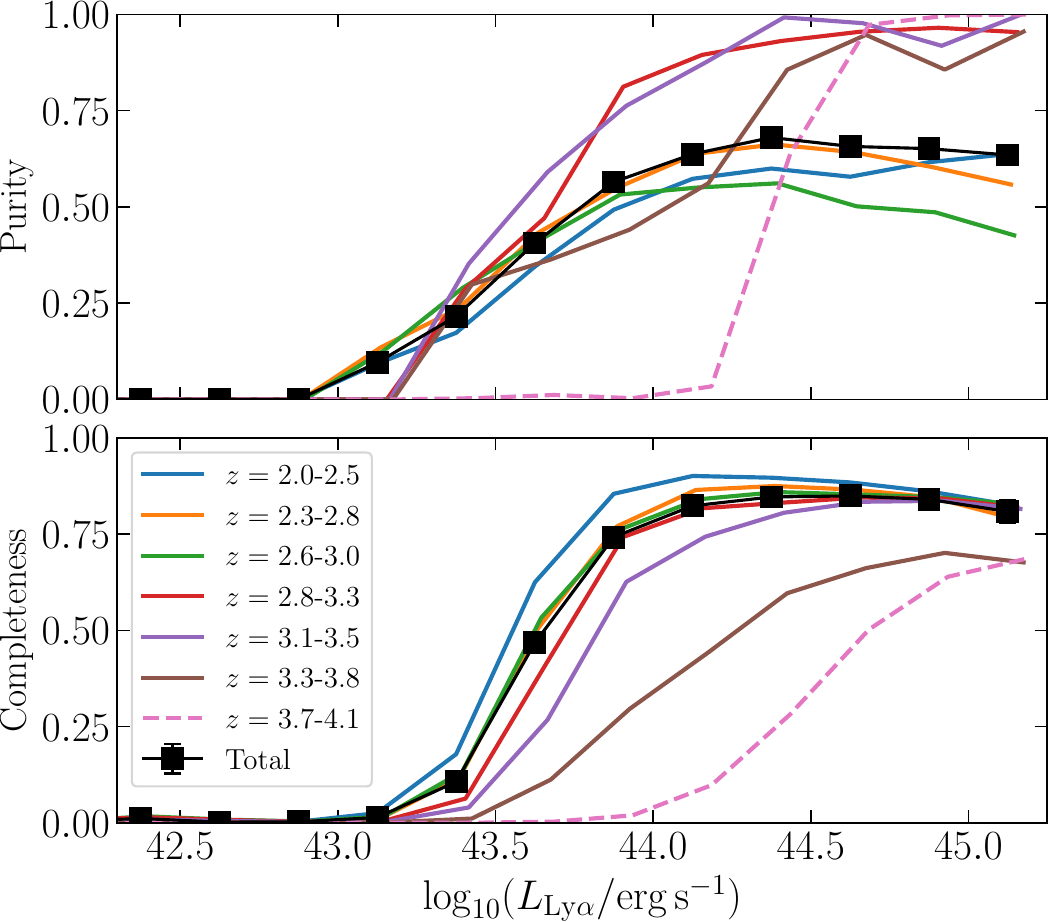}}
    \caption{Purity and completeness as a function of $L_{\mathrm{Ly}\alpha}$ of the full retrieved sample at $2.05<z<3.75$. We show the values of purity and completeness for the whole redshift range in black solid lines, and for each redshift interval used in colored solid lines. We represent an additional redshift interval to show the trend beyond the redshift range probed by this work (dashed line).
    }
    \label{fig:combined_puricomp}
\end{figure}

In Fig.~\ref{fig:combined_puricomp} we show the purity (top panel) and completeness (bottom panel) of our selection method as a function of the Ly$\alpha$\xspace luminosity for the whole sample ($r<24$, $2.05<z<3.75$). We also show the purity and completeness for the 6 bins of redshift used to compute the Ly$\alpha$\xspace LFs. The redshift intervals are composed of groups of 5 NBs, as listed in Table~\ref{tab:NB_z_intervals}. All the redshift bins exhibit a similar behaviour: the completeness increases with $L_{\mathrm{Ly}\alpha}$, reaching values of $\gtrsim 75\%$ for $\log_{10}(L_{\mathrm{Ly}\alpha} / \text{erg\,s$^{-1}$\xspace})\geq 44$ for $2\lesssim z\lesssim3.3$, and for {${\log_{10}(L_{\mathrm{Ly}\alpha} / \text{erg\,s$^{-1}$\xspace})\geq 44.5}$} at $3.1\leq z\leq3.8$. The sample purity also increases with $L_{\mathrm{Ly}\alpha}$ for all redshift bins, with a slight decline for the brightest luminosity in some intervals (see Fig.~\ref{fig:combined_puricomp}, top panel). The drop in purity at the bright-end can be explained by the overestimation of the line luminosity of the contaminants (for example, a \ion{C}{IV} emitter at $z=1.7$ with {${\log_{10}(L_{\ion{C}{IV}} / \text{erg\,s$^{-1}$\xspace})=44}$} will appear to have {${\log_{10}(L_{\ion{C}{IV}} / \text{erg\,s$^{-1}$\xspace})=44.4}$} if we assume its redshift to be $z=2.5$). This effect is increased by the rather high uncertainties on the line flux measurements in combination with the Eddington bias \citep{Eddington1913}. Interestingly, for $z\gtrsim 2.8$ the estimated purity reaches values very close to 1 for the brightest Ly$\alpha$\xspace luminosity. This can be explained by the fact that the potential contaminants in this luminosity regime are QSOs with $z>2$ for which the selected feature is the \ion{C}{IV}. We note that most of these sources are classified as LAEs at their correct redshift by our selection pipeline.

\begin{table}
    \centering
    \caption{Groups of NBs used in this work for the computation of the Ly$\alpha$\xspace LF, their associated redshift coverage of Ly$\alpha$\xspace , and the comoving volume of the Universe sampled by those redshift intervals for an area of $1.14$ deg$^2$. Each batch is composed by 5 contiguous NBs. Two subsequent batches have 2 NBs in common, thus their redshift coverage is partially overlapped.}
    \label{tab:NB_z_intervals}
    \resizebox{\linewidth}{!}{
    \begin{tabular}{cccc}
         \toprule
         \# & Filters & $\Delta z$ & \makecell{Volume \\ ($10^6$ Mpc$^3$)}\\
         \midrule
         1 & J0378, J0390, J0400, J0410, J0420 & 2.05--2.52 & $6.21$ \\
         2 & J0410, J0420, J0430, J0440, J0450 & 2.32--2.77 & $5.92$ \\
         3 & J0440, J0450, J0460, J0470, J0480 & 2.56--3.01 & $5.85$\\
         4 & J0470, J0480, J0490, J0500, J0510 & 2.81--3.25 & $5.79$\\
         5 & J0500, J0510, J0520, J0530, J0540 & 3.05--3.50 & $5.62$\\
         6 & J0530, J0540, J0550, J0560, J0570 & 3.29--3.75 & $5.61$\\
         \bottomrule
    \end{tabular}
    }
\end{table}

\subsection{2D purity and number counts correction maps}\label{sec:2dpuricomp}

We apply our selection method (see Sect.~\ref{sec:selection_method}) to the mock in order to characterize its performance as a function of $r$ magnitude and measured {$L_{\mathrm{Ly}\alpha}$}, in terms of purity and number count correction.
We build 2D maps of these two quantities over a grid of ($r$, $\log_{10}(L_{\mathrm{Ly}\alpha} / \text{erg\,s$^{-1}$\xspace})$) values, that will be used in the Ly$\alpha$\xspace LF computation.

As a first step, we compute the purity ($P^\mathrm{2D}$) of the sample in bins of ($\Delta r, \Delta\log_{10}L_{\mathrm{Ly}\alpha}$), according to Eq.~\ref{eq:purity}. We consider as true positives the objects detected inside a given interval of ($\Delta r, \Delta\log_{10}L_{\mathrm{Ly}\alpha}$) with a minimum Ly$\alpha$\xspace EW$_0=30$ \AA, whose redshift measurement is correct within a confidence interval of $\Delta z=0.12$\footnote{This interval is defined by the transmission FWHM of the NB filter of the detection}. As a second step, we define the number counts correction ($w^\mathrm{2D}$) as the ratio between the number of eligible LAEs inside a given interval of $r$ and $\log_{10}(L_{\mathrm{Ly}\alpha} / \text{erg\,s$^{-1}$\xspace})$ and the number of true positives retrieved by the selection inside that interval. This number-counts correction can be seen as the inverse of the completeness as a function of measured $L_{\mathrm{Ly}\alpha}$, defined in Eq.~\ref{eq:completeness}. Due to the uncertainties on the estimation of $L_{\mathrm{Ly}\alpha}$, some bins of this map contain values below 1, meaning that in some regimes we might get a larger number of true positives than the intrinsic number of LAEs.

We show an example of these correction maps in Appendix~\ref{app:2dmaps}. In general terms, the purity of the sample increases with $L_{\mathrm{Ly}\alpha}$ and $r$. This is because for a fixed value of $L_{\mathrm{Ly}\alpha}$, fainter magnitudes mean larger equivalent widths and therefore, our selection method is more successful in retrieving true LAEs at these regimes. For the brightest magnitudes the purity increases again due to the low relative errors of the photometry, which allow to reliably discern between true positives and contaminants.

\subsection{Computation of the Ly$\alpha$\xspace luminosity function}\label{sec:lyaLF_computation}

We compute our Ly$\alpha$\xspace LF through several realizations in order to take into account the various sources of uncertainty and variability. At each realization, we perturb the estimated values of $L_{\mathrm{Ly}\alpha}$ assuming Gaussian errors. Then, every selected LAE candidate is included in the current sub-sample with a probability based on the 2D purity, that is inferred from the $r$ magnitude and the perturbed $\log_{10}(L_{\mathrm{Ly}\alpha} / \text{erg\,s$^{-1}$})$ (see Sect.~\ref{sec:2dpuricomp}). For the candidate $j$, let us define {${p_j}$} as
\begin{equation}
    \begin{cases}
        p_j = 1 &\text{if}\quad \xi_j \leq P^\mathrm{2D},\\
        p_j = 0 &\text{if}\quad \xi_j > P^\mathrm{2D}\\
    \end{cases}\,,
\end{equation}
where $\xi_j$ is a random number drawn from a uniform distribution in the interval $\xi_j\in[0, 1]$.

Next, each source is weighted with a value $w_j$, computed as 
\begin{equation}
    w_j = C_\text{int}^{-1}\cdot w^\text{2D}\,,
\end{equation}
where $w^\text{2D}$ is the number count correction (as defined in Sect.~\ref{sec:2dpuricomp}) and $C_\mathrm{int}$ the intrinsic completeness of the survey. The value of $C_\mathrm{int}$ is computed for point-like and extended sources separately in miniJPAS (see \citealt{bonoli2020}). This process is done equivalently for J-NEP (see \citealt{Hernan-Caballero2023}). We assign every LAE candidate a value of $C_\mathrm{int}$ using the miniJPAS and J-NEP completeness curves, in terms of $r$ magnitude and the field in which the object was detected. Our target population are LAEs at $z>2$, these objects are expected to appear point-like in the BB images of miniJPAS\&J-NEP. Hence, we use the intrinsic completeness curves for point-like objects.

Each NB can probe \ion{Ly}{$\alpha$} in an effective range of redshift equal to $(\lambda_\text{NB} \pm 0.5\cdot\text{FWHM}_\text{NB}) / \lambda_0 - 1$. The volume ($V$) considered for the LF is the comoving volume comprised between those redshifts in the survey area, according to our cosmology. Given that the number count of candidates is not large enough to accurately estimate the LF at each NB independently, we combine several NBs to build the LF. The associated redshift range of each group of NBs goes from the minimum $z$ of the bluest filter to the maximum $z$ of the reddest. Adjacent miniJPAS NBs show significant overlap ($\sim45$ \AA). As explained in Sect.~\ref{sec:nbexsel}, in case of multiple line detections in adjacent NBs, we assign the Ly$\alpha$\xspace line to the NB with the highest measured flux. Therefore, the effective volume probed by a single NB in the wavelengths of the overlaps is halved.

The $i$th iteration of the LF $\Phi_i$ is computed as follows:
\begin{equation}
    \Phi_i\left[\log_{10}L_{\mathrm{Ly}\alpha}\right] = \frac{\sum\limits_j p_j \cdot w_j}{V\cdot\Delta\log_{10}L_{\mathrm{Ly}\alpha}}\,,
\end{equation}
where the sum extends to all the objects with a perturbed $L_{\mathrm{Ly}\alpha}$ falling inside a given luminosity bin. After performing 1000 realizations, our final LF ($\Phi$) is built with the median values of $\Phi_i$ for each luminosity bin.

For estimating the $\Phi$ uncertainties we have to take into account the contribution of: (i) the spatial variance of the candidates in the surveyed area (commonly referred to as ``cosmic variance''), (ii) the uncertainty of $L_{\mathrm{Ly}\alpha}$ estimation and (iii) the shot noise of the candidate sample. In order to estimate the contribution of (i), we divide our candidate sample in 5 sub-samples. First, we split the miniJPAS footprint in 4 regions of equal angular area. Then, the candidates are assigned to 4 different sub-samples according to the split region they belong. The 5th sub-sample is that of the J-NEP candidates. After that, we perform 1000 realizations of $\Phi_i$, each time using the candidates of 5 random sub-samples with repetition. On each realization, we re-sample the candidates of each sub-sample using the bootstrap technique and we also perturb $L_{\mathrm{Ly}\alpha}$ as explained above. The final uncertainties on our LFs are inferred via the 16th and 84th percentiles of the $\Phi_i$ distribution.

\subsection{Ly$\alpha$\xspace emitters candidate sample}\label{sec:LAEs_cat}

\begin{figure*}
    \centering
    \includegraphics[width=17cm]{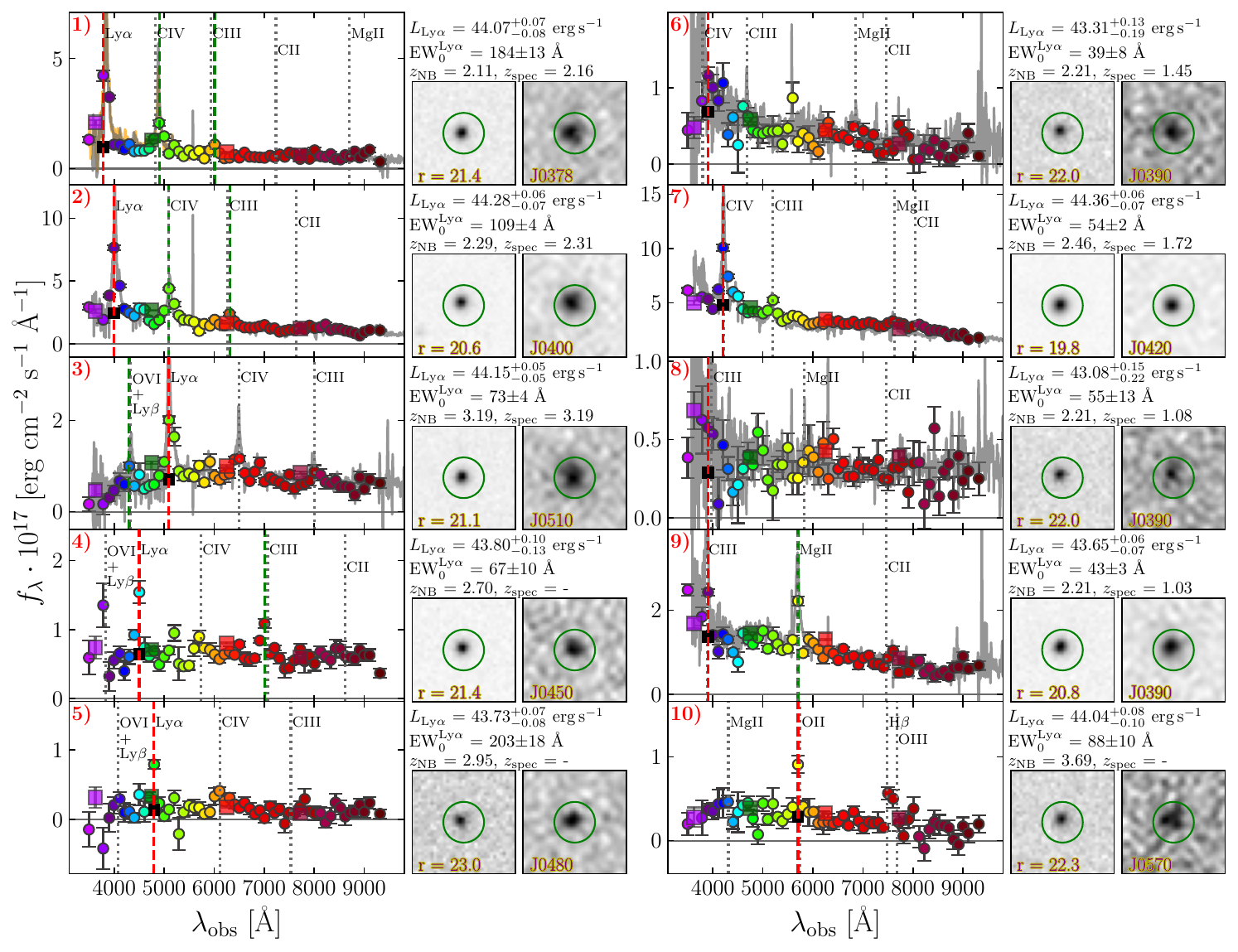}
    \caption{Examples of miniJPAS\&J-NEP LAE candidates. We show the fluxes of the 56 narrow medium (colored circles) and broad bands (colored squares) of the J-PAS filter set. The NB selected as Ly$\alpha$\xspace by our pipeline is marked with a red dashed line. The detected secondary QSO lines are marked with green dashed lines (see Sect.~\ref{sec:nbexsel}). Spectroscopically or visually identified emission lines of the objects are shown with a gray dotted line. We also show the images of each source in $r$ and the selected NB. The spectra of the SDSS DR16 (HETDEX) counterparts are shown in gray (orange) when available. The five objects in the left column are examples of LAEs identified by our method. The right column show five examples of possible contaminants.}
    \label{fig:LAEs_examples}
\end{figure*}

\begin{table*}
    \centering
    \caption{Number of candidates after applying a first cut in 3-$\sigma$ excess, and then the remaining candidates after applying every other cut described in section \ref{sec:nbexsel} separately. The last column shows the number count in the final sample, after the visual inspection (VI), used to estimate the Ly$\alpha$ LF.}
    \label{tab:N_sel}
    \resizebox{\linewidth}{!}{
    \begin{tabular}{ccccccc|ccc}
    \toprule
        Field & \makecell{Parent \\ sample} & \makecell{$3\sigma$ cut, \\ EW$_0>30$ \AA} & S/N > 6 & Mult. lines & Color & Morph. & \makecell{Total\\(no morph.)} &  \makecell{Total \\ + morph.} & \makecell{Total \\ + morph.\\ + VI} \\ 
        \midrule
        AEGIS001 & 7594 & 182(2.40\%) & 50(0.66\%) & 175(2.30\%) & 105(1.38\%) & 62(0.82\%) & 38(0.50\%) & 30(0.40\%) & 17(0.22\%) \\ 
        AEGIS002 & 6509 & 131(2.01\%) & 25(0.38\%) & 128(1.97\%) & 67(1.03\%) & 47(0.72\%) & 19(0.29\%) & 14(0.22\%) & 13(0.20\%) \\ 
        AEGIS003 & 7428 & 142(1.91\%) & 29(0.39\%) & 139(1.87\%) & 78(1.05\%) & 44(0.59\%) & 22(0.30\%) & 15(0.20\%) & 12(0.16\%) \\ 
        AEGIS004 & 6946 & 143(2.06\%) & 20(0.29\%) & 142(2.04\%) & 73(1.05\%) & 32(0.46\%) & 14(0.20\%) & 10(0.14\%) & 7(0.10\%) \\ 
        J-NEP & 7549 & 175(2.32\%) & 44(0.58\%) & 171(2.27\%) & 107(1.42\%) & 75(0.99\%) & 34(0.45\%) & 22(0.29\%) & 18(0.24\%) \\
        \midrule
        Total & 36026 & 773(2.15\%) & 168(0.47\%) & 755(2.10\%) & 430(1.19\%) & 260(0.72\%) & 127(0.35\%) & 91(0.25\%) & 67(0.19\%) \\ 
        \bottomrule
    \end{tabular}
    }
\end{table*}

The result of the preliminary selection is a sample of 135 candidates (38, 19, 22, 14 and 34 in AEGIS001, AEGIS002, AEGIS003, AEGIS004 and J-NEP, respectively) with redshifts between $2.05$ and $3.75$. Eight of these selected candidates were removed immediately after a first visual inspection, because their NB images were clearly affected by cosmic rays or artifacts, leaving a sample of 127 candidates.

In Fig.~\ref{fig:LAEs_examples} we show examples of miniJPAS\&J-NEP sources in order to illustrate the populations retrieved by our selection method. The five objects on the left column (1-5) are examples of genuine QSO LAEs. Three of them have SDSS/HETDEX spectroscopic confirmation. Candidates 1-4 have secondary QSO line detections that support the Ly$\alpha$\xspace redshift estimation (see Sect.~\ref{sec:nbexsel}). Candidate 5 lacks spectroscopical confirmation or other QSO line detection, however, through visual inspection we determined the presence of spectral features consistent with QSO emission lines, given the estimated Ly$\alpha$\xspace redshift (\ion{O}{VI}+\ion{Ly}{$\beta$}, \ion{Si}{IV}+\ion{O}{IV}, \ion{C}{IV}). Candidates 6-9 are examples of QSO contaminants selected because of their strong \ion{C}{IV} or \ion{C}{III}] emission. In the particular case of candidate 9, our method detects a secondary line consistent with \ion{Mg}{II}, given that the selected NB is spectroscopically identified as \ion{C}{III}] at $z=1.03$. Hence, candidate 9 is effectively not selected by our pipeline. Finally, candidate 10 is an example of a contaminant [\ion{O}{II}] emitter. A visual inspection reveals that this candidate shows a relevant feature consistent with \ion{H}{$\beta$} and [\ion{O}{III}] emission lines, if we assume the selected NB is [\ion{O}{II}] at $z=0.54$. Moreover, candidate 10 shows significant emission at bluer wavelengths than its strong emission line at $\lambda_{obs}\approx5800\text{\AA}$. Therefore, it is unlikely that this line is Ly$\alpha$\xspace, due to the absence of the expected decrease of flux due to the Ly$\alpha$\xspace forest, and beyond the Lyman limit break at $\lambda_0<912$ \AA\ ($\lambda_{obs}\ll5800\text{\AA}$ for the assumed Ly$\alpha$\xspace redshift of $z_{\mathrm{Ly}\alpha} = 3.69$; see the photometric drop at $\lambda_{obs}<4200\text{\AA}$ in the photospectrum of candidate 3, in the central left panel of Fig.~\ref{fig:LAEs_examples}).

\subsection{Sample contamination}

\begin{figure}
    \centering
    \resizebox{\hsize}{!}{\includegraphics{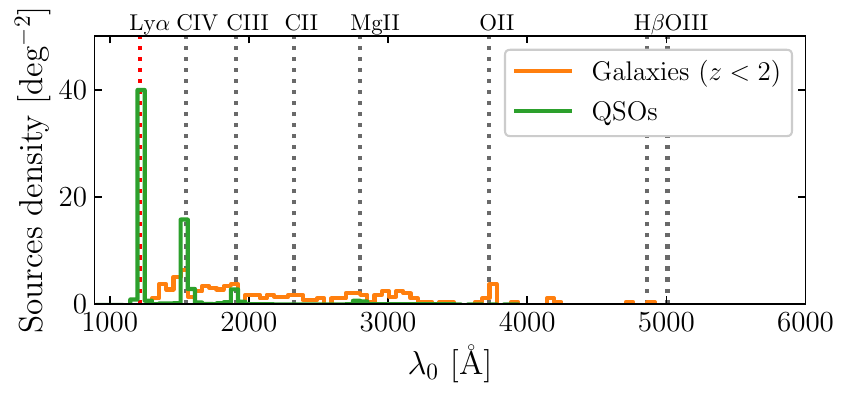}}
    \caption{Rest-frame wavelength of the selected features in the QSO and low-$z$ galaxy mock. The rest-frame wavelength is computed as $\lambda_0=\lambda_\mathrm{NB} / (1 + z)$, where $\lambda_\mathrm{NB}$ is the central wavelength of the selected NB and $z$ the true redshift of the mock object.}
    \label{fig:QSO&GAL_cont_hist}
\end{figure}

As discussed in Sect.~\ref{sec:mocks}, we expect the interlopers of our selection to be mainly low-$z$ galaxies and $z<2$ QSOs. Through the analysis of the selected sample in our mock, we can describe the predicted populations of contaminants.

Due to its typically high intrinsic luminosity, the \ion{C}{IV} line is the QSO feature which mainly contributes to the contamination of our samples  (see \citealt{VandenBerk2001}), followed by \ion{C}{III}], and in a lesser amount, \ion{Mg}{II} and \ion{O}{VI}. This can be clearly seen in Fig.~\ref{fig:QSO&GAL_cont_hist}, which presents the number of objects in the mock classified as LAEs by our method, as a function of the rest-frame wavelength of the selected feature. This is in line with the results of the spectroscopic follow-up presented in \cite{Spinoso2020}, which show that \ion{C}{IV} is the main source of contamination for samples of bright, NB-selected, LAE candidates. The contaminants whose NB wavelength does not correspond to any relevant QSO spectral feature are selected because of the scatter of NB fluxes due to random fluctuations. This causes either: (i) the flux of a NB to incidentally exceed our 3$\sigma$ detection limit or (ii) produce an under-estimation of the continuum.

Regarding low-$z$ galaxy interlopers, Fig.~\ref{fig:QSO&GAL_cont_hist} shows that several galaxies are selected as LAE candidates at a redshift which is not associated to any specific emission line, with the exception of a small peak at the [\ion{O}{II}] wavelength. Therefore, most of the low-$z$ galaxy contamination can be explained as false line detections caused by noise. On the other hand, many candidates in our observational sample might show extended BB emission, which classifies them as low-$z$ galaxies. These wrongly selected candidates can be easily removed via a posterior visual inspection (see Sect.~\ref{sec:visual_inspection}).

\subsection{Spectroscopic counterparts}\label{sec:spec_xmatch}

We cross-match the miniJPAS\&J-NEP catalogs with the spectroscopic catalogs of SDSS DR16 and HETDEX in order to characterize our candidate sample.

\begin{figure}

    \centering
    \resizebox{\hsize}{!}{\includegraphics{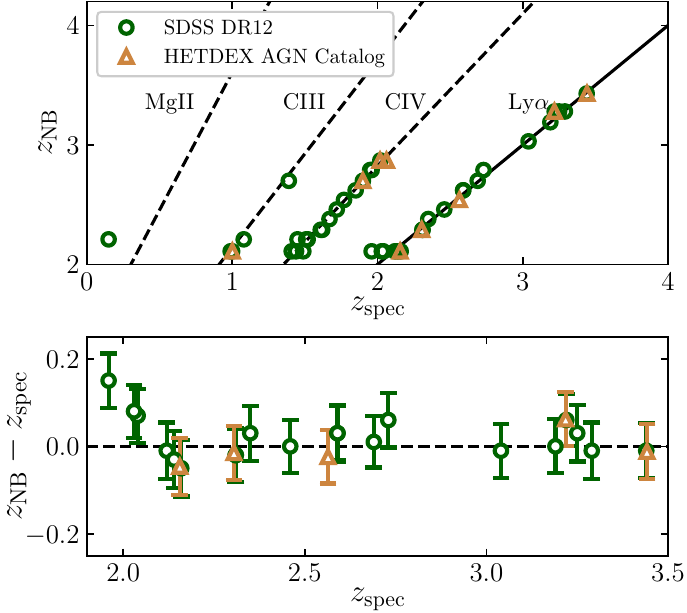}}
    \label{fig:zarr_zspec_c}
    
    \caption{Comparison between the NB and spectroscopic redshifts of the candidates. Top: Retrieved Ly$\alpha$\xspace redshift of the candidates with SDSS or HETDEX spectrum. The straight lines represent the redshift at which the most common QSO contaminant lines are selected as Ly$\alpha$\xspace . Bottom: Difference between the NB Ly$\alpha$\xspace redshift of the good candidates and the spectroscopic redshifts given by SDSS or HETDEX. The error bars show the redshift interval covered by the width of the NB in which the candidate is selected.
    }
    \label{fig:zarr_zspec}
\end{figure}

\subsubsection{Cross-match with SDSS DR16}\label{sec:sdss_xmatch}

We cross-match with the SDSS DR16 source catalog \citep{Lyke2020}, using a search radius of 1.5\arcsec\xspace among the whole catalog of miniJPAS. As a result, 32 (17 with $\log_{10}(L_{\mathrm{Ly}\alpha} / \text{erg\,s$^{-1}$\xspace}) > 44$) sources are identified as QSO LAEs by SDSS (i. e. sources with redshift in the range $z_\mathrm{spec}=2.1$--$4$, no redshift warnings and a significant Ly$\alpha$\xspace measurement). Among these 32 sources, 17 are selected by our method (53\%), and 13 out of 17 (76.5\%) within the ones with $\log_{10}(L_{\mathrm{Ly}\alpha} / \text{erg\,s$^{-1}$\xspace})>44$. This retrieval rate is in agreement with the completeness estimated by our mock (see Fig.~\ref{fig:combined_puricomp}), considering the statistical uncertainties and cosmic variance.

\subsubsection{Cross-match with HETDEX Public Source Catalog 1}\label{sec:HETDEX_xmatch}

We also cross-match the miniJPAS catalog with the HETDEX Public Source Catalog 1 \citep{Mentuch2023}. This catalog contains the spectra of 232\,650 sources observed by the HETDEX program \citep{Gebhardt2021} over 25 deg$^2$. The footprint of the HETDEX catalog partly overlaps with miniJPAS. We find 158 objects within a radius of 1.5\arcsec\xspace of any miniJPAS source with a reliable spectroscopic redshift measure according to the HETDEX catalog (\texttt{z\_hetdex\_conf} $>0.9$). Among these objects, 22 are labeled as AGNs by HETDEX and 12 have $z_\mathrm{spec} > 2$.  Within our selection, 10 objects have a HETDEX identification: 9 AGNs and one [\ion{O}{II}] emitter. Finally, 5 of the 9 spectroscopically confirmed AGNs have $z>2$ and clear Ly$\alpha$\xspace emission line measurements. This numbers translate into a $\sim(41\pm22)\%$ recovery rate of AGNs with $z>2$, and a purity of $\sim (56\pm30)\%$.

Moreover, the cross-match with HETDEX reveals the presence of 17 SFG LAEs (\texttt{z\_hetdex\_conf} $>0.9$) in the dual mode catalogs of miniJPAS. However, none of them is detected in our sample. This is because  all these SFG LAEs are too faint both in Ly$\alpha$\xspace luminosity and $r$ magnitude to be selected by our method. Indeed, they all show $\log_{10}(L_{\mathrm{Ly}\alpha} / \text{erg\,s$^{-1}$\xspace})\lesssim43.2$  and $r$SDSS magnitudes close or below the miniJPAS nominal depth (see Table~\ref{tab:filter_properties}). Hence, the signal-to-noise of these sources photometry is overall too low for them to be detected by our selection pipeline.

\subsubsection{Spectroscopic characterization of the LAE candidate sample}

Within our candidate sub-sample, we find an SDSS counterpart for 41 out of 127 LAE candidates, all of which are identified as QSOs at any redshift by SDSS. Fig.~\ref{fig:zarr_zspec} (upper panel) shows the spectroscopic redshift of those candidates with an SDSS or HETDEX counterpart, confirming that $z<2$ AGN emitting \ion{C}{IV} or \ion{C}{III}] (misclassified as Ly$\alpha$\xspace) are the main source of contamination for our pipeline results. There are no spectroscopically confirmed contaminants at $z_{\mathrm{Ly}\alpha}\geq 3$, which is in line with the high-purity we estimate for our samples at these high redshifts (Fig.~\ref{fig:combined_puricomp}). In the bottom panel of Fig.~\ref{fig:zarr_zspec} we show the offsets between the NB Ly$\alpha$\xspace redshift of our candidates and the SDSS spectroscopic redshift. For LAEs at $z\sim 2$, the observed Ly$\alpha$\xspace wavelength could lay slightly below the lower limits of our survey. In some of those cases, we still detect the redmost part of the line, often affected by the \ion{N}{V} flux. For this reason, we notice a small bias in the measured redshift for LAEs with $z_\mathrm{spec}\lesssim 2.1$ (see the bottom panel of Fig.~\ref{fig:zarr_zspec}). At higher redshifts, $z_\mathrm{NB}$ is a good estimator within the interval of confidence given by the width of the NBs. The mean offset of $z_\mathrm{NB}$ with respect to $z_\mathrm{spec}$ is $|\Delta z|\approx 0.013$, about a 20\% of the $z_\mathrm{NB}$ uncertainty.

In Fig.~\ref{fig:L_lya_contours} we show the comparison between the measured and spectroscopic $\log_{10}(L_{\mathrm{Ly}\alpha} / \text{erg\,s$^{-1}$\xspace})$ from SDSS DR16Q, compared to mock distributions. This figure { shows} that the estimation of $\log_{10}(L_{\mathrm{Ly}\alpha} / \text{erg\,s$^{-1}$\xspace})$ in the spectroscopic subsample of our candidates is consistent with the results in the mock.

\subsection{Photometric redshifts}

\begin{figure}
    \centering
    \resizebox{\hsize}{!}{\includegraphics{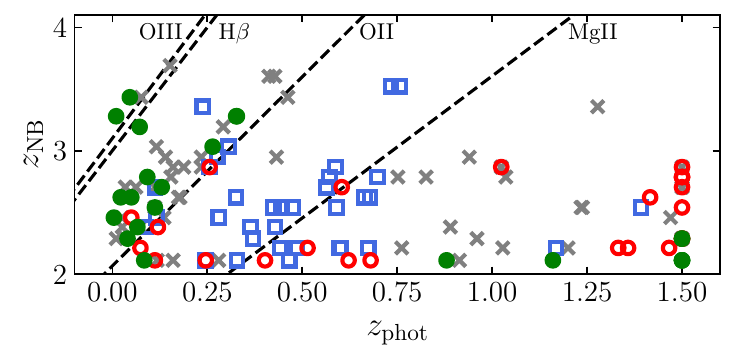}}
    \caption{Retrieved Ly$\alpha$\xspace redshift of the candidates as compared to the miniJPAS photometric redshifts { \protect\citep{Hernan-Caballero2021}}. We show spectroscopically confirmed LAEs (green filled circles) and contaminants (red empty circles), objects with extended morphology (\texttt{morph\_star\_prob}<0.1; blue empty squares), and sources without any spectral or morphology identification (gray crosses). { The dashed lines mark the confusion between Ly$\alpha$ and typical galactic emission lines.}}
    \label{fig:zphot_vs_znb}
\end{figure}

The miniJPAS and J-NEP dual-mode catalogs provide accurate photometric redshifts for galaxies in the interval $z=0$--$1.5$ \citep{Hernan-Caballero2021, Hernan-Caballero2023}. These photo-$z$ have been obtained using a template-fitting method which employs a sample of 50 galaxy templates. Fig.~\ref{fig:zphot_vs_znb} shows the photo-$z$ of the LAE candidates as compared to the redshifts obtained from the NB central wavelengths, assuming that the detected line of a candidate is Ly$\alpha$\xspace . From this figure it is not evident any clear pattern which may help to identify a systematic source of contamination; this is in agreement with the results of the contamination analysis of the mocks, that predicts a rather flat distribution in the selected rest-frame wavelengths of the galactic contaminants, with a small peak in the [\ion{O}{II}] line (Fig.~\ref{fig:QSO&GAL_cont_hist}). On the other hand, the current photo-$z$ code does not account for QSOs, hence they are not useful to confirm QSO LAEs or contaminants; most of the sources with $z_\text{phot}=1.5$ are likely to be QSOs with bad fit of the photo-$z$. { For the same reason, the photo-$z$ values exhibit arbitrary correspondence with the redshifts of our candidates having spectroscopic counterparts. (green filled circles in Fig.~\ref{fig:zphot_vs_znb})}. Analogously, the photo-$z$ are not useful to confirm SFG LAEs because their redshift ($z>2$) is far out of the working range of the miniJPAS photo-$z$ code.

\subsection{Morphology cut}\label{sec:LAEs_morph}

We notice that some of our candidates have visually evident extended emission in their BB images. The population of LAEs at $z>2$ is expected to appear point-like given the the expected observed size of either high-$z$ QSOs or SFG LAEs, and the average PSF of miniJPAS\&J-NEP. Hence, the candidates clearly showing BB extended morphologies are very likely to be low-$z$ contaminants. In order to remove this kind of objects from the Ly$\alpha$\xspace LF estimation, we make use of the star-galaxy estimator \texttt{morph\_prob\_star} \citep{Lopez-Sanjuan2019b}, available in the miniJPAS\&J-NEP catalogs. We only keep objects with \texttt{morph\_prob\_star} $>0.1$. With this cut we remove 36 extended objects (28.3\% of the selected sample), leaving a sample of 91 LAE candidates. To perform such morphology cut in the mock sample is not possible due to the lack of photometric images for the mock sources. However, the corrections for the Ly$\alpha$\xspace LF can be recomputed taking into account the morphology cut, and other posterior catalog cuts (see Sect.~\ref{sec:visual_inspection}).

On the other hand, LAEs often present NB extended emission in the Ly$\alpha$\xspace observed wavelength (see, e.g., \citealt{Haardt1996, Borisova2016, Battaia2016}) --not to be confused with extended BB continuum emission--. The recent work of \cite{Rahna2022} presented the Ly$\alpha$\xspace extended emission of two miniJPAS QSOs showing double-core Ly$\alpha$\xspace emission ($z=3.218, 3.287$); both objects are detected by our pipeline and included in our catalog.

\subsection{Visual inspection}\label{sec:visual_inspection}

We perform a visual inspection of the images and photospectra of the 127 initial candidates. We identify 39 objects in our sample as nearby galaxies either by their extended BB morphology or by their spectral features (e. g. emission lines not detected by our method, the presence of a 4000 \AA\ break etc.), 36 of which are already systematically removed by a cut in \texttt{morph\_prob\_star} $>0.1$ (see Sect.~\ref{sec:LAEs_morph}). Another 21 objects are clearly identified as contaminant QSOs at $z<2$. Finally, 32 objects are visually classified as secure QSOs with Ly$\alpha$\xspace emission. The remaining 35 objects do not have a secure classification due to having very noisy continua and/or unclear BB images. We remove the visually confirmed contaminants, leaving a sample of 67 objects. The visual inspection of the candidates is aided by the spectroscopic counterparts of SDSS and HETDEX (Sect. \ref{sec:spec_xmatch}).

After removing the visually selected contaminants, the purity of the final sample increases. Our mock selection predicts number counts of 53, 23, 59 and 4 deg$^{-2}$ for QSO LAEs, contaminant QSOs, low-$z$ galaxies and SFG LAEs, respectively, in the effective area of miniJPAS\&J-NEP. Hence, we conclude that after a visual inspection, we are able to remove { $\sim80\%$} of the QSO contaminants and { $\sim58\%$} of the low-$z$ galactic contaminants. The 35 unidentified objects are consistent with the remaining { $\sim42\%$} galaxies and the visually unidentified LAEs predicted by our mock selection. These purity estimates are reasonable within the sampling error of our method. We recompute the 2D purity and number count (see Sect.~\ref{sec:2dpuricomp}) of the remaining sample assuming the above fractions of removed galaxy and QSO contaminants. We underline that the Ly$\alpha$\xspace LFs we present in Sect.~\ref{sec:Lya_LF} are estimated using the sample of 67 candidates obtained after our visual inspection of the photospectra and NB images.

\subsection{miniJPAS\&J-NEP LAEs catalog}\label{sec:catalogs}

In Table~\ref{tab:N_sel} we show the number of candidates after applying every cut described in Sect.~\ref{sec:nbexsel}. The last three columns of this table display the number of candidates in three relevant sub-samples for this work with, respectively: 127, 91 and 67 candidates. The first sub-sample is the direct result of applying the selection method to the miniJPAS\&J-NEP catalogs, before the morphology cut. This first sub-sample will be used throughout Sect.~\ref{sec:results} to compare with the mock results. The second sub-sample, composed of 91 candidates, is obtained after applying the morphology cut to the previous one. In Table~\ref{tab:selected_catalog} we provide the catalog of sources in this sub-sample. Finally, we obtain the third sub-sample of 67 candidates, after performing a cross-match with available spectroscopic surveys and a visual inspection for further contamination removal. Nonetheless, in future J-PAS observations the available spectroscopic data can be limited. Also the volume of data can be large enough to make a visual inspection of all the candidates not feasible. The sample presented in Table~\ref{tab:selected_catalog} could therefore be intended as representative of what can be statistically obtained from any J-PAS data set.

\section{Results and discussion}\label{sec:results}

In this section, we describe the relevant features of the LAE candidate sample and we present the Ly$\alpha$\xspace LF. We also fit our Ly$\alpha$\xspace LF to a Schechter function and a power-law and give an estimation of the AGN/SFG fraction as a function of Ly$\alpha$\xspace luminosity. Finally, we discuss the expected performance of the method described through this work in future data releases of J-PAS.

\subsection{{EW$_0$} distribution}

We obtain the rest-frame Ly$\alpha$\xspace EW from the measured $F_{\mathrm{Ly}\alpha}$, following Eq.~\ref{eq:ew}. As stated in Sect.~\ref{sec:nbexsel}, one of the criteria of our candidates selection is a cut in Ly$\alpha$\xspace EW$_0>30$ \AA. However, the additional conditions on the NB-photometry S/N and on the line-excess significance (see Sect.~\ref{sec:nbexsel}) can override the condition on EW$_0$, effectively forcing a higher EW$_0$ limit (especially for faint sources and shallow NBs).

\begin{figure}
    \centering
    \resizebox{\hsize}{!}{\includegraphics{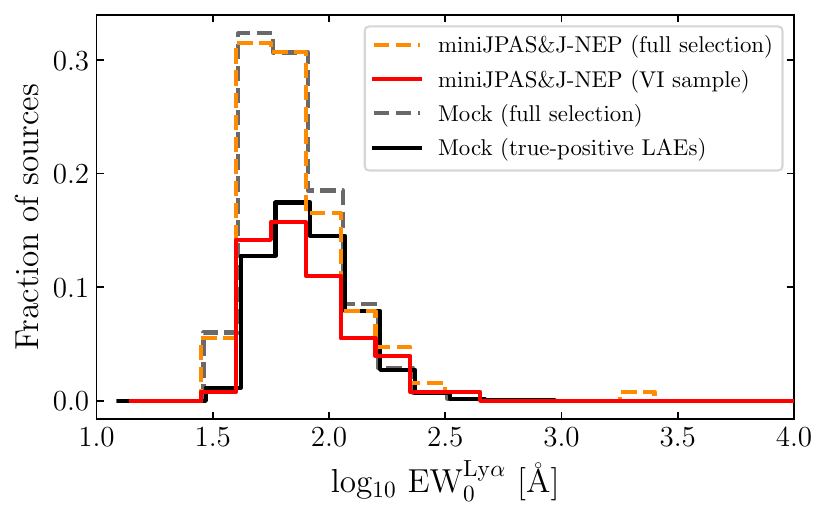}}
    \caption{Distribution of Ly$\alpha$\xspace EW$_0$ of the candidate sample of miniJPAS\&J-NEP. We compare the whole sample of candidates retrieved by our method in the mock (gray, dashed histogram) and in the observational data (orange, dashed histogram). We also compare the distribution of Ly$\alpha$\xspace EW$_0$ for the mock subsample of true-positive LAEs (solid, black histogram) and the visually inspected subsample of miniJPAS\&J-NEP (solid, red histogram) as defined in Sect.~\ref{sec:visual_inspection}. For $\rm log_{10}[EW_0^{Ly\alpha}/\text{\AA}]>2$, more than a 90\% of the sources retrieved from the mock are true LAEs.}
    \label{fig:ew_hist}
\end{figure}

The distribution of EW$_0$ retrieved from our 127 candidates sample (before applying the morphology cut, in order to match the mock; see Sect.~\ref{sec:catalogs}) is shown in Fig.~\ref{fig:ew_hist}. The miniJPAS\&J-NEP Ly$\alpha$\xspace EW$_0$ distribution ({ orange} dashed line) is in good agreement with the one resulting from applying our selection pipeline to our mock data ({ gray} dashed line).

Fig.~\ref{fig:ew_hist} also shows that the selected objects with $\log_{10}(EW_0^{Ly\alpha}/\text{\AA})>2$, are likely to be genuine LAEs, as the distribution of mock-LAEs ({ black} solid line) becomes closely comparable to the whole mock sample. Moreover, the EW$_0$ distribution of the selected LAEs in our mock is remarkably close to that of the observational sources visually classified as LAEs. 
Our retrieved Ly$\alpha$\xspace EW$_0$ distribution is also compatible with the determinations of \cite{Spinoso2020} and \cite{Liu2022a} for Ly$\alpha$\xspace lines of QSOs with $z\sim2$--$3.5$. All of our candidates are inside the range EW$_0=30$--$400$ \AA\ except for one candidate with an extremely large EW$_0$ of $2379 \pm 278$ \AA . However, this candidate has $r=23.8$, very close to the detection limit and the estimation of its continuum flux under the Ly$\alpha$\xspace line is likely to be underestimated (and its error overestimated). Furthermore, despite being in our selection, the purity assigned to this candidate by our method is $P^\mathrm{2D}=0$, making it irrelevant for the Ly$\alpha$\xspace LF estimation.

\subsection{Ly$\alpha$\xspace Luminosity Functions}\label{sec:Lya_LF}

\begin{figure*}
    \centering
    \includegraphics[width=17cm]{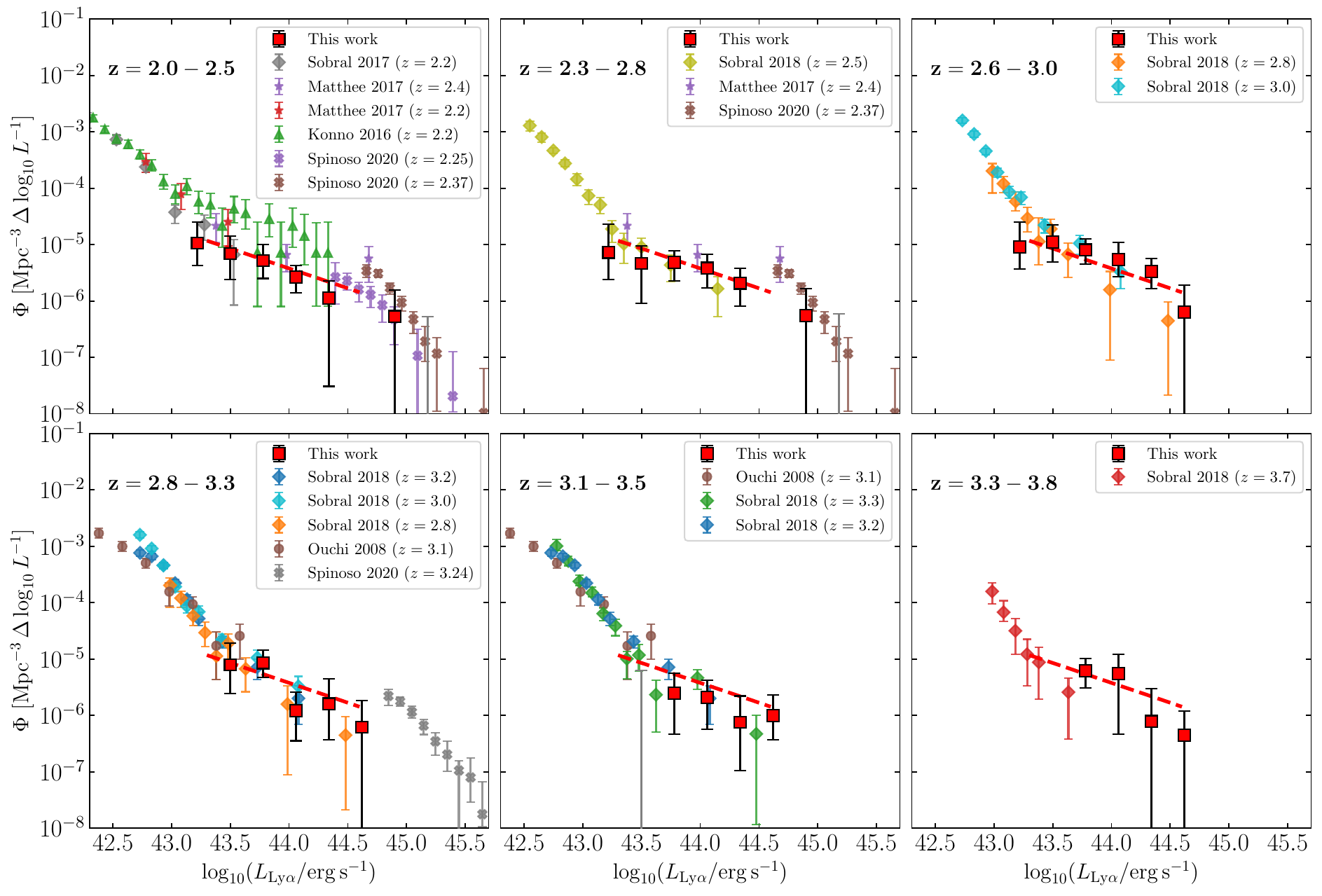}
    \caption{Ly$\alpha$\xspace LFs for six intervals in redshift. We show the full redshift range power-law fit for reference (dashed red line). The redshift bins showed in this figure overlap and are therefore correlated. The Ly$\alpha$\xspace LF shows no significant evolution with redshift within the given uncertainties.}
    \label{fig:multilf}
\end{figure*}

We compute the Ly$\alpha$\xspace LF for every redshift interval listed in Table~\ref{tab:NB_z_intervals} through the procedure explained in Sect.~\ref{sec:lyaLF_computation}. We use the candidate sample of 67 objects obtained after the visual inspection (see Sect.~\ref{sec:visual_inspection}).
Using this configuration, the full redshift range at which we probe the Ly$\alpha$\xspace LF is $2.05<z<3.75$. For $z\gtrsim 3.8$, the available QSO data in the SDSS DR16 starts to become scarce, thus limiting the effectiveness of our mock to compute the LF corrections (see Sect.~\ref{sec:QSO_mock}). Furthermore, the completeness of our sample drops drastically for $z\gtrsim 3.5$ (see Fig.~\ref{fig:combined_puricomp}). With the miniJPAS\&J-NEP dataset we are able to estimate the Ly$\alpha$\xspace LF in the intermediate luminosity regime ({ $\mathrm{43.5\gtrsim \log_{10}(L_{\mathrm{Ly}\alpha} / \text{erg\,s$^{-1}$\xspace})\lesssim 44.5}$}). This is the regime where the contribution of Ly$\alpha$\xspace emitting AGN begins to produce a clear deviation from a Schechter exponential decay of the Ly$\alpha$\xspace LF \citep[see e.g.,][]{Konno2016,Sobral2018,Zhang2021}. Our analysis at the faint end of the LF is limited by the depth of miniJPAS\&J-NEP (i.e., $r\sim 24$ at $5\sigma$), while at the bright-end ($\log_{10}(L_{\mathrm{Ly}\alpha} / \text{erg\,s$^{-1}$\xspace})\sim45$) our results are limited by low number counts and cosmic variance. In this regime, the determinations of \cite{Zhang2021} and \cite{Liu2022b} present an exponential decay.

As we explained in Sect.~\ref{sec:visual_inspection}, we confidently remove $\sim 75\%$ and $\sim 100\%$ of the contaminants coming from low-$z$ galaxies and QSOs,  respectively. For the estimation of the Ly$\alpha$\xspace LFs, we remove the securely identified contaminants from the candidate sample, and correct the 2D purity estimates according to the fraction of contaminants withdrawn after visual inspection.

\subsubsection{Evolution of the Ly$\alpha$\xspace LF with redshift}

We stress that our necessity to group NBs in order to increase the number counts in each $z$ bin is only due to the small area surveyed by miniJPAS. On the other hand, we expect that our method will be able to produce a reliable LF determination for each NB as soon as a wide-enough area of the J-PAS survey will be observed. This upcoming possibility will allow to study the Ly$\alpha$\xspace LF evolution with an unprecedented redshift detail. Therefore, the results presented in the following may be regarded as a proof-of-concept for these kind of tomographic analysis of the Ly$\alpha$\xspace LF (further discussion in Sect.~\ref{sec:jpas_forecast}).

In Fig.~\ref{fig:multilf} we show the Ly$\alpha$\xspace luminosity functions for different bins of redshift, ranging from $z=2.05$ to $z=3.75$. We compare our Ly$\alpha$\xspace LF estimates to previous determinations in the literature. Several works explore the faint and intermediate regime of the Ly$\alpha$\xspace LF ($43.3<\log_{10}(L_{\mathrm{Ly}\alpha} / \text{erg\,s$^{-1}$\xspace})<44$), at the transition between the population of SFG and AGN LAEs \citep{Ouchi2008, Blanc2011, Konno2016, Sobral2017, Matthee2017b}. Our measurements of the Ly$\alpha$\xspace LF at every redshift interval is compatible with all of these works.

\begin{figure}
    \centering
    \resizebox{\hsize}{!}{\includegraphics{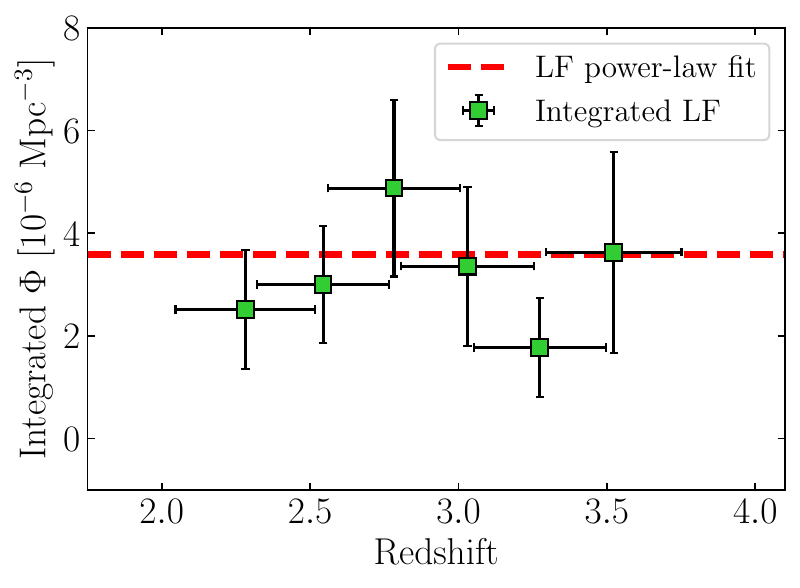}}
    \caption{Integrated LFs in $43.6<\log_{10}(L_{\mathrm{Ly}\alpha} / \text{erg\,s$^{-1}$\xspace})<44.8$ in the six intervals of redshift shown in Fig.~\ref{fig:multilf}. The chosen $L_{\mathrm{Ly}\alpha}$ interval is that where all six realizations of the Ly$\alpha$\xspace LF are well defined. The horizontal error bars represent the width of the redshift interval. The dashed red line represents the integral of the best power-law fit presented in Sect.~\ref{sec:sch_fit}}
    \label{fig:int_LF}
\end{figure}

Our data does not show evidence of evolution with redshift of the Ly$\alpha$\xspace LF within the given uncertainties. In Fig.~\ref{fig:int_LF} we show the integral in the range $43.6<\log_{10}(L_{\mathrm{Ly}\alpha} / \text{erg\,s$^{-1}$\xspace})<44.8$ of the Ly$\alpha$\xspace LFs estimated for each of the six intervals in redshift. The integral is computed as the sum of the LF bins multiplied by the width of the bins.

\begin{figure}
    \centering
    \resizebox{\hsize}{!}{\includegraphics{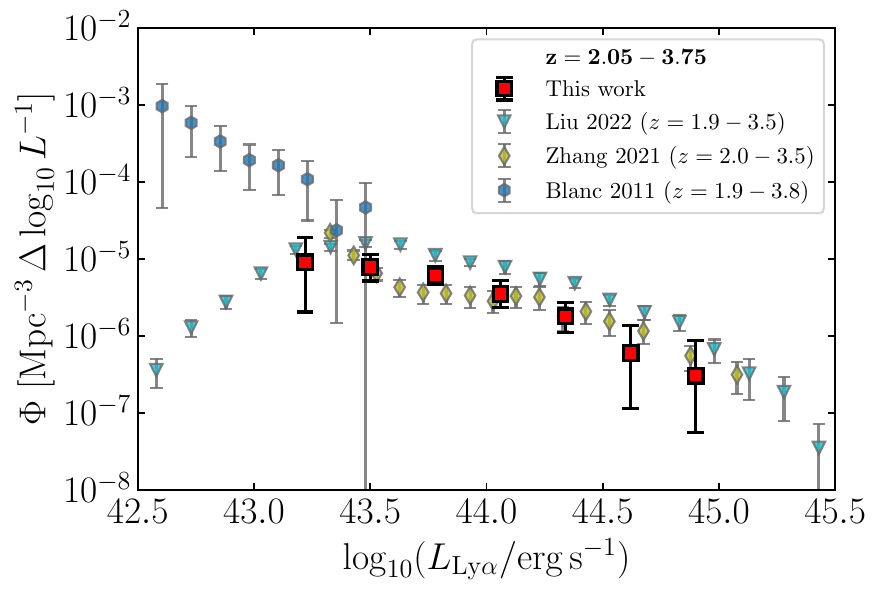}}
    \caption{Ly$\alpha$\xspace LF over the whole $2.05<z<3.75$ range (red squares and black error bars), obtained by combining data from all the NBs we employ. Our estimation covers the intermediate regime of the Ly$\alpha$\xspace LF, where the transition between the SFG and the QSO populations is expected.}
    \label{fig:LF}
\end{figure}

We also estimate the Ly$\alpha$\xspace LF in the full redshift range covered by our selection. Fig.~\ref{fig:LF} shows the Ly$\alpha$\xspace LF computed through the usual procedure (see Sect.~\ref{sec:lyaLF_computation}) but using all the realizations of the LF of every redshift bin. We compare our results to three past realizations of the Ly$\alpha$\xspace LF which cover similar redshift ranges \citep[i. e.][]{Blanc2011, Zhang2021, Liu2022b}.

\subsubsection{Schechter function and power-law fits}\label{sec:sch_fit}

\begin{figure*}
    \centering
    \includegraphics[width=0.49\linewidth]{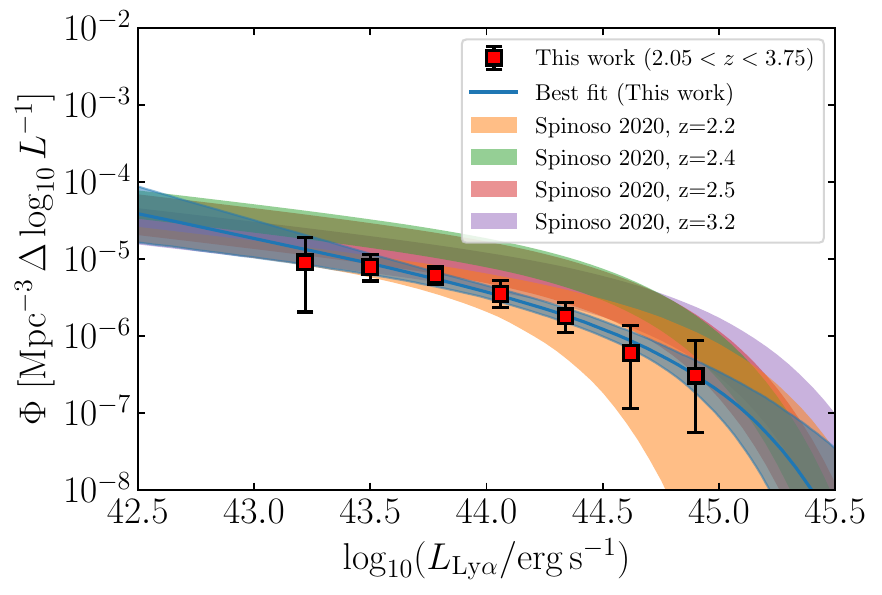}
    \includegraphics[width=0.49\linewidth]{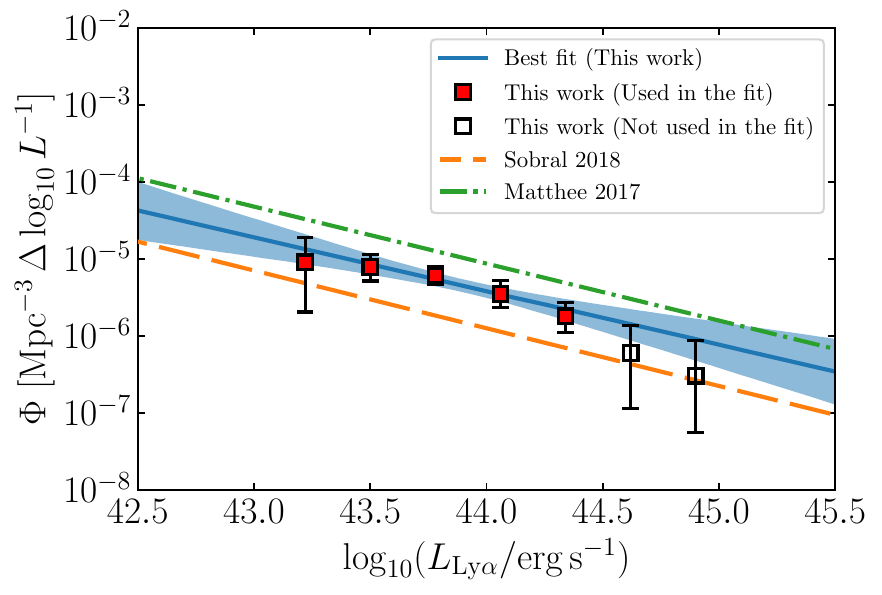}
    \caption{Schechter and power-law fits of the $2.05<z<3.75$ Ly$\alpha$ LF. Left panel: Schechter best fit and $1\sigma$ confidence region for the fit of the full $z$ range of this work (blue solid line and shaded area). We compare our result with the Schechter fits in \protect\cite{Spinoso2020} for 4 NBs of J-PLUS. Our Schechter fit is in line with the fits in \protect\cite{Spinoso2020} for $z=2.2, 2.5$ and $3.2$ at the faintest bins of our LF. We advise some caution when considering our results at the brightest luminosity (i.e., $\log_{10}(L_{\mathrm{Ly}\alpha} / \text{erg\,s$^{-1}$\xspace})>45.5$). Indeed our results in this regime are limited by our survey  area.
    Right panel: Power-law fit to our Ly$\alpha$\xspace LF at $\log_{10}(L_{\mathrm{Ly}\alpha} / \text{erg\,s$^{-1}$\xspace})<44.5$. The bins of the LF represented with empty squares are not used in this fit in particular. We compare to the power-law fits in \protect\cite{Matthee2017b, Sobral2018} in the same $L_{\mathrm{Ly}\alpha}$ regime. The shaded blue area marks the $1\sigma$ confidence region. Our constraint on the power-law slope $A$ for the faint-end of the AGN/QSO Ly$\alpha$\xspace LF is consistent with the other realizations shown in this plot within a $1\sigma$ confidence interval.}
    \label{fig:sch_pl_fit}
\end{figure*}

We use an MCMC algorithm in order to constrain the three parameters of a Schechter function,
\begin{equation}
    \Phi[L]\mathrm{d}\left(\log_{10} L\right)=\log 10\cdot\Phi^*\cdot\left(\frac{L}{L^*}\right)^{\alpha + 1}\cdot e^{-L/L^*}\mathrm{d}\left(\log_{10} L\right)\,,
\end{equation}
namely: the normalization $\Phi^*$, the characteristic Ly$\alpha$\xspace luminosity $L^*$ and the faint-end slope $\alpha$. As already discussed in Sect.~\ref{sec:Lya_LF}, our ability to reliably sample the bright-end of the Ly$\alpha$\xspace LF is limited by the surveyed area. Our data can measure the Ly$\alpha$\xspace LF up to $\log_{10}(L_{\mathrm{Ly}\alpha} / \text{erg\,s$^{-1}$\xspace})\sim 44.5$. Past works have shown that at higher luminosity than this, the Ly$\alpha$\xspace LF shows significant deviations from a simple power-law, either in the form of a Schechter exponential decay, or a ``broken power law'' (see \citealt{Spinoso2020, Zhang2021, Liu2022b}).
Due to this limitation, we use a gaussian with $\mu=44.65$ and $\sigma=0.7$ as prior distribution for $L^*$, based on the results of \cite{Spinoso2020} and \cite{Zhang2021}. We use wide, flat priors for the parameters $\Phi^*$ and $\alpha$. We obtain a best fitting Schechter function with $\log_{10} (\Phi^* / \text{Mpc$^{-3}$})=-6.30^{+0.48}_{-0.70}$, $\log_{10} (L^*/ \mathrm{erg\,s}^{-1})=44.85^{+0.50}_{-0.32}$ and $\alpha=-1.65^{+0.29}_{-0.27}$. Our constraints on the Schechter parameters are compatible within $1\sigma$ to those found by \cite{Spinoso2020} (see Table~\ref{tab:sch_parameters}). Our results are as well compatible within $1\sigma$ with the analog values obtained by \cite{Zhang2021} fitting the AGN/QSO Ly$\alpha$\xspace LF to a double power-law (DPL). In left panel of Fig.~\ref{fig:sch_pl_fit} we show our Schechter fit and compare it to the fits in \cite{Spinoso2020} for four NBs of J-PLUS.

\begin{table*}
    \centering
    \caption{Fit parameters obtained for the QSO/AGN Ly$\alpha$\xspace LF in the literature.}
    \label{tab:sch_parameters}
    \begin{tabular}{ccccc}
    \toprule
    Reference & Redshift & $\log_{10}(\Phi^*/\text{Mpc}^{-3})$ & $\log_{10}(L^*/\text{erg\,s}^{-1})$ & $\alpha$ \\
    \midrule
    This work & 2.05--3.75 & $-6.30^{+0.48}_{-0.70}$ & $44.85^{+0.50}_{-0.32}$ & $-1.65^{+0.29}_{-0.27}$ \\
    \cite{Spinoso2020} & 2.25 & $-5.73^{+0.52}_{-0.85}$ & $44.54^{+0.43}_{-0.35}$ & $-1.35$ (fixed) \\
    \cite{Spinoso2020} & 2.37 & $-5.33^{+0.36}_{-0.52}$ & $44.60^{+0.29}_{-0.21}$ & $-1.35$ (fixed) \\
    \cite{Spinoso2020} & 2.53 & $-5.44^{+0.34}_{-0.54}$ & $44.63^{+0.30}_{-0.22}$ & $-1.35$ (fixed) \\
    \cite{Spinoso2020} & 3.24 & $-5.67^{+0.42}_{-0.57}$ & $44.87^{+0.32}_{-0.26}$ & $-1.35$ (fixed) \\
    \cite{Zhang2021} & 2--3.2 & $-5.85^{+0.34}_{-0.36}$ & $44.60^{+0.32}_{-0.50}$ & $-1.2^{+0.5}_{-0.2}$ \\
    \bottomrule
    \end{tabular}
\end{table*}

The LF estimated by our data in the full redshift range is better described by a power-law. This can be interpreted as our data representing the faint-end of a Schechter function for the QSO/AGN component of the global Ly$\alpha$\xspace LF\footnote{A Schechter function can be approximated by a power-law for $L\ll L^*$}. We fit our Ly$\alpha$\xspace LF for $\log_{10}(L_{\mathrm{Ly}\alpha} / \text{erg\,s$^{-1}$\xspace}) < 44.5$ to a power-law of the form:
\begin{multline}
    \log_{10}\Phi_\text{PL}[\log_{10}(L_{\mathrm{Ly}\alpha} / \text{erg\,s$^{-1}$\xspace})] =\\
    A\cdot(\log_{10}(L_{\mathrm{Ly}\alpha} / \text{erg\,s$^{-1}$\xspace}) - 43.5) + B.
\end{multline}
We obtain the following results for the power-law fit: $A=-0.70^{+0.25}_{-0.25}$, $B=-5.07^{+0.14}_{-0.13}$. Past works that have fitted a power-law in the same regime of the AGN Ly$\alpha$\xspace LF: $A=-0.74\pm0.17$ \cite{Matthee2017b}; $A=-0.75\pm0.17$ \cite{Sobral2018}. Our estimation of the power-law slope $A$, is consistent with these past realizations.

\subsection{QSO/AGN fraction}\label{sec:qso_frac}

\begin{figure}
    \centering
    \resizebox{\hsize}{!}{\includegraphics{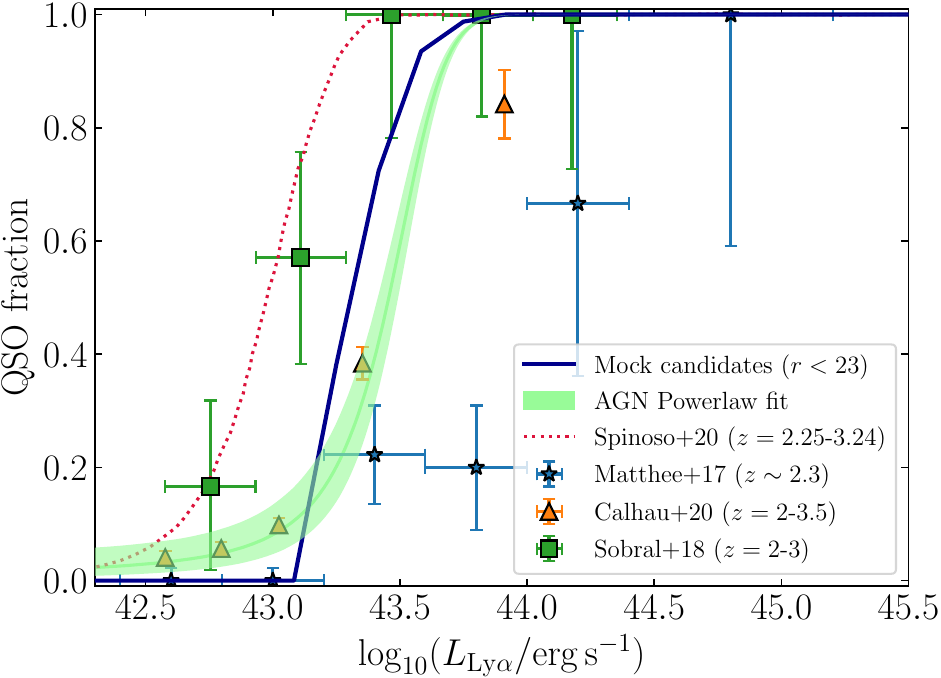}}
    \caption{Fraction of QSOs in the candidate sample from the mock at $2.05<z<3.75$. We represent our estimation of the QSO fraction (green solid line and green shaded area for 1$\sigma$ uncertainty) based on the power-law fit to our Ly$\alpha$\xspace LF and the SFG Ly$\alpha$\xspace LF of \protect\cite{Sobral2018}. We also show the QSO fraction extracted from our mocks (blue solid line). We compare our results to previous realizations from past works.
    }
    \label{fig:qso_frac}
\end{figure}

Fig.~\ref{fig:qso_frac} shows the fraction of AGN within the sample of sources with $r<23$ extracted from our mock data (blue solid line). At $\log_{10}(L_{\mathrm{Ly}\alpha} / \text{erg\,s$^{-1}$\xspace})\gtrsim 43.75$ the fraction of { AGN} is greater than $\sim 90\%$. This is in line with past works which analyzed high-$z$ samples of bright LAEs and found that the AGN fraction approaches to 100\% at $\log_{10}(L_{\mathrm{Ly}\alpha} / \text{erg\,s$^{-1}$\xspace})\gtrsim 44$ (e.g., \citealt{Matthee2017b, Wold2017, Sobral2018, Calhau2020}). By considering the depth of the miniJPAS and J-NEP observations, we conclude that our analysis can primarily focus on the luminosity range $\log_{10}(L_{\mathrm{Ly}\alpha} / \text{erg\,s$^{-1}$\xspace}) > 43$, where the class of AGN/QSOs is expected to be numerically dominant. Hence, if we assume that our power-law fit (see Sect.~\ref{sec:sch_fit}) describes a regime of the Ly$\alpha$\xspace LF populated entirely by QSOs, we can compute the intrinsic QSO fraction assuming that the SFG population is well described by the Ly$\alpha$\xspace LF presented in \cite{Sobral2018}, which is measured in the range $42.5<\log_{10}(L_{\mathrm{Ly}\alpha} / \text{erg\,s$^{-1}$\xspace})<44$  using a sample of $\sim 3700$ LAEs showing no X-ray emission. This result is shown as the green solid line in Fig.~\ref{fig:qso_frac}. The estimated intrinsic QSO fraction drops faster than the fraction in our candidate sample at $\log_{10}(L_{\mathrm{Ly}\alpha} / \text{erg\,s$^{-1}$\xspace})\sim 43.75$. This happens because SFGs typically have fainter continua, thus their detection is limited by the magnitude cut of our selection pipeline ($r=24$).

Nevertheless, our results can also populate the intermediate luminosity regime of $43.5 \lesssim \log_{10}(L_{\mathrm{Ly}\alpha} / \text{erg\,s$^{-1}$\xspace}) \lesssim 44.5$ which, to date, it is yet relatively unexplored \citep[see e. g.][]{Zhang2021, Liu2022b}. At this "intermediate" luminosity, bright SFG LAEs may still produce some minor contribution to the Ly$\alpha$\xspace LF. Consequently, we can speculate that our candidate samples contain a small fraction of SFGs. Nevertheless, systematic spectroscopic confirmation of our selected sources would be necessary to draw a definitive conclusion about this point.

\subsection{Expected results in J-PAS}\label{sec:jpas_forecast}

With the eventual release of the full J-PAS dataset, there will be a significantly larger source catalog available with very similar features to the one of miniJPAS\&J-NEP. Therefore, an analogous method to the one described in this work could be applied to the complete J-PAS dataset to build the Ly$\alpha$\xspace LF. The target depth for J-PAS is expected to be slightly shallower than miniJPAS and J-NEP, yet a much larger dataset will allow to describe the SFG-AGN/QSO transition of the Ly$\alpha$\xspace LF with better statistics. As to the bright-end, in this work we are limited by the intrinsic scarcity of extremely bright objects in Ly$\alpha$\xspace. In other words, due to the small area sampled by miniJPAS and J-NEP, our LF at $\log_{10}(L_{\mathrm{Ly}\alpha} / \text{erg\,s$^{-1}$\xspace})>44.5$  is dominated by shot noise and cosmic variance. On the other hand, a larger dataset will allow to estimate the Ly$\alpha$\xspace LF for every individual NB of the filter set. This has the potential to provide a precise and tomographic analysis of the Ly$\alpha$\xspace LF evolution with redshift.

Our work provides the means to infer the expected results for the first hundreds of square degrees of observed J-PAS data. By integrating the power-law fit of our Ly$\alpha$\xspace LF (see Sect.~\ref{sec:sch_fit}), we obtain a predicted number count of $72\pm25$ deg$^{-2}$ of LAEs in the range $43.5 < \log_{10}(L_{\mathrm{Ly}\alpha} / \text{erg\,s$^{-1}$\xspace}) < 44.5$  at $2.05<z<3.75$. The mean recovery rate of our method in this range is $C\sim60\%$. This results translate into $\sim4300$ recovered LAEs for the first 100 deg$^2$, and $\sim3.7\cdot 10^5$ in the full 8500 deg$^2$ expected at the completion of the J-PAS survey. On the other hand, our work cannot reliably provide a direct estimate for the number count of objects with $\log_{10}(L_{\mathrm{Ly}\alpha} / \text{erg\,s$^{-1}$\xspace})>44.5$. Therefore, for the analysis of the Ly$\alpha$\xspace LF bright-end, we integrate the best Schechter fit of \cite{Spinoso2020}. At this bright regime, the exponential decay component of the Schechter function dominates, rapidly decreasing the number of available candidates with increasing {$L_{\mathrm{Ly}\alpha}$}. The predicted number count of QSOs with Ly$\alpha$\xspace emission is $\sim 24.5$ deg$^{-2}$ for $\log_{10}(L_{\mathrm{Ly}\alpha} / \text{erg\,s$^{-1}$\xspace}) > 44.5$  and $\sim 1.7$ deg$^{-2}$ for $\log_{10}(L_{\mathrm{Ly}\alpha} / \text{erg\,s$^{-1}$\xspace})>45$. We define the upper limit $L_{\mathrm{Ly}\alpha}^\mathrm{max}$ as the maximum Ly$\alpha$\xspace luminosity for which the predicted average number of candidates with $\log_{10}(L_{\mathrm{Ly}\alpha} / \text{erg\,s$^{-1}$\xspace})>\log_{10}(L_{\mathrm{Ly}\alpha}^\mathrm{max} / \mathrm{erg\,s}^{-1})$ in a given survey area is $>1$ (at $2.05<z<3.75$). For 1 deg$^2$, this limit is $\log_{10}(L_{\mathrm{Ly}\alpha}^\mathrm{max} / \mathrm{erg\,s}^{-1})\sim45.07$, consistent with our candidate sample. This limit increases to $\log_{10}(L_{\mathrm{Ly}\alpha}^\mathrm{max} / \mathrm{erg\,s}^{-1})\sim45.44$  over 100 deg$^2$, $\log_{10}(L_{\mathrm{Ly}\alpha}^\mathrm{max} / \mathrm{erg\,s}^{-1})\sim45.56$  over 1000 deg$^2$, and $\log_{10}(L_{\mathrm{Ly}\alpha}^\mathrm{max} / \mathrm{erg\,s}^{-1})\sim 45.65$  over 8500 deg$^2$. At this bright-end regime, the estimated completeness of our method is $C\gtrsim80\%$ (see Sect.~\ref{sec:puricomp1d}).

Furthermore, as stated before, an increase in the volume of data will allow to measure the Ly$\alpha$\xspace LF in smaller bins of redshift. For example, with the release of the first 100 $\deg^2$ of J-PAS, it will be possible to estimate the Ly$\alpha$\xspace LF for each individual NB up to $z\sim 4$ (in intervals of $\Delta z\sim 0.12$) with $\log_{10}(L_{\mathrm{Ly}\alpha}^\mathrm{max} / \mathrm{erg\,s}^{-1})\sim45.2$  ($\log_{10}(L_{\mathrm{Ly}\alpha}^\mathrm{max} / \mathrm{erg\,s}^{-1})\sim45.54$  for 8500 deg$^2$). This will provide a remarkable measurement of the Ly$\alpha$\xspace LF evolution with redshift.

\section{Summary}\label{sec:summary}

In this work we developed a method to detect sources with Ly$\alpha$\xspace emission within the J-PAS filter system, and applied it to the J-PAS Pathfinder surveys: miniJPAS and J-NEP in order to estimate the Ly$\alpha$\xspace LF in the redshift range $2.05<z<3.75$. We summarize here our main results.

First, we build a mock catalog of LAEs and their contaminants in order to test and calibrate the accuracy of our LAE selection method, as well as to compute the corrections needed to estimate the Ly$\alpha$\xspace LF. { The mock is composed of four populations: (i) QSOs with $z>2$, which are potential LAE candidates; (ii) QSO interlopers, that is, $z<2$ QSOs detected via strong emission lines different than Ly$\alpha$\xspace ; (iii) SFG LAEs at $z>2$ and (iv) low-$z$ galaxies ($0<z<2$).} By studying the performance of our selection method on our mock, we are able to build 2D maps of purity and number count corrections in terms of {$L_{\mathrm{Ly}\alpha}$} and $r$ band magnitude { (see Fig.~\ref{fig:2dpuricomp})}. These 2D maps are used in order to compute the corrections for the Ly$\alpha$\xspace LF estimate.

Our method retrieves a sample of 127 LAE candidates with $2.05<z<3.75$. From our mock, we show that our sample is $>75\%$ complete and $>60\%$ pure at $\log_{10}(L_{\mathrm{Ly}\alpha} / \text{erg\,s$^{-1}$\xspace})\gtrsim 43.75$. This sample was obtained from a total effective area of $\sim 1.14$ deg$^{2}$. Through a visual inspection of the NB images and photospectra, we confirm 32 candidates as QSO LAEs, 39 as low-$z$ galaxies, 21 as QSOs with $z<2$. The remaining 35 candidates are left with no visual classification.
    
Using the data from our LAE candidate sample { (32 visually confirmed QSO LAEs and 35 candidates with no clear visual identification)} we are able to determine the Ly$\alpha$\xspace LF in the intermediate-bright luminosity range ($43.5\lesssim\log_{10}(L_{\mathrm{Ly}\alpha} / \text{erg\,s$^{-1}$\xspace})\lesssim 45$). At the faint end of this regime, we are limited by the depth of our survey ($r\sim 24$ at $5\sigma$), while at the brightest end we are limited by the survey area.

We fit Schechter function and power-law models to our estimated Ly$\alpha$\xspace LF in the whole redshift range used in this work $2.05<z<3.75$. The resulting Schechter parameters are: $\log_{10} (\Phi^* / \text{Mpc$^{-3}$})=-6.30^{+0.48}_{-0.70}$, $\log_{10} (L^*/ \mathrm{erg\,s}^{-1})=44.85^{+0.50}_{-0.32}$, $\alpha=-1.65^{+0.29}_{-0.27}$. The LF at $\log_{10}(L_{\mathrm{Ly}\alpha} / \text{erg\,s$^{-1}$\xspace}) < 44.5$  is fitted to a power-law of the form $A\cdot(\log_{10}(L_{\mathrm{Ly}\alpha} / \text{erg\,s$^{-1}$\xspace}) - 43.5) + B$ with $A=0.70^{+0.25}_{-0.25}$ and $B=-5.07^{+0.14}_{-0.13}$. These parameters are compatible to previous results and show the potential of our method when the larger survey area of J-PAS will be available.

Finally, we give predictions about the performance of the method in the eventual release of the J-PAS data. With the completion of the first hundreds of square degrees of J-PAS, it will already be possible to resolve the Ly$\alpha$\xspace LF in redshift bins of $\Delta z=0.12$ in the luminosity interval $43.5\lesssim\log_{10}(L_{\mathrm{Ly}\alpha} / \text{erg\,s$^{-1}$\xspace})\lesssim 45$.
{ This achievement will allow to study the evolution of the Ly$\alpha$ AGN LF with redshift with unprecedented precision.}


\begin{acknowledgements}
{We thank the anonymous scientific referee for their useful comments that have helped to refine the final version of the manuscript.}

We thank Mirjana Povi\'c for insightful discussion on the results of this work.

This work has been funded by project PID2019-109592GBI00/AEI/10.13039/501100011033 from the Spanish Ministerio de Ciencia e Innovación (MCIN)—Agencia Estatal de Investigación, by the Project of Excellence Prometeo/2020/085 from the Conselleria d’Innovació Universitats, Ciència i Societat Digital de la Generalitat Valenciana.

The authors acknowledge the financial support from the MCIN with funding from the European Union NextGenerationEU and Generalitat Valenciana in the call Programa de Planes Complementarios de I+D+i (PRTR 2022). Project (VAL-JPAS), reference ASFAE/2022/025.

D.S. acknowledges funding from National Key R\&D Program of China (grant no. 2018YFA0404503), the National Science Foundation of China (grant no. 12073014), the science research grants from the China Manned Space Project with no. CMS-CSST2021-A05 and Tsinghua University Initiative Scientific Research Program (no. 20223080023)

RGD acknowledges financial support from the grant CEX2021-001131-S funded by MCIN/AEI/10.13039/501100011033 and PID2019-109067-GB100.

I.M. acknowledges financial support from the Severo Ochoa grant CEX2021-001131-S funded by MCIN/AEI/10.13039/501100011033 and PID2019-106027GB-C41.

This work is based on observations made with the JST250 telescope and PathFinder camera for Mini J-PAS project at the Observatorio Astrofísico de Javalambre, in Teruel, owned, managed and operated by the Centro de Estudios de Física del Cosmos de Aragón. We acknowledge the OAJ Data Processing and Archiving Unit (UPAD) for reducing and calibrating the OAJ data used in this work.

Funding for the J-PAS Project has been provided by the Governments of Spain and Aragón through the Fondo de Inversión de Teruel, European FEDER funding and the Spanish Ministry of Science, Innovation and Universities, and by the Brazilian agencies FINEP, FAPESP, FAPERJ and by the National Observatory of Brazil. Additional funding was also provided by the Tartu Observatory and by the J-PAS Chinese Astronomical Consortium.

Funding for the Sloan Digital Sky Survey V has been provided by the Alfred P. Sloan Foundation, the Heising-Simons Foundation, the National Science Foundation, and the Participating Institutions. SDSS acknowledges support and resources from the Center for High-Performance Computing at the University of Utah. The SDSS web site is \url{www.sdss.org}.

SDSS is managed by the Astrophysical Research Consortium for the Participating Institutions of the SDSS Collaboration, including the Carnegie Institution for Science, Chilean National Time Allocation Committee (CNTAC) ratified researchers, the Gotham Participation Group, Harvard University, Heidelberg University, The Johns Hopkins University, L’Ecole polytechnique f\'{e}d\'{e}rale de Lausanne (EPFL), Leibniz-Institut f\"{u}r Astrophysik Potsdam (AIP), Max-Planck-Institut f\"{u}r Astronomie (MPIA Heidelberg), Max-Planck-Institut f\"{u}r Extraterrestrische Physik (MPE), Nanjing University, National Astronomical Observatories of China (NAOC), New Mexico State University, The Ohio State University, Pennsylvania State University, Smithsonian Astrophysical Observatory, Space Telescope Science Institute (STScI), the Stellar Astrophysics Participation Group, Universidad Nacional Aut\'{o}noma de M\'{e}xico, University of Arizona, University of Colorado Boulder, University of Illinois at Urbana-Champaign, University of Toronto, University of Utah, University of Virginia, Yale University, and Yunnan University.

HETDEX is led by the University of Texas at Austin McDonald Observatory and Department of Astronomy with participation from the Ludwig-Maximilians-Universität M\"unchen, Max-Planck-Institut f\"ur Extraterrestrische Physik (MPE), Leibniz-Institut f\"ur Astrophysik Potsdam (AIP), Texas A\&M University, Pennsylvania State University, Institut f\"ur Astrophysik Göttingen, The University of Oxford, Max-Planck-Institut f\"ur Astrophysik (MPA), The University of Tokyo and Missouri University of Science and Technology.

Observations for HETDEX were obtained with the Hobby-Eberly Telescope (HET), which is a joint project of the University of Texas at Austin, the Pennsylvania State University, Ludwig-Maximilians-Universität M\"unchen, and Georg-August-Universität Göttingen. The HET is named in honor of its principal benefactors, William P. Hobby and Robert E. Eberly. The Visible Integral-field Replicable Unit Spectrograph (VIRUS) was used for HETDEX observations. VIRUS is a joint project of the University of Texas at Austin, Leibniz-Institut f\"ur Astrophysik Potsdam (AIP), Texas A\&M University, Max-Planck-Institut f\"urExtraterrestrische Physik (MPE), Ludwig-Maximilians-Universität M\"unchen, Pennsylvania State University, Institut f\"ur Astrophysik Göttingen, University of Oxford, and the Max-Planck-Institut fur Astrophysik (MPA).

Funding for HETDEX has been provided by the partner institutions, the National Science Foundation, the State of Texas, the US Air Force, and by generous support from private individuals and foundations.

\end{acknowledgements}


\bibliographystyle{aa}
\bibliography{my_bibliography}

\appendix

\section{Candidates catalog}

Table \ref{tab:selected_catalog} lists the 91 candidates selected by our method as LAEs, containing only the sources with \texttt{morph\_prob\_star} < 0.1, as explained in { Sect.}~\ref{sec:LAEs_morph}. This entire catalog is representative of what can be directly obtained of any J-PAS data set. The sources used in the Ly$\alpha$ LF estimation are marked in blue. As explained in Sect. \ref{sec:visual_inspection}, the sub-sample of objects used for the Ly$\alpha$ LF are chosen after a visual inspection and cross-matches with SDSS DR16 and HETDEX Source Catalog 1.

The description of each column is the following:
\begin{itemize}
    \item Field: The sub-field of miniJPAS or J-NEP in which the source was detected by \texttt{SExtractor}.
    \item tile\_id: The ID identifying the selection tile of miniJPAS or J-NEP.
    \item number: Unique ID of each object within each tile in the dual-mode catalogs of miniJPAS or J-NEP.
    \item RA, DEC: Right ascension and declination.
    \item $z_\mathrm{NB}$: Ly$\alpha$ redshift of the central wavelength of the line detection. The provided error is inferred from the FWHM of the NB.
    \item Selected NB: Name of the NB of the Ly$\alpha$ line detection.
    \item SDSS spCL, SDSS $z_\mathrm{spec}$: Spectroscopic class and redshift of the counterpart in SDSS DR16.
    \item HETDEX spCl, HETDEX $z_\mathrm{spec}$: Spectroscopic class and redshift of the counterpart in HETDEX Source Catalog 1.
    \item VI class: Class assigned in the visual inspection (see Sect. \ref{sec:visual_inspection})
    \item $\log_{10}L_{\mathrm{Ly}\alpha}$: Estimated Ly$\alpha$ luminosity.
    \item EW$_0^{\mathrm{Ly}\alpha}$: Estimated rest-frame EW of the Ly$\alpha$ line.
    \item $P^\mathrm{2D}$: Purity extracted from the 2D correction maps according to the source $L_{\mathrm{Ly}\alpha}$ and $r$ magnitude (see Sect. \ref{sec:2dpuricomp}). This value can be seen as the probability of each source to be a true positive LAE.
\end{itemize}

{\onecolumn
\begin{landscape}

\begin{longtable}{ccccccccccccccc}
\toprule
Field & tile\textunderscore id & number & RA & DEC & \makecell{$z_\text{NB}$ \\ $(\pm 0.04)$} & Sel. NB & \makecell{SDSS \\ SpCl} & \makecell{SDSS \\ $z_\text{spec}$} & \makecell{HETDEX \\ SpCl} & \makecell{HETDEX \\ $z_\mathrm{spec}$} & VI class & \makecell{$\log_{10} L_{\mathrm{Ly}\alpha}$ \\ (erg\,s$^{-1}$)} & \makecell{EW$_0^{\mathrm{Ly}\alpha}$ \\ (\AA)} & $P(\text{LAE})$\\
\midrule
AEGIS001 & 2241 & 468 & 14h15m42.2s & 52\degr22\arcmin7.0\arcsec & 2.11 & J0378 & QSO & 1.00 & agn & 1.00 & Cont. QSO & 43.69$^{ + 0.11}_{ - 0.14}$ & 62$\pm$10 & 0.00 \\ 
        \color{blue}AEGIS001 & \color{blue}2241 & \color{blue}9344 & \color{blue}14h17m38.8s & \color{blue}52\degr23\arcmin33.1\arcsec & \color{blue}2.11 & \color{blue}J0378 & \color{blue}QSO & \color{blue}2.16 & \color{blue}agn & \color{blue}2.16 & \color{blue}QSO LAE & \color{blue}44.11$^{ + 0.06}_{ - 0.07}$ & \color{blue}184$\pm$13 & \color{blue}1.00 \\ 
        \color{blue}AEGIS001 & \color{blue}2241 & \color{blue}11097 & \color{blue}14h18m42.3s & \color{blue}52\degr36\arcmin43.9\arcsec & \color{blue}2.11 & \color{blue}J0378 & \color{blue}QSO & \color{blue}2.12 & \color{blue}- & \color{blue}- & \color{blue}- & \color{blue}43.63$^{ + 0.12}_{ - 0.16}$ & \color{blue}93$\pm$16 & \color{blue}0.98 \\ 
        AEGIS001 & 2241 & 13090 & 14h16m58.4s & 52\degr48\arcmin6.3\arcsec & 2.11 & J0378 & QSO & 1.96 & - & - & Cont. QSO & 43.73$^{ + 0.10}_{ - 0.13}$ & 37$\pm$6 & 0.00 \\ 
        AEGIS001 & 2241 & 14254 & 14h18m2.0s & 52\degr35\arcmin14.9\arcsec & 2.11 & J0378 & QSO & 1.49 & - & - & Cont. QSO & 43.92$^{ + 0.07}_{ - 0.09}$ & 37$\pm$4 & 0.00 \\ 
        \color{blue}AEGIS001 & \color{blue}2241 & \color{blue}17351 & \color{blue}14h18m32.8s & \color{blue}52\degr23\arcmin50.0\arcsec & \color{blue}2.11 & \color{blue}J0378 & \color{blue}QSO & \color{blue}2.04 & \color{blue}- & \color{blue}- & \color{blue}- & \color{blue}43.83$^{ + 0.08}_{ - 0.10}$ & \color{blue}42$\pm$5 & \color{blue}0.87 \\ 
        AEGIS001 & 2241 & 18457 & 14h17m36.0s & 52\degr30\arcmin29.8\arcsec & 2.11 & J0378 & QSO & 0.99 & - & - & Cont. QSO & 43.62$^{ + 0.12}_{ - 0.17}$ & 43$\pm$8 & 0.00 \\ 
        AEGIS001 & 2241 & 19989 & 14h19m11.4s & 52\degr32\arcmin34.7\arcsec & 2.11 & J0378 & QSO & 1.41 & - & - & Cont. QSO & 43.44$^{ + 0.16}_{ - 0.26}$ & 73$\pm$19 & 0.00 \\ 
        AEGIS001 & 2241 & 13288 & 14h17m37.3s & 52\degr42\arcmin36.7\arcsec & 2.21 & J0390 & QSO & 0.15 & - & - & Cont. QSO & 43.10$^{ + 0.14}_{ - 0.21}$ & 42$\pm$9 & 0.00 \\ 
        AEGIS001 & 2241 & 14038 & 14h18m42.9s & 52\degr29\arcmin19.0\arcsec & 2.21 & J0390 & QSO & 1.08 & - & - & Cont. QSO & 43.07$^{ + 0.15}_{ - 0.23}$ & 55$\pm$13 & 0.00 \\ 
        AEGIS001 & 2241 & 15333 & 14h18m13.1s & 52\degr31\arcmin10.1\arcsec & 2.21 & J0390 & - & - & - & - & Low-$z$ Gal. & 43.04$^{ + 0.16}_{ - 0.25}$ & 74$\pm$18 & 0.00 \\ 
        AEGIS001 & 2241 & 15615 & 14h17m23.9s & 52\degr38\arcmin6.0\arcsec & 2.21 & J0390 & QSO & 1.52 & - & - & Cont. QSO & 43.79$^{ + 0.05}_{ - 0.06}$ & 50$\pm$3 & 0.00 \\ 
        AEGIS001 & 2241 & 15867 & 14h18m16.2s & 52\degr29\arcmin40.7\arcsec & 2.29 & J0400 & QSO & 1.61 & - & - & Cont. QSO & 44.10$^{ + 0.05}_{ - 0.05}$ & 44$\pm$2 & 0.00 \\ 
        AEGIS001 & 2241 & 4536 & 14h16m19.6s & 52\degr24\arcmin58.8\arcsec & 2.46 & J0420 & - & - & - & - & Low-$z$ Gal. & 43.47$^{ + 0.15}_{ - 0.23}$ & 66$\pm$16 & 0.00 \\ 
        \color{blue}AEGIS001 & \color{blue}2241 & \color{blue}8524 & \color{blue}14h16m25.5s & \color{blue}52\degr32\arcmin42.6\arcsec & \color{blue}2.54 & \color{blue}J0430 & \color{blue}- & \color{blue}- & \color{blue}agn & \color{blue}2.56 & \color{blue}QSO LAE & \color{blue}43.93$^{ + 0.05}_{ - 0.05}$ & \color{blue}71$\pm$4 & \color{blue}0.99 \\ 
        AEGIS001 & 2241 & 20770 & 14h18m23.3s & 52\degr40\arcmin4.6\arcsec & 2.54 & J0430 & QSO & 1.77 & - & - & Cont. QSO & 44.13$^{ + 0.04}_{ - 0.05}$ & 91$\pm$3 & 0.00 \\ 
        \color{blue}AEGIS001 & \color{blue}2241 & \color{blue}4481 & \color{blue}14h17m23.1s & \color{blue}52\degr15\arcmin14.7\arcsec & \color{blue}2.62 & \color{blue}J0440 & \color{blue}QSO & \color{blue}2.59 & \color{blue}- & \color{blue}- & \color{blue}QSO LAE & \color{blue}44.16$^{ + 0.05}_{ - 0.05}$ & \color{blue}43$\pm$2 & \color{blue}1.00 \\ 
        AEGIS001 & 2241 & 2717 & 14h16m50.0s & 52\degr16\arcmin40.2\arcsec & 2.70 & J0450 & - & - & - & - & Low-$z$ Gal. & 43.79$^{ + 0.14}_{ - 0.21}$ & 53$\pm$13 & 0.00 \\ 
        \color{blue}AEGIS001 & \color{blue}2241 & \color{blue}7775 & \color{blue}14h16m9.0s & \color{blue}52\degr33\arcmin23.1\arcsec & \color{blue}2.79 & \color{blue}J0460 & \color{blue}- & \color{blue}- & \color{blue}- & \color{blue}- & \color{blue}- & \color{blue}43.23$^{ + 0.14}_{ - 0.20}$ & \color{blue}96$\pm$21 & \color{blue}0.95 \\ 
        \color{blue}AEGIS001 & \color{blue}2241 & \color{blue}14404 & \color{blue}14h17m35.3s & \color{blue}52\degr38\arcmin51.4\arcsec & \color{blue}2.79 & \color{blue}J0460 & \color{blue}QSO & \color{blue}1.96 & \color{blue}- & \color{blue}- & \color{blue}- & \color{blue}44.43$^{ + 0.04}_{ - 0.04}$ & \color{blue}66$\pm$1 & \color{blue}0.97 \\ 
        \color{blue}AEGIS001 & \color{blue}2241 & \color{blue}15255 & \color{blue}14h17m47.3s & \color{blue}52\degr35\arcmin10.6\arcsec & \color{blue}2.79 & \color{blue}J0460 & \color{blue}- & \color{blue}- & \color{blue}- & \color{blue}- & \color{blue}QSO LAE & \color{blue}43.57$^{ + 0.08}_{ - 0.09}$ & \color{blue}81$\pm$8 & \color{blue}0.91 \\ 
        \color{blue}AEGIS001 & \color{blue}2241 & \color{blue}15524 & \color{blue}14h16m51.6s & \color{blue}52\degr43\arcmin12.4\arcsec & \color{blue}2.79 & \color{blue}J0460 & \color{blue}- & \color{blue}- & \color{blue}- & \color{blue}- & \color{blue}- & \color{blue}43.19$^{ + 0.15}_{ - 0.23}$ & \color{blue}53$\pm$13 & \color{blue}0.83 \\ 
        \color{blue}AEGIS001 & \color{blue}2241 & \color{blue}6762 & \color{blue}14h17m42.0s & \color{blue}52\degr17\arcmin7.3\arcsec & \color{blue}2.87 & \color{blue}J0470 & \color{blue}- & \color{blue}- & \color{blue}agn & \color{blue}2.06 & \color{blue}- & \color{blue}43.44$^{ + 0.13}_{ - 0.19}$ & \color{blue}90$\pm$18 & \color{blue}0.89 \\ 
        \color{blue}AEGIS001 & \color{blue}2241 & \color{blue}20626 & \color{blue}14h18m40.6s & \color{blue}52\degr37\arcmin16.4\arcsec & \color{blue}2.87 & \color{blue}J0470 & \color{blue}- & \color{blue}- & \color{blue}- & \color{blue}- & \color{blue}- & \color{blue}43.49$^{ + 0.12}_{ - 0.16}$ & \color{blue}49$\pm$9 & \color{blue}0.84 \\ 
        \color{blue}AEGIS001 & \color{blue}2241 & \color{blue}4983 & \color{blue}14h16m15.3s & \color{blue}52\degr26\arcmin26.5\arcsec & \color{blue}2.95 & \color{blue}J0480 & \color{blue}- & \color{blue}- & \color{blue}- & \color{blue}- & \color{blue}- & \color{blue}43.52$^{ + 0.13}_{ - 0.18}$ & \color{blue}47$\pm$9 & \color{blue}0.57 \\ 
        \color{blue}AEGIS001 & \color{blue}2241 & \color{blue}20297 & \color{blue}14h18m9.7s & \color{blue}52\degr43\arcmin0.2\arcsec & \color{blue}3.19 & \color{blue}J0510 & \color{blue}QSO & \color{blue}3.19 & \color{blue}- & \color{blue}- & \color{blue}QSO LAE & \color{blue}44.17$^{ + 0.04}_{ - 0.05}$ & \color{blue}73$\pm$4 & \color{blue}0.98 \\ 
        \color{blue}AEGIS001 & \color{blue}2241 & \color{blue}5837 & \color{blue}14h16m11.9s & \color{blue}52\degr28\arcmin44.5\arcsec & \color{blue}3.28 & \color{blue}J0520 & \color{blue}QSO & \color{blue}3.25 & \color{blue}- & \color{blue}- & \color{blue}QSO LAE & \color{blue}43.95$^{ + 0.12}_{ - 0.17}$ & \color{blue}68$\pm$13 & \color{blue}1.00 \\ 
        \color{blue}AEGIS001 & \color{blue}2241 & \color{blue}14553 & \color{blue}14h17m55.3s & \color{blue}52\degr35\arcmin32.8\arcsec & \color{blue}3.36 & \color{blue}J0530 & \color{blue}- & \color{blue}- & \color{blue}- & \color{blue}- & \color{blue}- & \color{blue}43.47$^{ + 0.12}_{ - 0.16}$ & \color{blue}164$\pm$27 & \color{blue}0.91 \\ 
        \color{blue}AEGIS001 & \color{blue}2241 & \color{blue}9742 & \color{blue}14h16m25.3s & \color{blue}52\degr35\arcmin38.3\arcsec & \color{blue}3.60 & \color{blue}J0560 & \color{blue}- & \color{blue}- & \color{blue}- & \color{blue}- & \color{blue}- & \color{blue}44.07$^{ + 0.08}_{ - 0.09}$ & \color{blue}377$\pm$38 & \color{blue}0.98 \\ 
        \color{blue}AEGIS001 & \color{blue}2241 & \color{blue}13746 & \color{blue}14h17m24.2s & \color{blue}52\degr41\arcmin50.8\arcsec & \color{blue}3.69 & \color{blue}J0570 & \color{blue}- & \color{blue}- & \color{blue}- & \color{blue}- & \color{blue}- & \color{blue}44.02$^{ + 0.08}_{ - 0.10}$ & \color{blue}88$\pm$10 & \color{blue}0.97 \\ 
        \color{blue}AEGIS002 & \color{blue}2243 & \color{blue}2395 & \color{blue}14h18m18.5s & \color{blue}52\degr43\arcmin56.1\arcsec & \color{blue}2.11 & \color{blue}J0378 & \color{blue}QSO & \color{blue}2.14 & \color{blue}- & \color{blue}- & \color{blue}QSO LAE & \color{blue}43.92$^{ + 0.06}_{ - 0.08}$ & \color{blue}131$\pm$12 & \color{blue}1.00 \\ 
        \color{blue}AEGIS002 & \color{blue}2243 & \color{blue}5085 & \color{blue}14h19m18.1s & \color{blue}52\degr41\arcmin58.4\arcsec & \color{blue}2.11 & \color{blue}J0378 & \color{blue}QSO & \color{blue}2.03 & \color{blue}- & \color{blue}- & \color{blue}QSO LAE & \color{blue}43.99$^{ + 0.06}_{ - 0.07}$ & \color{blue}64$\pm$5 & \color{blue}0.98 \\ 
        \color{blue}AEGIS002 & \color{blue}2243 & \color{blue}4295 & \color{blue}14h18m14.3s & \color{blue}52\degr49\arcmin13.1\arcsec & \color{blue}2.21 & \color{blue}J0390 & \color{blue}- & \color{blue}- & \color{blue}- & \color{blue}- & \color{blue}- & \color{blue}43.14$^{ + 0.17}_{ - 0.29}$ & \color{blue}67$\pm$19 & \color{blue}0.89 \\ 
        \color{blue}AEGIS002 & \color{blue}2243 & \color{blue}7958 & \color{blue}14h20m16.8s & \color{blue}52\degr42\arcmin2.8\arcsec & \color{blue}2.21 & \color{blue}J0390 & \color{blue}- & \color{blue}- & \color{blue}- & \color{blue}- & \color{blue}- & \color{blue}43.29$^{ + 0.13}_{ - 0.18}$ & \color{blue}150$\pm$29 & \color{blue}1.00 \\ 
        \color{blue}AEGIS002 & \color{blue}2243 & \color{blue}7718 & \color{blue}14h19m29.0s & \color{blue}52\degr49\arcmin59.3\arcsec & \color{blue}2.29 & \color{blue}J0400 & \color{blue}- & \color{blue}- & \color{blue}agn & \color{blue}2.31 & \color{blue}QSO LAE & \color{blue}43.60$^{ + 0.10}_{ - 0.13}$ & \color{blue}162$\pm$23 & \color{blue}1.00 \\ 
        AEGIS002 & 2243 & 15119 & 14h20m16.0s & 52\degr51\arcmin1.3\arcsec & 2.29 & J0400 & QSO & 1.62 & - & - & Cont. QSO & 43.78$^{ + 0.07}_{ - 0.09}$ & 96$\pm$10 & 0.00 \\ 
        \color{blue}AEGIS002 & \color{blue}2243 & \color{blue}12352 & \color{blue}14h19m52.8s & \color{blue}53\degr2\arcmin4.2\arcsec & \color{blue}2.70 & \color{blue}J0450 & \color{blue}QSO & \color{blue}1.90 & \color{blue}agn & \color{blue}1.90 & \color{blue}- & \color{blue}44.29$^{ + 0.06}_{ - 0.08}$ & \color{blue}52$\pm$4 & \color{blue}1.00 \\ 
        \color{blue}AEGIS002 & \color{blue}2243 & \color{blue}14558 & \color{blue}14h19m26.8s & \color{blue}53\degr0\arcmin3.5\arcsec & \color{blue}2.95 & \color{blue}J0480 & \color{blue}- & \color{blue}- & \color{blue}- & \color{blue}- & \color{blue}- & \color{blue}43.75$^{ + 0.13}_{ - 0.19}$ & \color{blue}59$\pm$13 & \color{blue}0.97 \\ 
        \color{blue}AEGIS002 & \color{blue}2243 & \color{blue}4755 & \color{blue}14h19m23.1s & \color{blue}52\degr40\arcmin13.4\arcsec & \color{blue}3.03 & \color{blue}J0490 & \color{blue}- & \color{blue}- & \color{blue}- & \color{blue}- & \color{blue}- & \color{blue}43.85$^{ + 0.12}_{ - 0.17}$ & \color{blue}50$\pm$10 & \color{blue}0.99 \\ 
        \color{blue}AEGIS002 & \color{blue}2243 & \color{blue}5404 & \color{blue}14h18m13.4s & \color{blue}52\degr52\arcmin40.0\arcsec & \color{blue}3.28 & \color{blue}J0520 & \color{blue}QSO & \color{blue}3.29 & \color{blue}- & \color{blue}- & \color{blue}QSO LAE & \color{blue}44.40$^{ + 0.06}_{ - 0.07}$ & \color{blue}140$\pm$9 & \color{blue}1.00 \\ 
        \color{blue}AEGIS002 & \color{blue}2243 & \color{blue}14873 & \color{blue}14h19m35.6s & \color{blue}52\degr57\arcmin10.7\arcsec & \color{blue}3.28 & \color{blue}J0520 & \color{blue}QSO & \color{blue}3.22 & \color{blue}agn & \color{blue}3.22 & \color{blue}QSO LAE & \color{blue}44.54$^{ + 0.05}_{ - 0.05}$ & \color{blue}83$\pm$4 & \color{blue}1.00 \\ 
        \color{blue}AEGIS002 & \color{blue}2243 & \color{blue}673 & \color{blue}14h19m14.3s & \color{blue}52\degr30\arcmin50.3\arcsec & \color{blue}3.43 & \color{blue}J0540 & \color{blue}- & \color{blue}- & \color{blue}- & \color{blue}- & \color{blue}- & \color{blue}43.86$^{ + 0.10}_{ - 0.12}$ & \color{blue}134$\pm$20 & \color{blue}0.95 \\ 
        \color{blue}AEGIS002 & \color{blue}2243 & \color{blue}5769 & \color{blue}14h17m56.9s & \color{blue}52\degr56\arcmin15.4\arcsec & \color{blue}3.43 & \color{blue}J0540 & \color{blue}- & \color{blue}- & \color{blue}- & \color{blue}- & \color{blue}- & \color{blue}43.75$^{ + 0.12}_{ - 0.16}$ & \color{blue}63$\pm$11 & \color{blue}0.90 \\ 
        \color{blue}AEGIS002 & \color{blue}2243 & \color{blue}15610 & \color{blue}14h19m43.6s & \color{blue}52\degr54\arcmin31.3\arcsec & \color{blue}3.43 & \color{blue}J0540 & \color{blue}QSO & \color{blue}3.44 & \color{blue}agn & \color{blue}3.44 & \color{blue}QSO LAE & \color{blue}44.52$^{ + 0.04}_{ - 0.04}$ & \color{blue}191$\pm$6 & \color{blue}0.99 \\ 
        AEGIS003 & 2406 & 6731 & 14h22m5.7s & 53\degr10\arcmin54.1\arcsec & 2.21 & J0390 & QSO & 1.45 & - & - & Cont. QSO & 43.30$^{ + 0.13}_{ - 0.20}$ & 39$\pm$8 & 0.00 \\ 
        \color{blue}AEGIS003 & \color{blue}2406 & \color{blue}1224 & \color{blue}14h21m18.0s & \color{blue}52\degr53\arcmin46.0\arcsec & \color{blue}2.29 & \color{blue}J0400 & \color{blue}QSO & \color{blue}2.31 & \color{blue}- & \color{blue}- & \color{blue}QSO LAE & \color{blue}44.36$^{ + 0.05}_{ - 0.05}$ & \color{blue}109$\pm$4 & \color{blue}1.00 \\ 
        \color{blue}AEGIS003 & \color{blue}2406 & \color{blue}1482 & \color{blue}14h20m6.0s & \color{blue}53\degr5\arcmin9.4\arcsec & \color{blue}2.46 & \color{blue}J0420 & \color{blue}- & \color{blue}- & \color{blue}- & \color{blue}- & \color{blue}- & \color{blue}43.69$^{ + 0.14}_{ - 0.21}$ & \color{blue}67$\pm$16 & \color{blue}0.99 \\ 
        AEGIS003 & 2406 & 7049 & 14h20m43.7s & 53\degr22\arcmin6.3\arcsec & 2.46 & J0420 & QSO & 1.72 & - & - & Cont. QSO & 44.45$^{ + 0.05}_{ - 0.06}$ & 54$\pm$2 & 0.00 \\ 
        \color{blue}AEGIS003 & \color{blue}2406 & \color{blue}11608 & \color{blue}14h23m6.1s & \color{blue}53\degr15\arcmin29.0\arcsec & \color{blue}2.46 & \color{blue}J0420 & \color{blue}QSO & \color{blue}2.46 & \color{blue}- & \color{blue}- & \color{blue}QSO LAE & \color{blue}45.00$^{ + 0.04}_{ - 0.05}$ & \color{blue}71$\pm$1 & \color{blue}1.00 \\ 
        \color{blue}AEGIS003 & \color{blue}2406 & \color{blue}6169 & \color{blue}14h20m46.4s & \color{blue}53\degr12\arcmin24.5\arcsec & \color{blue}2.54 & \color{blue}J0430 & \color{blue}- & \color{blue}- & \color{blue}- & \color{blue}- & \color{blue}QSO LAE & \color{blue}43.86$^{ + 0.08}_{ - 0.10}$ & \color{blue}90$\pm$11 & \color{blue}0.99 \\ 
        \color{blue}AEGIS003 & \color{blue}2406 & \color{blue}14091 & \color{blue}14h21m32.5s & \color{blue}53\degr22\arcmin44.2\arcsec & \color{blue}2.54 & \color{blue}J0430 & \color{blue}- & \color{blue}- & \color{blue}- & \color{blue}- & \color{blue}- & \color{blue}43.56$^{ + 0.13}_{ - 0.20}$ & \color{blue}81$\pm$17 & \color{blue}0.99 \\ 
        \color{blue}AEGIS003 & \color{blue}2406 & \color{blue}4342 & \color{blue}14h20m10.5s & \color{blue}53\degr12\arcmin23.9\arcsec & \color{blue}2.62 & \color{blue}J0440 & \color{blue}QSO & \color{blue}2.59 & \color{blue}- & \color{blue}- & \color{blue}QSO LAE & \color{blue}44.31$^{ + 0.04}_{ - 0.05}$ & \color{blue}56$\pm$2 & \color{blue}0.99 \\ 
        AEGIS003 & 2406 & 4658 & 14h20m25.6s & 53\degr11\arcmin5.2\arcsec & 2.62 & J0440 & QSO & 1.85 & - & - & Cont. QSO & 43.50$^{ + 0.12}_{ - 0.17}$ & 62$\pm$11 & 0.00 \\ 
        \color{blue}AEGIS003 & \color{blue}2406 & \color{blue}3657 & \color{blue}14h20m18.8s & \color{blue}53\degr9\arcmin11.3\arcsec & \color{blue}2.79 & \color{blue}J0460 & \color{blue}- & \color{blue}- & \color{blue}- & \color{blue}- & \color{blue}QSO LAE & \color{blue}43.74$^{ + 0.07}_{ - 0.08}$ & \color{blue}114$\pm$11 & \color{blue}0.92 \\ 
        \color{blue}AEGIS003 & \color{blue}2406 & \color{blue}8977 & \color{blue}14h21m13.3s & \color{blue}53\degr12\arcmin18.6\arcsec & \color{blue}2.79 & \color{blue}J0460 & \color{blue}QSO & \color{blue}1.95 & \color{blue}- & \color{blue}- & \color{blue}- & \color{blue}44.07$^{ + 0.04}_{ - 0.05}$ & \color{blue}55$\pm$3 & \color{blue}0.97 \\ 
        \color{blue}AEGIS003 & \color{blue}2406 & \color{blue}11219 & \color{blue}14h21m60.0s & \color{blue}53\degr26\arcmin9.2\arcsec & \color{blue}2.79 & \color{blue}J0460 & \color{blue}QSO & \color{blue}2.73 & \color{blue}- & \color{blue}- & \color{blue}QSO LAE & \color{blue}44.08$^{ + 0.05}_{ - 0.05}$ & \color{blue}47$\pm$3 & \color{blue}0.99 \\ 
        \color{blue}AEGIS003 & \color{blue}2406 & \color{blue}14869 & \color{blue}14h21m36.5s & \color{blue}53\degr20\arcmin14.2\arcsec & \color{blue}2.87 & \color{blue}J0470 & \color{blue}QSO & \color{blue}2.02 & \color{blue}agn & \color{blue}2.02 & \color{blue}QSO LAE & \color{blue}43.89$^{ + 0.07}_{ - 0.09}$ & \color{blue}53$\pm$6 & \color{blue}0.95 \\ 
        \color{blue}AEGIS003 & \color{blue}2406 & \color{blue}4964 & \color{blue}14h21m54.8s & \color{blue}52\degr58\arcmin41.9\arcsec & \color{blue}2.95 & \color{blue}J0480 & \color{blue}- & \color{blue}- & \color{blue}- & \color{blue}- & \color{blue}- & \color{blue}43.84$^{ + 0.07}_{ - 0.09}$ & \color{blue}54$\pm$6 & \color{blue}0.91 \\ 
        \color{blue}AEGIS003 & \color{blue}2406 & \color{blue}12752 & \color{blue}14h21m47.1s & \color{blue}53\degr24\arcmin5.8\arcsec & \color{blue}3.03 & \color{blue}J0490 & \color{blue}QSO & \color{blue}3.04 & \color{blue}- & \color{blue}- & \color{blue}QSO LAE & \color{blue}44.40$^{ + 0.06}_{ - 0.06}$ & \color{blue}66$\pm$4 & \color{blue}0.99 \\ 
        AEGIS004 & 2470 & 8781 & 14h15m51.6s & 52\degr0\arcmin25.6\arcsec & 2.21 & J0390 & QSO & 1.51 & - & - & Cont. QSO & 44.48$^{ + 0.05}_{ - 0.05}$ & 41$\pm$1 & 0.00 \\ 
        \color{blue}AEGIS004 & \color{blue}2470 & \color{blue}2363 & \color{blue}14h15m11.8s & \color{blue}51\degr52\arcmin55.8\arcsec & \color{blue}2.29 & \color{blue}J0400 & \color{blue}QSO & \color{blue}2.31 & \color{blue}- & \color{blue}- & \color{blue}QSO LAE & \color{blue}44.37$^{ + 0.05}_{ - 0.05}$ & \color{blue}57$\pm$2 & \color{blue}0.98 \\ 
        \color{blue}AEGIS004 & \color{blue}2470 & \color{blue}4455 & \color{blue}14h13m47.9s & \color{blue}52\degr12\arcmin5.0\arcsec & \color{blue}2.38 & \color{blue}J0410 & \color{blue}QSO & \color{blue}2.35 & \color{blue}- & \color{blue}- & \color{blue}QSO LAE & \color{blue}44.01$^{ + 0.07}_{ - 0.09}$ & \color{blue}97$\pm$11 & \color{blue}1.00 \\ 
        AEGIS004 & 2470 & 9749 & 14h13m47.7s & 52\degr16\arcmin46.3\arcsec & 2.38 & J0410 & QSO & 1.67 & - & - & Cont. QSO & 43.64$^{ + 0.16}_{ - 0.25}$ & 45$\pm$12 & 0.00 \\ 
        \color{blue}AEGIS004 & \color{blue}2470 & \color{blue}13007 & \color{blue}14h14m59.3s & \color{blue}52\degr24\arcmin25.0\arcsec & \color{blue}2.38 & \color{blue}J0410 & \color{blue}- & \color{blue}- & \color{blue}- & \color{blue}- & \color{blue}QSO LAE & \color{blue}43.77$^{ + 0.11}_{ - 0.15}$ & \color{blue}127$\pm$23 & \color{blue}0.99 \\ 
        \color{blue}AEGIS004 & \color{blue}2470 & \color{blue}3723 & \color{blue}14h14m28.2s & \color{blue}52\degr3\arcmin47.2\arcsec & \color{blue}2.70 & \color{blue}J0450 & \color{blue}QSO & \color{blue}2.69 & \color{blue}- & \color{blue}- & \color{blue}QSO LAE & \color{blue}43.97$^{ + 0.10}_{ - 0.13}$ & \color{blue}87$\pm$15 & \color{blue}0.99 \\ 
        AEGIS004 & 2470 & 13064 & 14h15m56.9s & 52\degr16\arcmin7.2\arcsec & 2.70 & J0450 & QSO & 1.39 & - & - & Cont. QSO & 43.87$^{ + 0.12}_{ - 0.17}$ & 47$\pm$10 & 0.00 \\ 
        \color{blue}AEGIS004 & \color{blue}2470 & \color{blue}12623 & \color{blue}14h14m50.9s & \color{blue}52\degr26\arcmin40.4\arcsec & \color{blue}2.87 & \color{blue}J0470 & \color{blue}- & \color{blue}- & \color{blue}- & \color{blue}- & \color{blue}QSO LAE & \color{blue}43.52$^{ + 0.12}_{ - 0.17}$ & \color{blue}44$\pm$8 & \color{blue}0.89 \\ 
        \color{blue}AEGIS004 & \color{blue}2470 & \color{blue}15095 & \color{blue}14h16m17.8s & \color{blue}52\degr7\arcmin18.3\arcsec & \color{blue}2.95 & \color{blue}J0480 & \color{blue}- & \color{blue}- & \color{blue}- & \color{blue}- & \color{blue}- & \color{blue}43.77$^{ + 0.09}_{ - 0.11}$ & \color{blue}49$\pm$7 & \color{blue}0.88 \\ 
        \color{blue}AEGIS004 & \color{blue}2470 & \color{blue}6481 & \color{blue}14h15m42.7s & \color{blue}52\degr9\arcmin27.2\arcsec & \color{blue}3.19 & \color{blue}J0510 & \color{blue}- & \color{blue}- & \color{blue}- & \color{blue}- & \color{blue}QSO LAE & \color{blue}44.01$^{ + 0.05}_{ - 0.06}$ & \color{blue}257$\pm$17 & \color{blue}0.98 \\ 
        \color{blue}J-NEP & \color{blue}2520 & \color{blue}3222 & \color{blue}17h24m40.4s & \color{blue}65\degr35\arcmin0.2\arcsec & \color{blue}2.11 & \color{blue}J0378 & \color{blue}- & \color{blue}- & \color{blue}- & \color{blue}- & \color{blue}- & \color{blue}43.58$^{ + 0.16}_{ - 0.27}$ & \color{blue}65$\pm$17 & \color{blue}0.94 \\ 
        J-NEP & 2520 & 6815 & 17h23m43.8s & 65\degr40\arcmin32.8\arcsec & 2.11 & J0378 & - & - & - & - & Cont. QSO & 43.56$^{ + 0.17}_{ - 0.29}$ & 60$\pm$17 & 0.00 \\ 
        J-NEP & 2520 & 9243 & 17h20m45.2s & 65\degr43\arcmin42.0\arcsec & 2.11 & J0378 & - & - & - & - & Cont. QSO & 43.66$^{ + 0.14}_{ - 0.22}$ & 50$\pm$12 & 0.00 \\ 
        J-NEP & 2520 & 12771 & 17h23m14.1s & 65\degr47\arcmin46.2\arcsec & 2.11 & J0378 & QSO & 1.44 & - & - & Cont. QSO & 45.09$^{ + 0.05}_{ - 0.06}$ & 58$\pm$1 & 0.00 \\ 
        J-NEP & 2520 & 22098 & 17h24m13.6s & 65\degr59\arcmin55.7\arcsec & 2.11 & J0378 & - & - & - & - & Cont. QSO & 44.29$^{ + 0.06}_{ - 0.07}$ & 37$\pm$2 & 0.00 \\ 
        \color{blue}J-NEP & \color{blue}2520 & \color{blue}10356 & \color{blue}17h20m59.5s & \color{blue}65\degr44\arcmin57.8\arcsec & \color{blue}2.29 & \color{blue}J0400 & \color{blue}- & \color{blue}- & \color{blue}- & \color{blue}- & \color{blue}QSO LAE & \color{blue}43.72$^{ + 0.07}_{ - 0.09}$ & \color{blue}109$\pm$10 & \color{blue}0.99 \\ 
        \color{blue}J-NEP & \color{blue}2520 & \color{blue}12539 & \color{blue}17h22m8.9s & \color{blue}65\degr47\arcmin43.1\arcsec & \color{blue}2.29 & \color{blue}J0400 & \color{blue}- & \color{blue}- & \color{blue}- & \color{blue}- & \color{blue}QSO LAE & \color{blue}44.09$^{ + 0.05}_{ - 0.06}$ & \color{blue}100$\pm$4 & \color{blue}1.00 \\ 
        \color{blue}J-NEP & \color{blue}2520 & \color{blue}15690 & \color{blue}17h24m13.3s & \color{blue}65\degr55\arcmin28.8\arcsec & \color{blue}2.29 & \color{blue}J0400 & \color{blue}- & \color{blue}- & \color{blue}- & \color{blue}- & \color{blue}- & \color{blue}43.30$^{ + 0.15}_{ - 0.23}$ & \color{blue}39$\pm$9 & \color{blue}0.93 \\ 
        \color{blue}J-NEP & \color{blue}2520 & \color{blue}2518 & \color{blue}17h24m19.5s & \color{blue}65\degr33\arcmin59.7\arcsec & \color{blue}2.38 & \color{blue}J0410 & \color{blue}- & \color{blue}- & \color{blue}- & \color{blue}- & \color{blue}- & \color{blue}43.55$^{ + 0.08}_{ - 0.10}$ & \color{blue}66$\pm$7 & \color{blue}0.98 \\ 
        \color{blue}J-NEP & \color{blue}2520 & \color{blue}5302 & \color{blue}17h22m51.4s & \color{blue}65\degr38\arcmin20.5\arcsec & \color{blue}2.38 & \color{blue}J0410 & \color{blue}- & \color{blue}- & \color{blue}- & \color{blue}- & \color{blue}- & \color{blue}43.33$^{ + 0.13}_{ - 0.19}$ & \color{blue}71$\pm$14 & \color{blue}0.97 \\ 
        \color{blue}J-NEP & \color{blue}2520 & \color{blue}10708 & \color{blue}17h20m33.0s & \color{blue}65\degr45\arcmin18.7\arcsec & \color{blue}2.46 & \color{blue}J0420 & \color{blue}- & \color{blue}- & \color{blue}- & \color{blue}- & \color{blue}- & \color{blue}44.13$^{ + 0.05}_{ - 0.06}$ & \color{blue}42$\pm$2 & \color{blue}0.98 \\ 
        \color{blue}J-NEP & \color{blue}2520 & \color{blue}6012 & \color{blue}17h22m28.2s & \color{blue}65\degr39\arcmin23.5\arcsec & \color{blue}2.62 & \color{blue}J0440 & \color{blue}- & \color{blue}- & \color{blue}- & \color{blue}- & \color{blue}- & \color{blue}43.22$^{ + 0.15}_{ - 0.23}$ & \color{blue}43$\pm$10 & \color{blue}0.93 \\ 
        \color{blue}J-NEP & \color{blue}2520 & \color{blue}12726 & \color{blue}17h24m38.0s & \color{blue}65\degr47\arcmin48.6\arcsec & \color{blue}2.62 & \color{blue}J0440 & \color{blue}- & \color{blue}- & \color{blue}- & \color{blue}- & \color{blue}- & \color{blue}43.23$^{ + 0.15}_{ - 0.23}$ & \color{blue}61$\pm$14 & \color{blue}0.96 \\ 
        \color{blue}J-NEP & \color{blue}2520 & \color{blue}19539 & \color{blue}17h23m9.4s & \color{blue}65\degr50\arcmin13.7\arcsec & \color{blue}2.62 & \color{blue}J0440 & \color{blue}- & \color{blue}- & \color{blue}- & \color{blue}- & \color{blue}- & \color{blue}43.18$^{ + 0.16}_{ - 0.25}$ & \color{blue}60$\pm$16 & \color{blue}0.91 \\ 
        \color{blue}J-NEP & \color{blue}2520 & \color{blue}1652 & \color{blue}17h21m13.3s & \color{blue}65\degr32\arcmin45.2\arcsec & \color{blue}2.70 & \color{blue}J0450 & \color{blue}- & \color{blue}- & \color{blue}- & \color{blue}- & \color{blue}QSO LAE & \color{blue}43.85$^{ + 0.09}_{ - 0.12}$ & \color{blue}67$\pm$10 & \color{blue}0.99 \\ 
        \color{blue}J-NEP & \color{blue}2520 & \color{blue}4903 & \color{blue}17h21m22.6s & \color{blue}65\degr37\arcmin45.3\arcsec & \color{blue}2.70 & \color{blue}J0450 & \color{blue}- & \color{blue}- & \color{blue}- & \color{blue}- & \color{blue}- & \color{blue}44.22$^{ + 0.05}_{ - 0.06}$ & \color{blue}57$\pm$3 & \color{blue}0.99 \\ 
        \color{blue}J-NEP & \color{blue}2520 & \color{blue}5247 & \color{blue}17h23m26.1s & \color{blue}65\degr38\arcmin16.3\arcsec & \color{blue}2.70 & \color{blue}J0450 & \color{blue}- & \color{blue}- & \color{blue}- & \color{blue}- & \color{blue}- & \color{blue}44.20$^{ + 0.05}_{ - 0.06}$ & \color{blue}60$\pm$4 & \color{blue}0.99 \\ 
        \color{blue}J-NEP & \color{blue}2520 & \color{blue}6520 & \color{blue}17h24m11.7s & \color{blue}65\degr40\arcmin4.7\arcsec & \color{blue}2.70 & \color{blue}J0450 & \color{blue}- & \color{blue}- & \color{blue}- & \color{blue}- & \color{blue}- & \color{blue}43.66$^{ + 0.12}_{ - 0.17}$ & \color{blue}85$\pm$16 & \color{blue}0.99 \\ 
        \color{blue}J-NEP & \color{blue}2520 & \color{blue}6636 & \color{blue}17h22m39.9s & \color{blue}65\degr40\arcmin17.1\arcsec & \color{blue}2.87 & \color{blue}J0470 & \color{blue}- & \color{blue}- & \color{blue}- & \color{blue}- & \color{blue}QSO LAE & \color{blue}43.55$^{ + 0.10}_{ - 0.14}$ & \color{blue}48$\pm$7 & \color{blue}0.92 \\ 
        \color{blue}J-NEP & \color{blue}2520 & \color{blue}14925 & \color{blue}17h21m10.7s & \color{blue}65\degr56\arcmin27.4\arcsec & \color{blue}2.87 & \color{blue}J0470 & \color{blue}- & \color{blue}- & \color{blue}- & \color{blue}- & \color{blue}QSO LAE & \color{blue}44.48$^{ + 0.04}_{ - 0.04}$ & \color{blue}43$\pm$1 & \color{blue}0.98 \\ 
        \color{blue}J-NEP & \color{blue}2520 & \color{blue}8395 & \color{blue}17h23m55.8s & \color{blue}65\degr42\arcmin46.0\arcsec & \color{blue}2.95 & \color{blue}J0480 & \color{blue}- & \color{blue}- & \color{blue}- & \color{blue}- & \color{blue}QSO LAE & \color{blue}43.76$^{ + 0.06}_{ - 0.08}$ & \color{blue}203$\pm$18 & \color{blue}0.86 \\ 
        \color{blue}J-NEP & \color{blue}2520 & \color{blue}14697 & \color{blue}17h22m17.8s & \color{blue}66\degr0\arcmin3.9\arcsec & \color{blue}3.60 & \color{blue}J0560 & \color{blue}- & \color{blue}- & \color{blue}- & \color{blue}- & \color{blue}- & \color{blue}44.08$^{ + 0.08}_{ - 0.09}$ & \color{blue}127$\pm$13 & \color{blue}1.00 \\ 
\bottomrule
\caption{Catalog of sources selected as LAEs.}
\label{tab:selected_catalog}
\end{longtable}

\end{landscape}
}

\twocolumn

\section{2D purity and number count corrections}\label{app:2dmaps}

\begin{figure*}
    \centering
    \begin{subfigure}{0.485\linewidth}
    \centering
    \includegraphics[width=\linewidth]{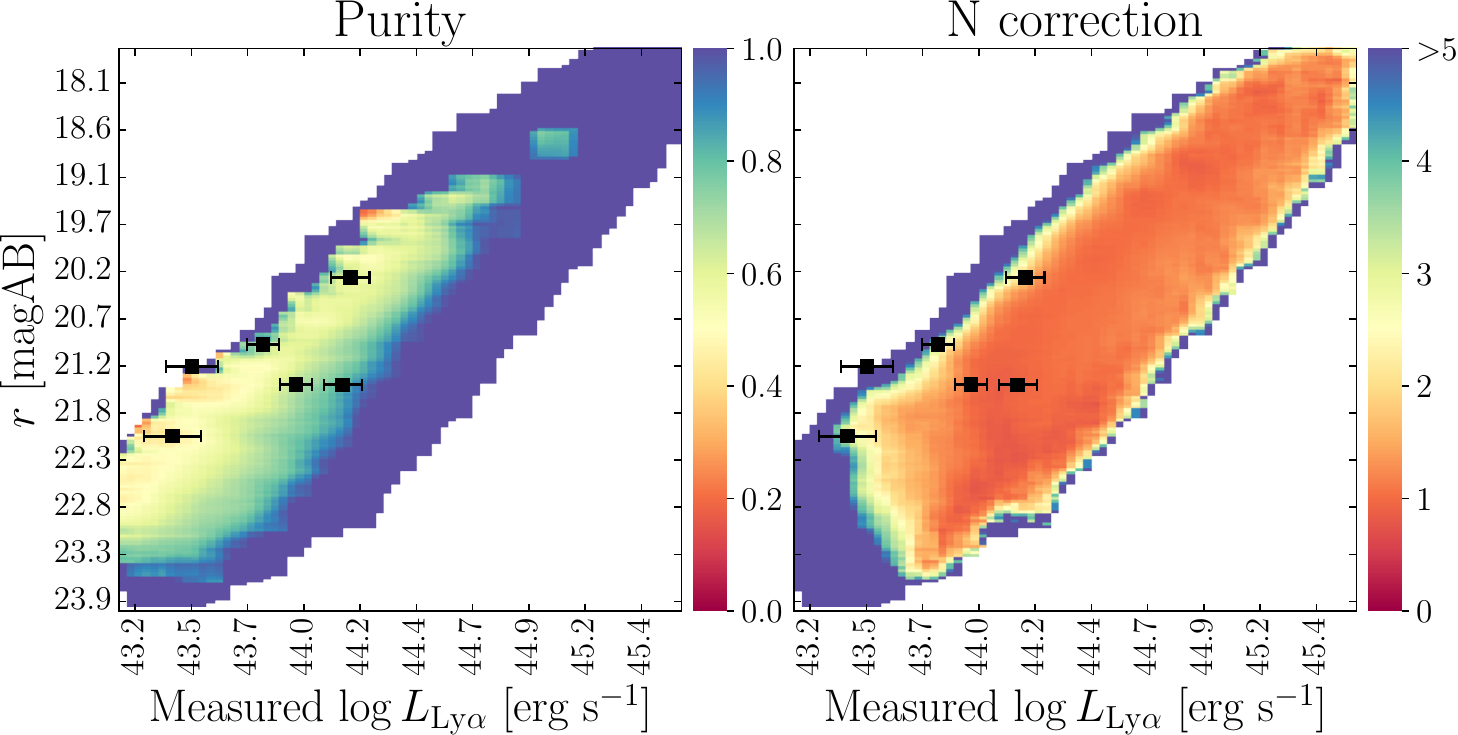}
    \caption{AEGIS001}
    \end{subfigure}
    \begin{subfigure}{0.485\linewidth}
    \centering
    \includegraphics[width=\linewidth]{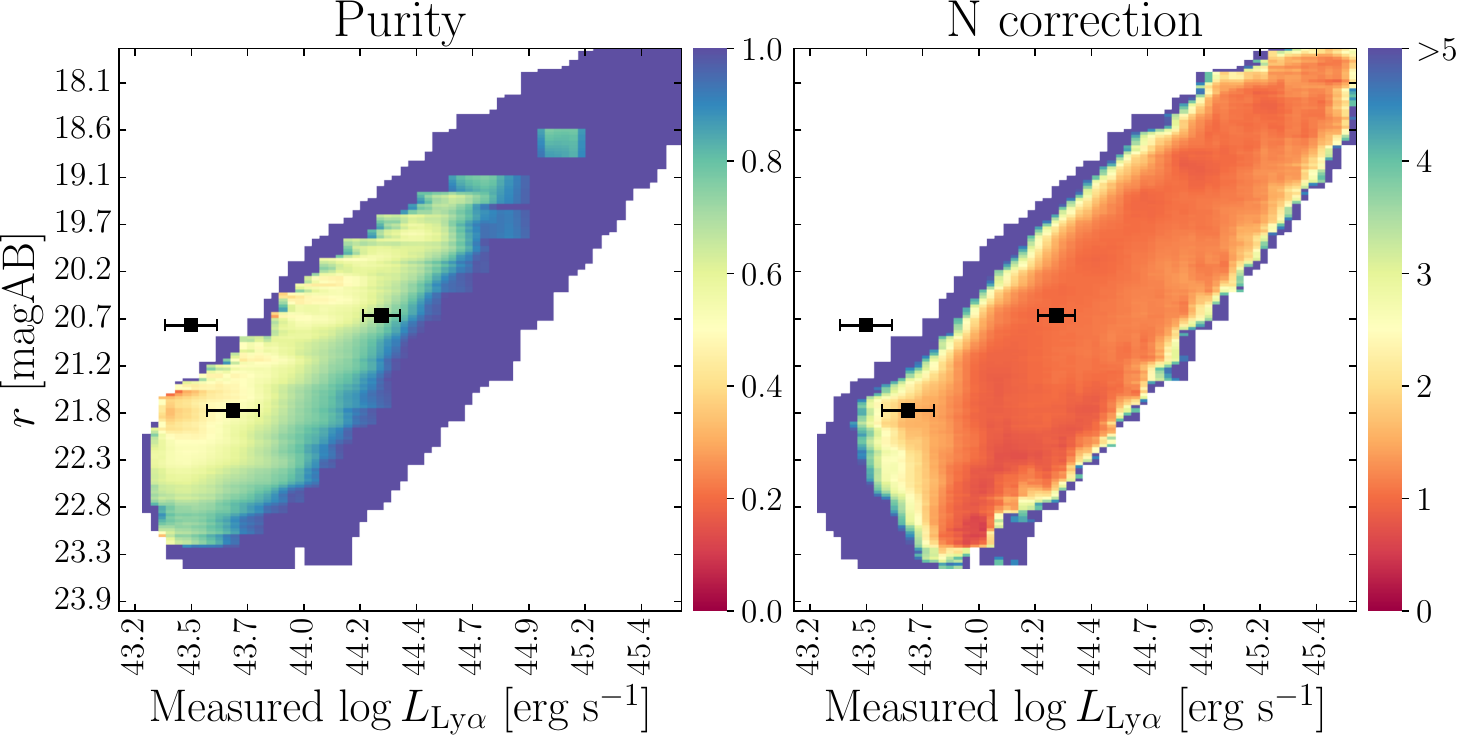}
    \caption{AEGIS002}
    \end{subfigure}
    \begin{subfigure}{0.485\linewidth}
    \centering
    \includegraphics[width=\linewidth]{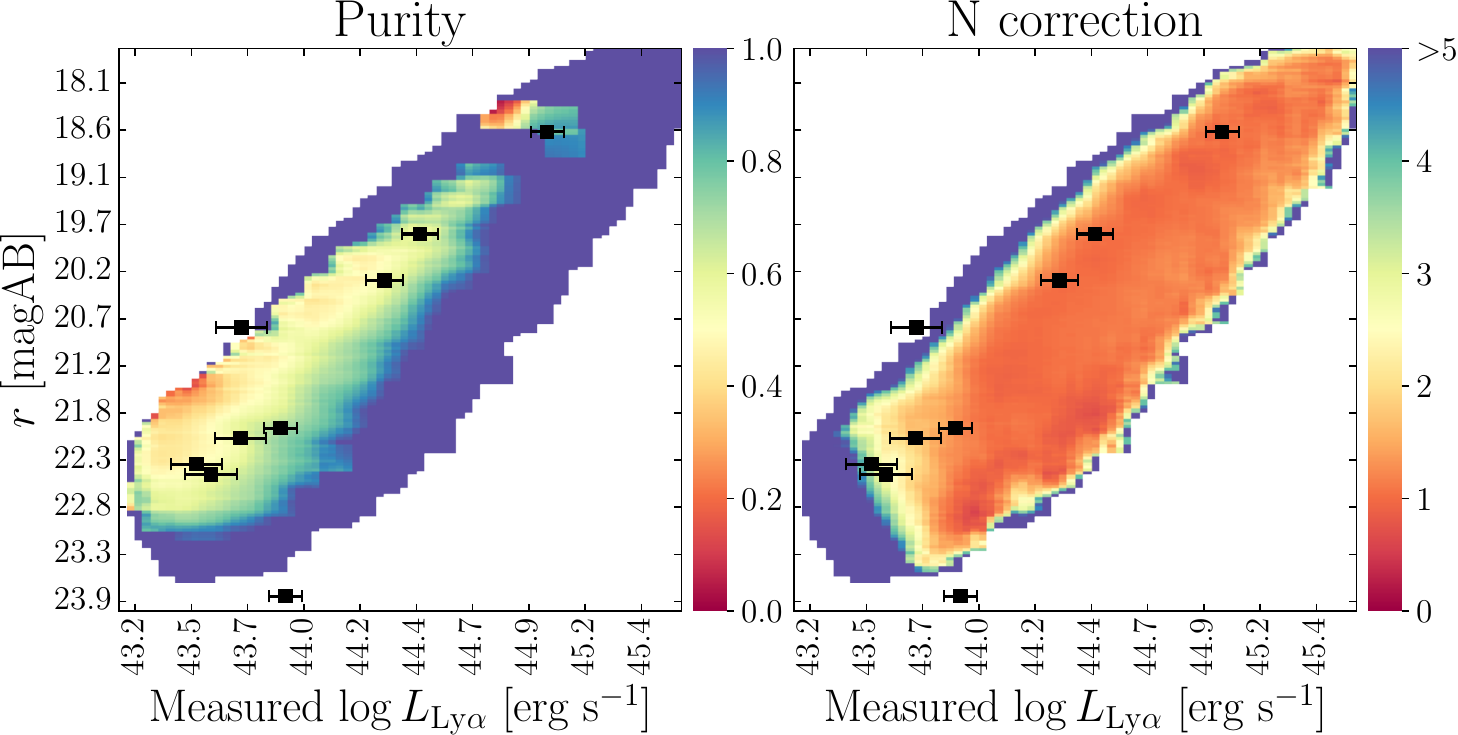}
    \caption{AEGIS003}
    \end{subfigure}
    \begin{subfigure}{0.485\linewidth}
    \centering
    \includegraphics[width=\linewidth]{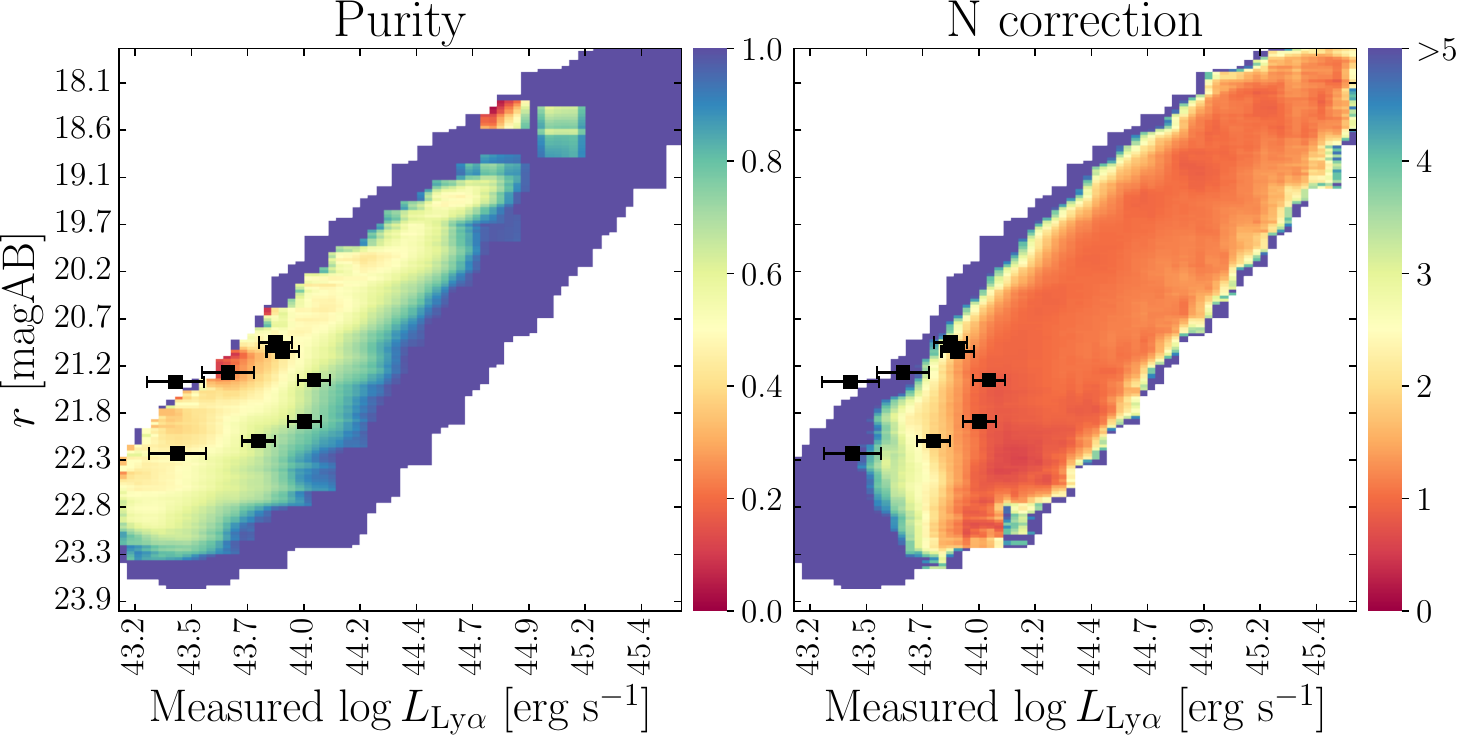}
    \caption{AEGIS004}
    \end{subfigure}
    \begin{subfigure}{0.485\linewidth}
    \centering
    \includegraphics[width=\linewidth]{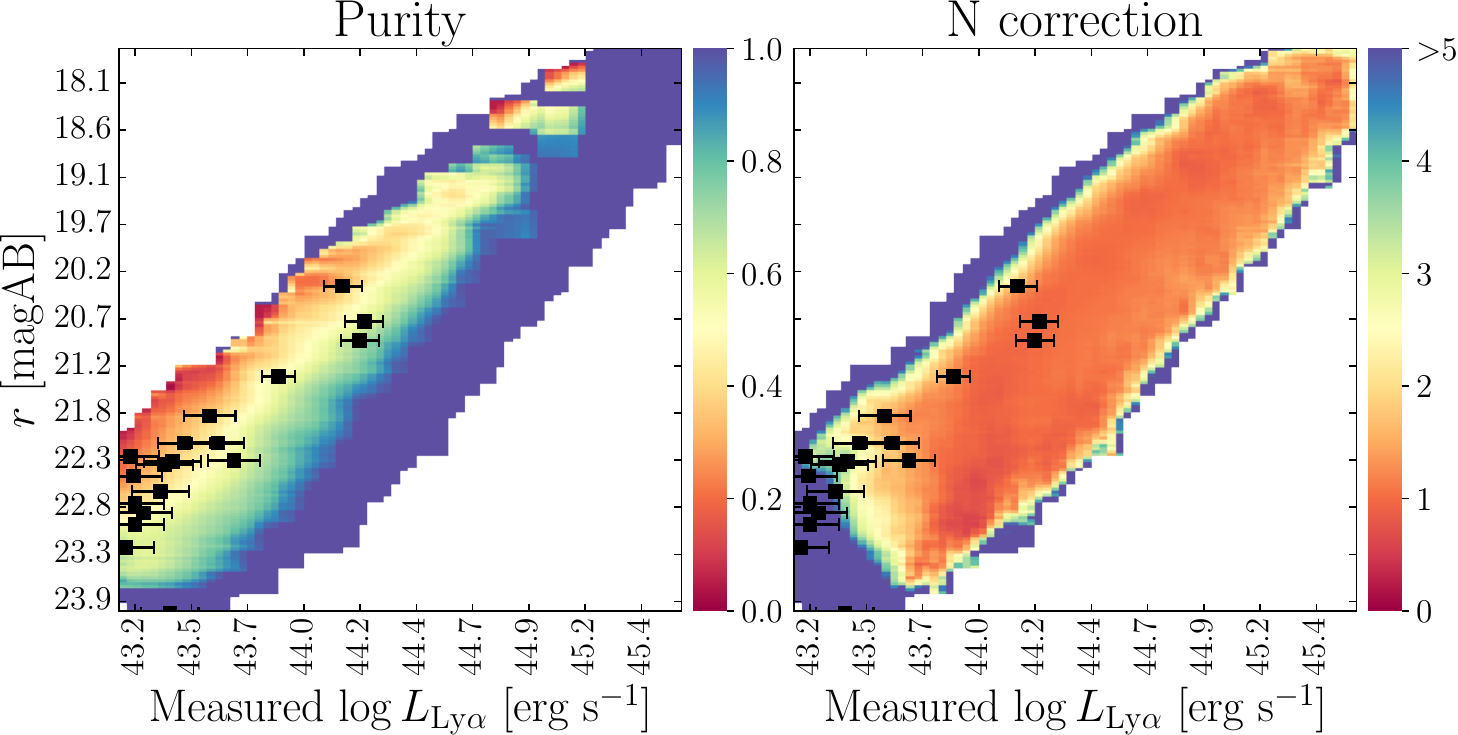}
    \caption{J-NEP}
    \end{subfigure}
        
    \caption{2D maps of the purity and number count correction for the four miniJPAS fields and J-NEP in the redshift interval $z=2.3$--$2.8$. The maps are computed for every field in miniJPAS and J-NEP and every interval of redshift used in this work, only one $z$ interval is shown for brevity. The black squares and error bars represent the LAE candidates of each field in the chosen interval of redshift.}
    \label{fig:2dpuricomp}
\end{figure*}

In Fig.~\ref{fig:2dpuricomp} we show example representations of the 2D correction maps described in Sect.~\ref{sec:2dpuricomp}, for $2.3<z<2.8$. These maps are computed for every redshift interval used in this work, for every field of miniJPAS and J-NEP. { The non-colored regions in Fig.~\ref{fig:2dpuricomp} correspond to combinations of parameters incompatible with the LAEs of our mock, thus the candidates laying in this areas of the 2D map are assigned $P^\mathrm{2D}=w^\mathrm{2D}=0$}.

\section{Purity calibration}\label{sec:prior_LF_choice}

\begin{figure}
    \centering
    \includegraphics[width=\linewidth]{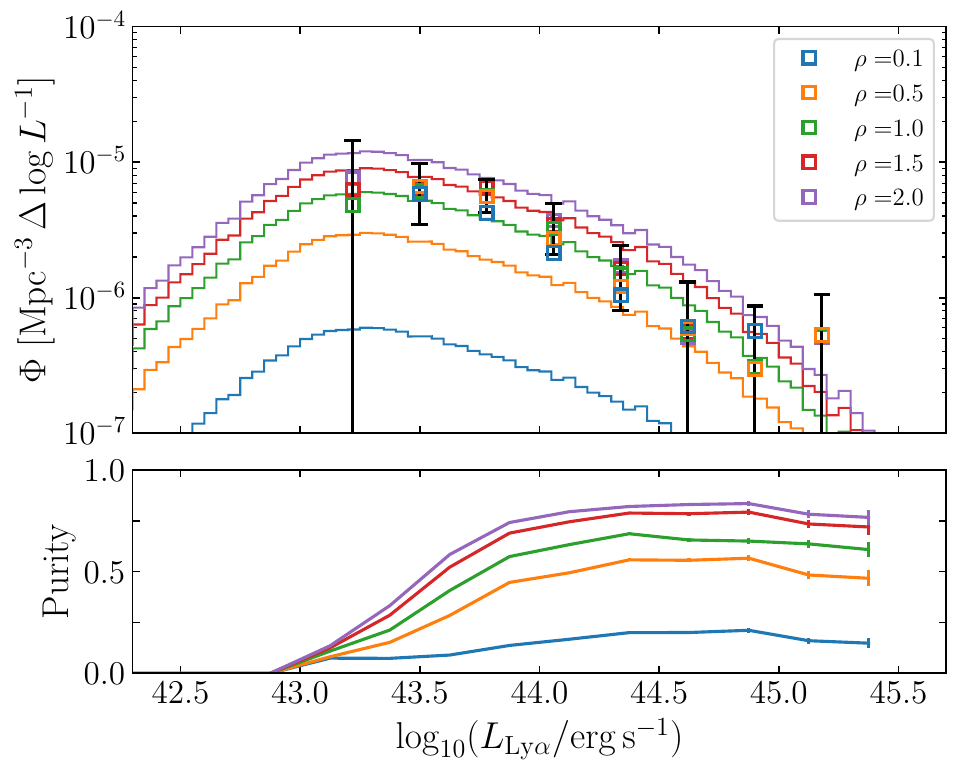}
    \caption{Effects of varying the sample purity on the estimated Ly$\alpha$ LF. {Top:} Colored squares show different realizations of our Ly$\alpha$ LF in the range $z=2.0-3.8$, obtained after scaling the number of true positives in the purity computation by a factor $\rho=0.1, 0.5, 1.0, 1.5, 2.0$. We also show the mock QSO LF for each $\rho$ (solid lines). We show the uncertainties of the LF with $\rho=1$. The variation of the LF is inside the $1\sigma$ errors for all the shown values of $\rho$. { Bottom: Estimated purity for the different values of $\rho$.}}
    \label{fig:rho_LF}
\end{figure}

As introduced in Sect. \ref{sec:mocks}, the corrections of our LF are likely to be biased by the underlying luminosity distribution we impose to our mock data. The number count correction estimate does not depend on the parameter space distribution of the mocks, as it is only a measure on how likely is to select a source as a function of its intrinsic $L_{\mathrm{Ly}\alpha}$ and EW. Nonetheless, the purity estimate is strongly dependent on the relative abundances of the objects from the target population in relation to that of the contaminants.

To check the robustness of our corrections in the QSO regime, we alter the purity estimate by introducing a factor $\rho$ to the true positive count. We recalculate the 2D purity as
\begin{equation}
    p^\mathrm{2D}=\frac{\rho\cdot\text{TP}}{\rho\cdot\text{TP} + \text{FP}} ,
\end{equation}
and compute the Ly$\alpha$ LF of miniJPAS\&J-NEP, using this new definition of the purity, for different values of $\rho$. Since the effect of increasing/decreasing the value of $\rho$ is equivalent to increasing/decreasing the mock QSO { number density}, this procedure tests the effect of under/overestimating the purity of our selected sample. We estimate the Ly$\alpha$ LF for different values of $\rho$ using the whole selected sample of 127 candidates (see Sec. \ref{sec:LAEs_cat}). The upper panel of Fig. \ref{fig:rho_LF} shows the recomputed LFs for different values of $\rho$ (as detailed by the plot label). Solid lines show the resulting $L_{\mathrm{Ly}\alpha}$ distribution of our QSO mock, for each $\rho$ value. The comparison between the different LF realizations (colored squares and error-bars) show that the change produced onto the LF by varying $\rho$ is small, especially when compared to the $1\sigma$ uncertainties for the $\rho=1$ realization (black error bars). In addition, the value of $\rho$ that produces an input LF comparable to the estimated is close to $\rho\sim 1$. With the eventual release of larger catalog of J-PAS, it will be possible to accurately calibrate the purity of the LAE candidate sample.

As already discussed in Sect. \ref{sec:qso_frac}, our LAE candidate sample is expected to be vastly dominated by AGN. Consequently, any change in the prior LF used for the construction of the SFG mock produces a minimum change onto our final results. Although we predict a non-zero contribution of SFG in our sample, the observational data is insufficient for calibrating the purity at the SFG regime ($\log_{10}(L_{\mathrm{Ly}\alpha} / \mathrm{erg\,s}^{-1})\lesssim 43.5$) with the current selection method.

\end{document}